\documentclass[traditabstract,longauth]{aa}
\pdfoutput=1
\usepackage{epsfig}
\usepackage{graphicx}
\usepackage{url}
\usepackage{color}
\usepackage[nonamebreak]{natbib}
\usepackage{ifthen}
\usepackage{amssymb,,amsmath}
\usepackage{longtable}
\usepackage{multirow}
\usepackage{breakurl}
\usepackage{txfonts}
\usepackage[usenames,svgnames]{xcolor}
\bibpunct{(}{)}{;}{a}{}{,}
\usepackage{ifthen}
\definecolor{linkcolor}{rgb}{0.6,0,0}
\definecolor{citecolor}{rgb}{0,0,0.75}
\definecolor{urlcolor}{rgb}{0.12,0.46,0.7}
\usepackage[breaklinks,colorlinks,urlcolor=urlcolor,linkcolor=linkcolor,citecolor=citecolor,pdfencoding=auto]{hyperref}
\hypersetup{linktocpage}

\def\lsim{\mathrel{\raise .4ex\hbox{\rlap{$<$}\lower 1.2ex\hbox{$\sim$}}}}
\def\gsim{\mathrel{\raise .4ex\hbox{\rlap{$>$}\lower 1.2ex\hbox{$\sim$}}}}
\def\arcm{\ifmmode {^{\scriptscriptstyle\prime}}
          \else $^{\scriptscriptstyle\prime}$\fi}

\def\WMAP{WMAP}

\def\setsymbol#1#2{\expandafter\def\csname #1\endcsname{#2}}
\def\getsymbol#1{\csname #1\endcsname}

\def\Planck{\textit{Planck}}

\def\allearlypapers{\nocite{planck2011-1.1, planck2011-1.3, planck2011-1.4, planck2011-1.5, planck2011-1.6, planck2011-1.7, planck2011-1.10, planck2011-1.10sup, planck2011-5.1a, planck2011-5.1b, planck2011-5.2a, planck2011-5.2b, planck2011-5.2c, planck2011-6.1, planck2011-6.2, planck2011-6.3a, planck2011-6.4a, planck2011-6.4b, planck2011-6.6, planck2011-7.0, planck2011-7.2, planck2011-7.3, planck2011-7.7a, planck2011-7.7b, planck2011-7.12, planck2011-7.13}}

\def\alltwentythirteenresultspapers{\nocite{planck2013-p01, planck2013-p02, planck2013-p02a, planck2013-p02d, planck2013-p02b, planck2013-p03, planck2013-p03c, planck2013-p03f, planck2013-p03d, planck2013-p03e, planck2013-p01a, planck2013-p06, planck2013-p03a, planck2013-pip88, planck2013-p08, planck2013-p11, planck2013-p12, planck2013-p13, planck2013-p14, planck2013-p15, planck2013-p05b, planck2013-p17, planck2013-p09, planck2013-p09a, planck2013-p20, planck2013-p19, planck2013-pipaberration, planck2013-p05, planck2013-p05a, planck2013-pip56, planck2013-p06b, planck2013-p01a}}

\def\alltwentyfifteenresultspapers{\nocite{planck2014-a01, planck2014-a03, planck2014-a04, planck2014-a05, planck2014-a06, planck2014-a07, planck2014-a08, planck2014-a09, planck2014-a11, planck2014-a12, planck2014-a13, planck2014-a14, planck2014-a15, planck2014-a16, planck2014-a17, planck2014-a18, planck2014-a19, planck2014-a20, planck2014-a22, planck2014-a24, planck2014-a26, planck2014-a28, planck2014-a29, planck2014-a30, planck2014-a31, planck2014-a35, planck2014-a36, planck2014-a37, planck2014-ES}}

\newbox\tablebox    \newdimen\tablewidth
\def\leaderfil{\leaders\hbox to 5pt{\hss.\hss}\hfil}
\def\endPlancktable{\tablewidth=\columnwidth 
    $$\hss\copy\tablebox\hss$$
    \vskip-\lastskip\vskip -2pt}
\def\endPlancktablewide{\tablewidth=\textwidth 
    $$\hss\copy\tablebox\hss$$
    \vskip-\lastskip\vskip -2pt}
\def\tablenote#1 #2\par{\begingroup \parindent=0.8em
    \abovedisplayshortskip=0pt\belowdisplayshortskip=0pt
    \noindent
    $$\hss\vbox{\hsize\tablewidth \hangindent=\parindent \hangafter=1 \noindent
    \hbox to \parindent{$^#1$\hss}\strut#2\strut\par}\hss$$
    \endgroup}
\def\doubleline{\vskip 3pt\hrule \vskip 1.5pt \hrule \vskip 5pt}

\def\L2{\ifmmode L_2\else $L_2$\fi}

\def\DeltaT{\ifmmode \Delta T\else $\Delta T$\fi}
\def\deltat{\ifmmode \Delta t\else $\Delta t$\fi}
\def\fknee{\ifmmode f_{\rm knee}\else $f_{\rm knee}$\fi}
\def\Fmax{\ifmmode F_{\rm max}\else $F_{\rm max}$\fi}
\def\solar{\ifmmode{\rm M}_{\mathord\odot}\else${\rm M}_{\mathord\odot}$\fi}
\def\Msolar{\ifmmode{\rm M}_{\mathord\odot}\else${\rm M}_{\mathord\odot}$\fi}
\def\Lsolar{\ifmmode{\rm L}_{\mathord\odot}\else${\rm L}_{\mathord\odot}$\fi}
\def\inv{\ifmmode^{-1}\else$^{-1}$\fi}
\def\mo{\ifmmode^{-1}\else$^{-1}$\fi}
\def\sup#1{\ifmmode ^{\rm #1}\else $^{\rm #1}$\fi}
\def\expo#1{\ifmmode \times 10^{#1}\else $\times 10^{#1}$\fi}
\def\,{\thinspace}
\def\lsim{\mathrel{\raise .4ex\hbox{\rlap{$<$}\lower 1.2ex\hbox{$\sim$}}}}
\def\gsim{\mathrel{\raise .4ex\hbox{\rlap{$>$}\lower 1.2ex\hbox{$\sim$}}}}

\def\simprop{\mathrel{\raise .4ex\hbox{\rlap{$\propto$}\lower 1.2ex\hbox{$\sim$}}}}
\def\deg{\ifmmode^\circ\else$^\circ$\fi}
\def\pdeg{\ifmmode $\setbox0=\hbox{$^{\circ}$}\rlap{\hskip.11\wd0 .}$^{\circ}
          \else \setbox0=\hbox{$^{\circ}$}\rlap{\hskip.11\wd0 .}$^{\circ}$\fi}
\def\arcs{\ifmmode {^{\scriptstyle\prime\prime}}
          \else $^{\scriptstyle\prime\prime}$\fi}
\def\arcm{\ifmmode {^{\scriptstyle\prime}}
          \else $^{\scriptstyle\prime}$\fi}
\newdimen\sa  \newdimen\sb
\def\parcs{\sa=.07em \sb=.03em
     \ifmmode \hbox{\rlap{.}}^{\scriptstyle\prime\kern -\sb\prime}\hbox{\kern -\sa}
     \else \rlap{.}$^{\scriptstyle\prime\kern -\sb\prime}$\kern -\sa\fi}
\def\parcm{\sa=.08em \sb=.03em
     \ifmmode \hbox{\rlap{.}\kern\sa}^{\scriptstyle\prime}\hbox{\kern-\sb}
     \else \rlap{.}\kern\sa$^{\scriptstyle\prime}$\kern-\sb\fi}
\def\ra[#1 #2 #3.#4]{#1\sup{h}#2\sup{m}#3\sup{s}\llap.#4}
\def\dec[#1 #2 #3.#4]{#1\deg#2\arcm#3\arcs\llap.#4}
\def\deco[#1 #2 #3]{#1\deg#2\arcm#3\arcs}
\def\rra[#1 #2]{#1\sup{h}#2\sup{m}}

\def\dots{\relax\ifmmode \ldots\else $\ldots$\fi}
\def\WHzsr{\ifmmode $W\,Hz\mo\,sr\mo$\else W\,Hz\mo\,sr\mo\fi}
\def\mHz{\ifmmode $\,mHz$\else \,mHz\fi}
\def\GHz{\ifmmode $\,GHz$\else \,GHz\fi}
\def\mKs{\ifmmode $\,mK\,s$^{1/2}\else \,mK\,s$^{1/2}$\fi}
\def\muKs{\ifmmode \,\mu$K\,s$^{1/2}\else \,$\mu$K\,s$^{1/2}$\fi}
\def\muKRJs{\ifmmode \,\mu$K$_{\rm RJ}$\,s$^{1/2}\else \,$\mu$K$_{\rm RJ}$\,s$^{1/2}$\fi}
\def\muKHz{\ifmmode \,\mu$K\,Hz$^{-1/2}\else \,$\mu$K\,Hz$^{-1/2}$\fi}
\def\MJysr{\ifmmode \,$MJy\,sr\mo$\else \,MJy\,sr\mo\fi}
\def\MJysrmK{\ifmmode \,$MJy\,sr\mo$\,mK$_{\rm CMB}\mo\else \,MJy\,sr\mo\,mK$_{\rm CMB}\mo$\fi}
\def\microns{\ifmmode \,\mu$m$\else \,$\mu$m\fi}

\def\muK{\ifmmode \,\mu$K$\else \,$\mu$\hbox{K}\fi}
\def\microK{\ifmmode \,\mu$K$\else \,$\mu$\hbox{K}\fi}
\def\muW{\ifmmode \,\mu$W$\else \,$\mu$\hbox{W}\fi}
\def\kms{\ifmmode $\,km\,s$^{-1}\else \,km\,s$^{-1}$\fi}
\def\kmsMpc{\ifmmode $\,\kms\,Mpc\mo$\else \,\kms\,Mpc\mo\fi}
\providecommand{\sorthelp}[1]{}

\setsymbol{LFI:center:frequency:70GHz:units}{$70.4$\,GHz}
\setsymbol{LFI:center:frequency:44GHz:units}{$44.1$\,GHz}
\setsymbol{LFI:center:frequency:30GHz:units}{$28.4$\,GHz}

\setsymbol{LFI:center:frequency:70GHz}{$70.4$}
\setsymbol{LFI:center:frequency:44GHz}{$44.1$}
\setsymbol{LFI:center:frequency:30GHz}{$28.4$}
\setsymbol{LFI:bandwidth:70GHz}{$14.90$}
\setsymbol{LFI:bandwidth:44GHz}{$10.72$}
\setsymbol{LFI:bandwidth:30GHz}{$9.89$}

\setsymbol{LFI:white:noise:sensitivity:LFI18:Rad:M:units}{512.5\muKs}
\setsymbol{LFI:white:noise:sensitivity:LFI19:Rad:M:units}{578.9\muKs}
\setsymbol{LFI:white:noise:sensitivity:LFI20:Rad:M:units}{586.9\muKs}
\setsymbol{LFI:white:noise:sensitivity:LFI21:Rad:M:units}{450.4\muKs}
\setsymbol{LFI:white:noise:sensitivity:LFI22:Rad:M:units}{490.1\muKs}
\setsymbol{LFI:white:noise:sensitivity:LFI23:Rad:M:units}{503.9\muKs}
\setsymbol{LFI:white:noise:sensitivity:LFI24:Rad:M:units}{462.8\muKs}
\setsymbol{LFI:white:noise:sensitivity:LFI25:Rad:M:units}{415.2\muKs}
\setsymbol{LFI:white:noise:sensitivity:LFI26:Rad:M:units}{482.6\muKs}
\setsymbol{LFI:white:noise:sensitivity:LFI27:Rad:M:units}{281.5\muKs}
\setsymbol{LFI:white:noise:sensitivity:LFI28:Rad:M:units}{317.9\muKs}
\setsymbol{LFI:white:noise:sensitivity:LFI18:Rad:S:units}{466.7\muKs}
\setsymbol{LFI:white:noise:sensitivity:LFI19:Rad:S:units}{554.2\muKs}
\setsymbol{LFI:white:noise:sensitivity:LFI20:Rad:S:units}{620.0\muKs}
\setsymbol{LFI:white:noise:sensitivity:LFI21:Rad:S:units}{559.8\muKs}
\setsymbol{LFI:white:noise:sensitivity:LFI22:Rad:S:units}{530.9\muKs}
\setsymbol{LFI:white:noise:sensitivity:LFI23:Rad:S:units}{539.1\muKs}
\setsymbol{LFI:white:noise:sensitivity:LFI24:Rad:S:units}{400.5\muKs}
\setsymbol{LFI:white:noise:sensitivity:LFI25:Rad:S:units}{395.0\muKs}
\setsymbol{LFI:white:noise:sensitivity:LFI26:Rad:S:units}{422.9\muKs}
\setsymbol{LFI:white:noise:sensitivity:LFI27:Rad:S:units}{302.8\muKs}
\setsymbol{LFI:white:noise:sensitivity:LFI28:Rad:S:units}{286.1\muKs}

\setsymbol{LFI:white:noise:sensitivity:uncertainty:LFI18:Rad:M:units}{2.0\muKs} 
\setsymbol{LFI:white:noise:sensitivity:uncertainty:LFI19:Rad:M:units}{2.1\muKs} 
\setsymbol{LFI:white:noise:sensitivity:uncertainty:LFI20:Rad:M:units}{1.9\muKs} 
\setsymbol{LFI:white:noise:sensitivity:uncertainty:LFI21:Rad:M:units}{1.4\muKs} 
\setsymbol{LFI:white:noise:sensitivity:uncertainty:LFI22:Rad:M:units}{1.3\muKs} 
\setsymbol{LFI:white:noise:sensitivity:uncertainty:LFI23:Rad:M:units}{1.6\muKs} 
\setsymbol{LFI:white:noise:sensitivity:uncertainty:LFI24:Rad:M:units}{1.3\muKs} 
\setsymbol{LFI:white:noise:sensitivity:uncertainty:LFI25:Rad:M:units}{1.3\muKs} 
\setsymbol{LFI:white:noise:sensitivity:uncertainty:LFI26:Rad:M:units}{1.9\muKs} 
\setsymbol{LFI:white:noise:sensitivity:uncertainty:LFI27:Rad:M:units}{2.0\muKs} 
\setsymbol{LFI:white:noise:sensitivity:uncertainty:LFI28:Rad:M:units}{2.4\muKs} 
\setsymbol{LFI:white:noise:sensitivity:uncertainty:LFI18:Rad:S:units}{2.3\muKs} 
\setsymbol{LFI:white:noise:sensitivity:uncertainty:LFI19:Rad:S:units}{2.2\muKs} 
\setsymbol{LFI:white:noise:sensitivity:uncertainty:LFI20:Rad:S:units}{2.6\muKs} 
\setsymbol{LFI:white:noise:sensitivity:uncertainty:LFI21:Rad:S:units}{1.8\muKs} 
\setsymbol{LFI:white:noise:sensitivity:uncertainty:LFI22:Rad:S:units}{2.2\muKs} 
\setsymbol{LFI:white:noise:sensitivity:uncertainty:LFI23:Rad:S:units}{1.7\muKs} 
\setsymbol{LFI:white:noise:sensitivity:uncertainty:LFI24:Rad:S:units}{1.1\muKs} 
\setsymbol{LFI:white:noise:sensitivity:uncertainty:LFI25:Rad:S:units}{3.1\muKs} 
\setsymbol{LFI:white:noise:sensitivity:uncertainty:LFI26:Rad:S:units}{2.5\muKs} 
\setsymbol{LFI:white:noise:sensitivity:uncertainty:LFI27:Rad:S:units}{1.8\muKs} 
\setsymbol{LFI:white:noise:sensitivity:uncertainty:LFI28:Rad:S:units}{2.1\muKs} 

\setsymbol{LFI:white:noise:sensitivity:LFI18:Rad:M}{512.5}
\setsymbol{LFI:white:noise:sensitivity:LFI19:Rad:M}{578.9}
\setsymbol{LFI:white:noise:sensitivity:LFI20:Rad:M}{586.9}
\setsymbol{LFI:white:noise:sensitivity:LFI21:Rad:M}{450.4}
\setsymbol{LFI:white:noise:sensitivity:LFI22:Rad:M}{490.1}
\setsymbol{LFI:white:noise:sensitivity:LFI23:Rad:M}{503.9}
\setsymbol{LFI:white:noise:sensitivity:LFI24:Rad:M}{462.8}
\setsymbol{LFI:white:noise:sensitivity:LFI25:Rad:M}{415.2}
\setsymbol{LFI:white:noise:sensitivity:LFI26:Rad:M}{482.6}
\setsymbol{LFI:white:noise:sensitivity:LFI27:Rad:M}{281.5}
\setsymbol{LFI:white:noise:sensitivity:LFI28:Rad:M}{317.9}
\setsymbol{LFI:white:noise:sensitivity:LFI18:Rad:S}{466.7}
\setsymbol{LFI:white:noise:sensitivity:LFI19:Rad:S}{554.2}
\setsymbol{LFI:white:noise:sensitivity:LFI20:Rad:S}{620.0}
\setsymbol{LFI:white:noise:sensitivity:LFI21:Rad:S}{559.8}
\setsymbol{LFI:white:noise:sensitivity:LFI22:Rad:S}{530.9}
\setsymbol{LFI:white:noise:sensitivity:LFI23:Rad:S}{539.1}
\setsymbol{LFI:white:noise:sensitivity:LFI24:Rad:S}{400.5}
\setsymbol{LFI:white:noise:sensitivity:LFI25:Rad:S}{395.0}
\setsymbol{LFI:white:noise:sensitivity:LFI26:Rad:S}{422.9}
\setsymbol{LFI:white:noise:sensitivity:LFI27:Rad:S}{302.8}
\setsymbol{LFI:white:noise:sensitivity:LFI28:Rad:S}{286.1}

\setsymbol{LFI:white:noise:sensitivity:uncertainty:LFI18:Rad:M}{2.0} 
\setsymbol{LFI:white:noise:sensitivity:uncertainty:LFI19:Rad:M}{2.1} 
\setsymbol{LFI:white:noise:sensitivity:uncertainty:LFI20:Rad:M}{1.9} 
\setsymbol{LFI:white:noise:sensitivity:uncertainty:LFI21:Rad:M}{1.4} 
\setsymbol{LFI:white:noise:sensitivity:uncertainty:LFI22:Rad:M}{1.3} 
\setsymbol{LFI:white:noise:sensitivity:uncertainty:LFI23:Rad:M}{1.6} 
\setsymbol{LFI:white:noise:sensitivity:uncertainty:LFI24:Rad:M}{1.3} 
\setsymbol{LFI:white:noise:sensitivity:uncertainty:LFI25:Rad:M}{1.3} 
\setsymbol{LFI:white:noise:sensitivity:uncertainty:LFI26:Rad:M}{1.9} 
\setsymbol{LFI:white:noise:sensitivity:uncertainty:LFI27:Rad:M}{2.0} 
\setsymbol{LFI:white:noise:sensitivity:uncertainty:LFI28:Rad:M}{2.4} 
\setsymbol{LFI:white:noise:sensitivity:uncertainty:LFI18:Rad:S}{2.3} 
\setsymbol{LFI:white:noise:sensitivity:uncertainty:LFI19:Rad:S}{2.2} 
\setsymbol{LFI:white:noise:sensitivity:uncertainty:LFI20:Rad:S}{2.7} 
\setsymbol{LFI:white:noise:sensitivity:uncertainty:LFI21:Rad:S}{1.8} 
\setsymbol{LFI:white:noise:sensitivity:uncertainty:LFI22:Rad:S}{2.2} 
\setsymbol{LFI:white:noise:sensitivity:uncertainty:LFI23:Rad:S}{1.7} 
\setsymbol{LFI:white:noise:sensitivity:uncertainty:LFI24:Rad:S}{1.1} 
\setsymbol{LFI:white:noise:sensitivity:uncertainty:LFI25:Rad:S}{3.1} 
\setsymbol{LFI:white:noise:sensitivity:uncertainty:LFI26:Rad:S}{2.5} 
\setsymbol{LFI:white:noise:sensitivity:uncertainty:LFI27:Rad:S}{1.8} 
\setsymbol{LFI:white:noise:sensitivity:uncertainty:LFI28:Rad:S}{2.1}

\setsymbol{LFI:knee:frequency:LFI18:Rad:M:units}{14.8\mHz}
\setsymbol{LFI:knee:frequency:LFI19:Rad:M:units}{11.7\mHz}
\setsymbol{LFI:knee:frequency:LFI20:Rad:M:units}{7.9\mHz}
\setsymbol{LFI:knee:frequency:LFI21:Rad:M:units}{37.9\mHz}
\setsymbol{LFI:knee:frequency:LFI22:Rad:M:units}{9.5\mHz}
\setsymbol{LFI:knee:frequency:LFI23:Rad:M:units}{29.6\mHz}
\setsymbol{LFI:knee:frequency:LFI24:Rad:M:units}{26.9\mHz}
\setsymbol{LFI:knee:frequency:LFI25:Rad:M:units}{19.7\mHz}
\setsymbol{LFI:knee:frequency:LFI26:Rad:M:units}{64.4\mHz}
\setsymbol{LFI:knee:frequency:LFI27:Rad:M:units}{173.7\mHz}
\setsymbol{LFI:knee:frequency:LFI28:Rad:M:units}{128.5\mHz}
\setsymbol{LFI:knee:frequency:LFI18:Rad:S:units}{17.8\mHz}
\setsymbol{LFI:knee:frequency:LFI19:Rad:S:units}{13.7\mHz}
\setsymbol{LFI:knee:frequency:LFI20:Rad:S:units}{5.6\mHz}
\setsymbol{LFI:knee:frequency:LFI21:Rad:S:units}{13.1\mHz}
\setsymbol{LFI:knee:frequency:LFI22:Rad:S:units}{14.3\mHz}
\setsymbol{LFI:knee:frequency:LFI23:Rad:S:units}{58.9\mHz}
\setsymbol{LFI:knee:frequency:LFI24:Rad:S:units}{89.6\mHz}
\setsymbol{LFI:knee:frequency:LFI25:Rad:S:units}{46.8\mHz}
\setsymbol{LFI:knee:frequency:LFI26:Rad:S:units}{70.7\mHz}
\setsymbol{LFI:knee:frequency:LFI27:Rad:S:units}{109.6\mHz}
\setsymbol{LFI:knee:frequency:LFI28:Rad:S:units}{44.1\mHz}

\setsymbol{LFI:knee:frequency:uncertainty:LFI18:Rad:M:units}{2.1\mHz}
\setsymbol{LFI:knee:frequency:uncertainty:LFI19:Rad:M:units}{1.1\mHz}
\setsymbol{LFI:knee:frequency:uncertainty:LFI20:Rad:M:units}{1.5\mHz}
\setsymbol{LFI:knee:frequency:uncertainty:LFI21:Rad:M:units}{5.7\mHz}
\setsymbol{LFI:knee:frequency:uncertainty:LFI22:Rad:M:units}{1.7\mHz}
\setsymbol{LFI:knee:frequency:uncertainty:LFI23:Rad:M:units}{1.1\mHz}
\setsymbol{LFI:knee:frequency:uncertainty:LFI24:Rad:M:units}{1.0\mHz}
\setsymbol{LFI:knee:frequency:uncertainty:LFI25:Rad:M:units}{1.0\mHz}
\setsymbol{LFI:knee:frequency:uncertainty:LFI26:Rad:M:units}{1.8\mHz}
\setsymbol{LFI:knee:frequency:uncertainty:LFI27:Rad:M:units}{3.1\mHz}
\setsymbol{LFI:knee:frequency:uncertainty:LFI28:Rad:M:units}{10.9\mHz}
\setsymbol{LFI:knee:frequency:uncertainty:LFI18:Rad:S:units}{1.5\mHz}
\setsymbol{LFI:knee:frequency:uncertainty:LFI19:Rad:S:units}{1.3\mHz}
\setsymbol{LFI:knee:frequency:uncertainty:LFI20:Rad:S:units}{1.0\mHz}
\setsymbol{LFI:knee:frequency:uncertainty:LFI21:Rad:S:units}{1.3\mHz}
\setsymbol{LFI:knee:frequency:uncertainty:LFI22:Rad:S:units}{6.7\mHz}
\setsymbol{LFI:knee:frequency:uncertainty:LFI23:Rad:S:units}{1.6\mHz}
\setsymbol{LFI:knee:frequency:uncertainty:LFI24:Rad:S:units}{13.8\mHz}
\setsymbol{LFI:knee:frequency:uncertainty:LFI25:Rad:S:units}{1.9\mHz}
\setsymbol{LFI:knee:frequency:uncertainty:LFI26:Rad:S:units}{18.7\mHz}
\setsymbol{LFI:knee:frequency:uncertainty:LFI27:Rad:S:units}{2.5\mHz}
\setsymbol{LFI:knee:frequency:uncertainty:LFI28:Rad:S:units}{2.2\mHz}

\setsymbol{LFI:knee:frequency:LFI18:Rad:M}{14.8}
\setsymbol{LFI:knee:frequency:LFI19:Rad:M}{11.7}
\setsymbol{LFI:knee:frequency:LFI20:Rad:M}{7.9}
\setsymbol{LFI:knee:frequency:LFI21:Rad:M}{37.9}
\setsymbol{LFI:knee:frequency:LFI22:Rad:M}{9.5}
\setsymbol{LFI:knee:frequency:LFI23:Rad:M}{29.6}
\setsymbol{LFI:knee:frequency:LFI24:Rad:M}{26.9}
\setsymbol{LFI:knee:frequency:LFI25:Rad:M}{19.7}
\setsymbol{LFI:knee:frequency:LFI26:Rad:M}{64.4}
\setsymbol{LFI:knee:frequency:LFI27:Rad:M}{173.7}
\setsymbol{LFI:knee:frequency:LFI28:Rad:M}{128.5}
\setsymbol{LFI:knee:frequency:LFI18:Rad:S}{17.8}
\setsymbol{LFI:knee:frequency:LFI19:Rad:S}{13.7}
\setsymbol{LFI:knee:frequency:LFI20:Rad:S}{5.6}
\setsymbol{LFI:knee:frequency:LFI21:Rad:S}{13.1}
\setsymbol{LFI:knee:frequency:LFI22:Rad:S}{14.3}
\setsymbol{LFI:knee:frequency:LFI23:Rad:S}{58.9}
\setsymbol{LFI:knee:frequency:LFI24:Rad:S}{89.6}
\setsymbol{LFI:knee:frequency:LFI25:Rad:S}{46.8}
\setsymbol{LFI:knee:frequency:LFI26:Rad:S}{70.7}
\setsymbol{LFI:knee:frequency:LFI27:Rad:S}{109.6}
\setsymbol{LFI:knee:frequency:LFI28:Rad:S}{44.1}

\setsymbol{LFI:knee:frequency:uncertainty:LFI18:Rad:M}{2.1}
\setsymbol{LFI:knee:frequency:uncertainty:LFI19:Rad:M}{1.1}
\setsymbol{LFI:knee:frequency:uncertainty:LFI20:Rad:M}{1.5}
\setsymbol{LFI:knee:frequency:uncertainty:LFI21:Rad:M}{5.7}
\setsymbol{LFI:knee:frequency:uncertainty:LFI22:Rad:M}{1.7}
\setsymbol{LFI:knee:frequency:uncertainty:LFI23:Rad:M}{1.1}
\setsymbol{LFI:knee:frequency:uncertainty:LFI24:Rad:M}{1.0}
\setsymbol{LFI:knee:frequency:uncertainty:LFI25:Rad:M}{1.0}
\setsymbol{LFI:knee:frequency:uncertainty:LFI26:Rad:M}{1.8}
\setsymbol{LFI:knee:frequency:uncertainty:LFI27:Rad:M}{3.1}
\setsymbol{LFI:knee:frequency:uncertainty:LFI28:Rad:M}{10.9}
\setsymbol{LFI:knee:frequency:uncertainty:LFI18:Rad:S}{1.5}
\setsymbol{LFI:knee:frequency:uncertainty:LFI19:Rad:S}{1.3}
\setsymbol{LFI:knee:frequency:uncertainty:LFI20:Rad:S}{1.0}
\setsymbol{LFI:knee:frequency:uncertainty:LFI21:Rad:S}{1.3}
\setsymbol{LFI:knee:frequency:uncertainty:LFI22:Rad:S}{6.7}
\setsymbol{LFI:knee:frequency:uncertainty:LFI23:Rad:S}{1.6}
\setsymbol{LFI:knee:frequency:uncertainty:LFI24:Rad:S}{13.8}
\setsymbol{LFI:knee:frequency:uncertainty:LFI25:Rad:S}{1.9}
\setsymbol{LFI:knee:frequency:uncertainty:LFI26:Rad:S}{18.7}
\setsymbol{LFI:knee:frequency:uncertainty:LFI27:Rad:S}{2.5}
\setsymbol{LFI:knee:frequency:uncertainty:LFI28:Rad:S}{2.2}

\setsymbol{LFI:slope:LFI18:Rad:M}{$-1.06$}
\setsymbol{LFI:slope:LFI19:Rad:M}{$-1.21$}
\setsymbol{LFI:slope:LFI20:Rad:M}{$-1.19$}
\setsymbol{LFI:slope:LFI21:Rad:M}{$-1.24$}
\setsymbol{LFI:slope:LFI22:Rad:M}{$-1.41$}
\setsymbol{LFI:slope:LFI23:Rad:M}{$-1.07$}
\setsymbol{LFI:slope:LFI24:Rad:M}{$-0.94$}
\setsymbol{LFI:slope:LFI25:Rad:M}{$-0.85$}
\setsymbol{LFI:slope:LFI26:Rad:M}{$-0.92$}
\setsymbol{LFI:slope:LFI27:Rad:M}{$-0.93$}
\setsymbol{LFI:slope:LFI28:Rad:M}{$-0.93$}
\setsymbol{LFI:slope:LFI18:Rad:S}{$-1.18$}
\setsymbol{LFI:slope:LFI19:Rad:S}{$-1.10$}
\setsymbol{LFI:slope:LFI20:Rad:S}{$-1.30$}
\setsymbol{LFI:slope:LFI21:Rad:S}{$-1.20$}
\setsymbol{LFI:slope:LFI22:Rad:S}{$-1.23$}
\setsymbol{LFI:slope:LFI23:Rad:S}{$-1.21$}
\setsymbol{LFI:slope:LFI24:Rad:S}{$-0.91$}
\setsymbol{LFI:slope:LFI25:Rad:S}{$-0.90$}
\setsymbol{LFI:slope:LFI26:Rad:S}{$-0.75$}
\setsymbol{LFI:slope:LFI27:Rad:S}{$-0.91$}
\setsymbol{LFI:slope:LFI28:Rad:S}{$-0.90$}

\setsymbol{LFI:slope:uncertainty:LFI18:Rad:M}{$0.08$}
\setsymbol{LFI:slope:uncertainty:LFI19:Rad:M}{$0.23$}
\setsymbol{LFI:slope:uncertainty:LFI20:Rad:M}{$0.27$}
\setsymbol{LFI:slope:uncertainty:LFI21:Rad:M}{$0.08$}
\setsymbol{LFI:slope:uncertainty:LFI22:Rad:M}{$0.25$}
\setsymbol{LFI:slope:uncertainty:LFI23:Rad:M}{$0.02$}
\setsymbol{LFI:slope:uncertainty:LFI24:Rad:M}{$0.01$}
\setsymbol{LFI:slope:uncertainty:LFI25:Rad:M}{$0.01$}
\setsymbol{LFI:slope:uncertainty:LFI26:Rad:M}{$0.01$}
\setsymbol{LFI:slope:uncertainty:LFI27:Rad:M}{$0.01$}
\setsymbol{LFI:slope:uncertainty:LFI28:Rad:M}{$0.01$}
\setsymbol{LFI:slope:uncertainty:LFI18:Rad:S}{$0.10$}
\setsymbol{LFI:slope:uncertainty:LFI19:Rad:S}{$0.13$}
\setsymbol{LFI:slope:uncertainty:LFI20:Rad:S}{$0.33$}
\setsymbol{LFI:slope:uncertainty:LFI21:Rad:S}{$0.08$}
\setsymbol{LFI:slope:uncertainty:LFI22:Rad:S}{$0.27$}
\setsymbol{LFI:slope:uncertainty:LFI23:Rad:S}{$0.02$}
\setsymbol{LFI:slope:uncertainty:LFI24:Rad:S}{$0.01$}
\setsymbol{LFI:slope:uncertainty:LFI25:Rad:S}{$0.01$}
\setsymbol{LFI:slope:uncertainty:LFI26:Rad:S}{$0.07$}
\setsymbol{LFI:slope:uncertainty:LFI27:Rad:S}{$0.01$}
\setsymbol{LFI:slope:uncertainty:LFI28:Rad:S}{$0.02$}

\setsymbol{LFI:systematic:effects:p2p:uncertainty:70GHz}{7.87}
\setsymbol{LFI:systematic:effects:p2p:uncertainty:44GHz}{5.61}
\setsymbol{LFI:systematic:effects:p2p:uncertainty:30GHz}{21.02}

\begin{document}
\title{\textit{Planck} 2018 results. II. Low Frequency Instrument data
 processing}

%This author list corresponds to \title{Author list for L02\_LFI\_Data\_Processing}
%Prepared by M. Lopez-Caniego (Marcos.Lopez.Caniego@sciops.esa.int), ESAC/ESA
%This version is from Tue Feb  6 16:33:33 2018 CET
%\subtitle{There are 146 co-authors in this list}
\author{\small
Planck Collaboration: Y.~Akrami\inst{54, 56}
\and
F.~Arg\"{u}eso\inst{14}
\and
M.~Ashdown\inst{63, 4}
\and
J.~Aumont\inst{87}
\and
C.~Baccigalupi\inst{74}
\and
M.~Ballardini\inst{18, 41}
\and
A.~J.~Banday\inst{87, 6}
\and
R.~B.~Barreiro\inst{58}
\and
N.~Bartolo\inst{24, 59}
\and
S.~Basak\inst{51}
% ...
}
\institute{\small
APC, AstroParticule et Cosmologie, Universit\'{e} Paris Diderot, CNRS/IN2P3, CEA/lrfu, Observatoire de Paris, Sorbonne Paris Cit\'{e}, 10, rue Alice Domon et L\'{e}onie Duquet, 75205 Paris Cedex 13, France\goodbreak
\and
African Institute for Mathematical Sciences, 6-8 Melrose Road, Muizenberg, Cape Town, South Africa\goodbreak
\and
Agenzia Spaziale Italiana, Via del Politecnico snc, 00133, Roma, Italy\goodbreak
\and
Astrophysics Group, Cavendish Laboratory, University of Cambridge, J J Thomson Avenue, Cambridge CB3 0HE, U.K.\goodbreak
\and
CITA, University of Toronto, 60 St. George St., Toronto, ON M5S 3H8, Canada\goodbreak
\and
CNRS, IRAP, 9 Av. colonel Roche, BP 44346, F-31028 Toulouse cedex 4, France\goodbreak
\and
Cahill Center for Astronomy and Astrophysics, California Institute of Technology, Pasadena CA,  91125, USA\goodbreak
\and
California Institute of Technology, Pasadena, California, U.S.A.\goodbreak
\and
Computational Cosmology Center, Lawrence Berkeley National Laboratory, Berkeley, California, U.S.A.\goodbreak
\and
D\'{e}partement de Physique Th\'{e}orique, Universit\'{e} de Gen\`{e}ve, 24, Quai E. Ansermet,1211 Gen\`{e}ve 4, Switzerland\goodbreak
\and
Departamento de Astrof\'{i}sica, Universidad de La Laguna (ULL), E-38206 La Laguna, Tenerife, Spain\goodbreak
\and
Departamento de F\'{\i}sica, Universidad de Oviedo, C/ Federico Garc\'{\i}a Lorca, 18 , Oviedo, Spain\goodbreak
\and
Departamento de F\'{i}sica Matematica, Instituto de F\'{i}sica, Universidade de S\~{a}o Paulo, Rua do Mat\~{a}o 1371, S\~{a}o Paulo, Brazil\goodbreak
\and
Departamento de Matem\'{a}ticas, Universidad de Oviedo, C/ Federico Garc\'{\i}a Lorca, 18, Oviedo, Spain\goodbreak
\and
Department of Astrophysics/IMAPP, Radboud University, P.O. Box 9010, 6500 GL Nijmegen, The Netherlands\goodbreak
\and
Department of Mathematics, University of Stellenbosch, Stellenbosch 7602, South Africa\goodbreak
\and
Department of Physics \& Astronomy, University of British Columbia, 6224 Agricultural Road, Vancouver, British Columbia, Canada\goodbreak
\and
Department of Physics \& Astronomy, University of the Western Cape, Cape Town 7535, South Africa\goodbreak
\and
Department of Physics and Astronomy, Dana and David Dornsife College of Letter, Arts and Sciences, University of Southern California, Los Angeles, CA 90089, U.S.A.\goodbreak
\and
Department of Physics, Gustaf H\"{a}llstr\"{o}min katu 2a, University of Helsinki, Helsinki, Finland\goodbreak
\and
Department of Physics, Princeton University, Princeton, New Jersey, U.S.A.\goodbreak
\and
Department of Physics, University of California, Santa Barbara, California, U.S.A.\goodbreak
\and
Department of Physics, University of Illinois at Urbana-Champaign, 1110 West Green Street, Urbana, Illinois, U.S.A.\goodbreak
\and
Dipartimento di Fisica e Astronomia G. Galilei, Universit\`{a} degli Studi di Padova, via Marzolo 8, 35131 Padova, Italy\goodbreak
\and
Dipartimento di Fisica e Scienze della Terra, Universit\`{a} di Ferrara, Via Saragat 1, 44122 Ferrara, Italy\goodbreak
\and
Dipartimento di Fisica, Universit\`{a} La Sapienza, P. le A. Moro 2, Roma, Italy\goodbreak
\and
Dipartimento di Fisica, Universit\`{a} degli Studi di Milano, Via Celoria, 16, Milano, Italy\goodbreak
\and
Dipartimento di Fisica, Universit\`{a} degli Studi di Trieste, via A. Valerio 2, Trieste, Italy\goodbreak
\and
Dipartimento di Fisica, Universit\`{a} di Roma Tor Vergata, Via della Ricerca Scientifica, 1, Roma, Italy\goodbreak
\and
European Space Agency, ESAC, Planck Science Office, Camino bajo del Castillo, s/n, Urbanizaci\'{o}n Villafranca del Castillo, Villanueva de la Ca\~{n}ada, Madrid, Spain\goodbreak
\and
European Space Agency, ESTEC, Keplerlaan 1, 2201 AZ Noordwijk, The Netherlands\goodbreak
\and
Gran Sasso Science Institute, INFN, viale F. Crispi 7, 67100 L'Aquila, Italy\goodbreak
\and
HGSFP and University of Heidelberg, Theoretical Physics Department, Philosophenweg 16, 69120, Heidelberg, Germany\goodbreak
\and
Haverford College Astronomy Department, 370 Lancaster Avenue, Haverford, Pennsylvania, U.S.A.\goodbreak
\and
Helsinki Institute of Physics, Gustaf H\"{a}llstr\"{o}min katu 2, University of Helsinki, Helsinki, Finland\goodbreak
\and
INAF - OAS Bologna, Istituto Nazionale di Astrofisica - Osservatorio di Astrofisica e Scienza dello Spazio di Bologna, Area della Ricerca del CNR, Via Gobetti 101, 40129, Bologna, Italy\goodbreak
\and
INAF - Osservatorio Astronomico di Padova, Vicolo dell'Osservatorio 5, Padova, Italy\goodbreak
\and
INAF - Osservatorio Astronomico di Trieste, Via G.B. Tiepolo 11, Trieste, Italy\goodbreak
\and
INAF Istituto di Radioastronomia, Via P. Gobetti 101, 40129 Bologna, Italy\goodbreak
\and
INAF, Istituto di Radioastronomia, Via Piero Gobetti 101, I-40129 Bologna, Italy\goodbreak
\and
INAF/IASF Bologna, Via Gobetti 101, Bologna, Italy\goodbreak
\and
INAF/IASF Milano, Via E. Bassini 15, Milano, Italy\goodbreak
\and
INFN - CNAF, viale Berti Pichat 6/2, 40127 Bologna, Italy\goodbreak
\and
INFN, Sezione di Bologna, viale Berti Pichat 6/2, 40127 Bologna, Italy\goodbreak
\and
INFN, Sezione di Ferrara, Via Saragat 1, 44122 Ferrara, Italy\goodbreak
\and
INFN, Sezione di Milano, Via Celoria 16, Milano, Italy\goodbreak
\and
INFN, Sezione di Roma 1, Universit\`{a} di Roma Sapienza, Piazzale Aldo Moro 2, 00185, Roma, Italy\goodbreak
\and
INFN, Sezione di Roma 2, Universit\`{a} di Roma Tor Vergata, Via della Ricerca Scientifica, 1, Roma, Italy\goodbreak
\and
INFN/National Institute for Nuclear Physics, Via Valerio 2, I-34127 Trieste, Italy\goodbreak
\and
Imperial College London, Astrophysics group, Blackett Laboratory, Prince Consort Road, London, SW7 2AZ, U.K.\goodbreak
\and
Indian Institute of Science Education and Research, Vithura, Thiruvananthapuram - 695551, India\goodbreak
\and
Institut d'Astrophysique Spatiale, CNRS, Univ. Paris-Sud, Universit\'{e} Paris-Saclay, B\^{a}t. 121, 91405 Orsay cedex, France\goodbreak
\and
Institut d'Astrophysique de Paris, CNRS (UMR7095), 98 bis Boulevard Arago, F-75014, Paris, France\goodbreak
\and
Institute Lorentz, Leiden University, PO Box 9506, Leiden 2300 RA, The Netherlands\goodbreak
\and
Institute of Astronomy, University of Cambridge, Madingley Road, Cambridge CB3 0HA, U.K.\goodbreak
\and
Institute of Theoretical Astrophysics, University of Oslo, Blindern, Oslo, Norway\goodbreak
\and
Instituto de Astrof\'{\i}sica de Canarias, C/V\'{\i}a L\'{a}ctea s/n, La Laguna, Tenerife, Spain\goodbreak
\and
Instituto de F\'{\i}sica de Cantabria (CSIC-Universidad de Cantabria), Avda. de los Castros s/n, Santander, Spain\goodbreak
\and
Istituto Nazionale di Fisica Nucleare, Sezione di Padova, via Marzolo 8, I-35131 Padova, Italy\goodbreak
\and
Jet Propulsion Laboratory, California Institute of Technology, 4800 Oak Grove Drive, Pasadena, California, U.S.A.\goodbreak
\and
Jodrell Bank Centre for Astrophysics, Alan Turing Building, School of Physics and Astronomy, The University of Manchester, Oxford Road, Manchester, M13 9PL, U.K.\goodbreak
\and
Kavli Institute for Cosmological Physics, University of Chicago, Chicago, IL 60637, USA\goodbreak
\and
Kavli Institute for Cosmology Cambridge, Madingley Road, Cambridge, CB3 0HA, U.K.\goodbreak
\and
LERMA, CNRS, Observatoire de Paris, 61 Avenue de l'Observatoire, Paris, France\goodbreak
\and
Laboratoire de Physique Subatomique et Cosmologie, Universit\'{e} Grenoble-Alpes, CNRS/IN2P3, 53, rue des Martyrs, 38026 Grenoble Cedex, France\goodbreak
\and
Laboratoire de Physique Th\'{e}orique, Universit\'{e} Paris-Sud 11 \& CNRS, B\^{a}timent 210, 91405 Orsay, France\goodbreak
\and
Lawrence Berkeley National Laboratory, Berkeley, California, U.S.A.\goodbreak
\and
Low Temperature Laboratory, Department ofÊApplied Physics, Aalto University, Espoo, FI-00076 AALTO, Finland\goodbreak
\and
Max-Planck-Institut f\"{u}r Astrophysik, Karl-Schwarzschild-Str. 1, 85741 Garching, Germany\goodbreak
\and
Mullard Space Science Laboratory, University College London, Surrey RH5 6NT, U.K.\goodbreak
\and
Nicolaus Copernicus Astronomical Center, Polish Academy of Sciences, Bartycka 18, 00-716 Warsaw, Poland\goodbreak
\and
Nordita (Nordic Institute for Theoretical Physics), Roslagstullsbacken 23, SE-106 91 Stockholm, Sweden\goodbreak
\and
Purple Mountain Observatory, Chinese Academy of Sciences, Nanjing 210008, China\goodbreak
\and
SISSA, Astrophysics Sector, via Bonomea 265, 34136, Trieste, Italy\goodbreak
\and
San Diego Supercomputer Center, University of California, San Diego,Ê9500 Gilman Drive, La Jolla, CA 92093, USA\goodbreak
\and
School of Chemistry and Physics, University of KwaZulu-Natal, Westville Campus, Private Bag X54001, Durban, 4000, South Africa\goodbreak
\and
School of Physical Sciences, National Institute of Science Education and Research, HBNI, Jatni-752050, Odissa, India\goodbreak
\and
School of Physics and Astronomy, Cardiff University, Queens Buildings, The Parade, Cardiff, CF24 3AA, U.K.\goodbreak
\and
School of Physics and Astronomy, Sun Yat-Sen University, 135 Xingang Xi Road, Guangzhou, China\goodbreak
\and
School of Physics and Astronomy, University of Nottingham, Nottingham NG7 2RD, U.K.\goodbreak
\and
Simon Fraser University, Department of Physics, 8888 University Drive, Burnaby BC, Canada\goodbreak
\and
Sorbonne Universit\'{e}-UPMC, UMR7095, Institut d'Astrophysique de Paris, 98 bis Boulevard Arago, F-75014, Paris, France\goodbreak
\and
Space Science Data Center - Agenzia Spaziale Italiana, Via del Politecnico snc, 00133, Roma, Italy\goodbreak
\and
Space Sciences Laboratory, University of California, Berkeley, California, U.S.A.\goodbreak
\and
The Oskar Klein Centre for Cosmoparticle Physics, Department of Physics,Stockholm University, AlbaNova, SE-106 91 Stockholm, Sweden\goodbreak
\and
UPMC Univ Paris 06, UMR7095, 98 bis Boulevard Arago, F-75014, Paris, France\goodbreak
\and
Universit\'{e} de Toulouse, UPS-OMP, IRAP, F-31028 Toulouse cedex 4, France\goodbreak
% ...
}

\titlerunning{LFI data processing}
\authorrunning{Planck Collaboration}

\abstract{
  We present a final description of the data-processing pipeline for the
  \Planck\, Low Frequency Instrument (LFI), implemented for the 2018 data release.
  Several improvements have been made with respect to the previous release, especially
  in the calibration process and in the correction of instrumental features such as the
  effects of nonlinearity in the response of the analogue-to-digital converters.  We provide a brief pedagogical
  introduction to the complete pipeline, as well as a detailed description of the important
  changes implemented. Self-consistency of the pipeline is demonstrated using dedicated
  simulations and null tests.  We present the final version of the LFI full sky maps at
  30, 44, and 70\GHz, both in temperature and polarization, together with a refined estimate
  of the solar dipole and a final assessment of the main LFI  instrumental parameters.}

\maketitle
\allearlypapers

\tableofcontents

\section{Introduction}
\label{sec_introduction}

This paper is part of the 2018 data release (`PR3') of the
\Planck\footnotemark\footnotetext{\Planck\, ({\url{http://www.esa.int/Planck}}) is a project
of the European Space Agency (ESA) with instruments provided by two scientific consortia
funded by ESA member states and let by Principal Investigators from France and Italy,
telescope reflectors provided through a collaboration between ESA and a scientific
consortium led and funded by Denmark, and additional contributions from NASA (USA).}
mission, and reports on the Low Frequency Instrument (LFI) data processing for the legacy
data products and cosmological analysis. The 2018 release is based on the same data set
as the previous release (`PR2') in 2015, in other words, a total of 48~months of observation
(eight full-sky Surveys), more than three times the nominal mission length of 15.5
months originally planned \citep{planck2011-1.1}.

This paper describes in detail the complete data flow through the LFI scientific pipeline
as it was actually implemented in the LFI data-processing centre (DPC), starting from
the basic steps of handling raw telemetry (for both scientific and house-keeping data),
and ending with the creation of frequency maps and validation of the released data
products (similar information for the High Frequency Instrument [HFI] can be found
in \citealt{planck2016-l03}).  Since this is the last Planck Collaboration paper on the LFI
data analysis, in this introduction we provide a pedagogical description of all
the data-processing steps in the pipeline.  Later sections report in greater detail on
those pipeline steps that have been updated, modified, or improved with respect to
the previous data release. For the many steps that remain unchanged, the interested
reader should consult \citet{planck2014-a03}.

\begin{figure*}[htpb]
\centerline{
\includegraphics[width=20cm]{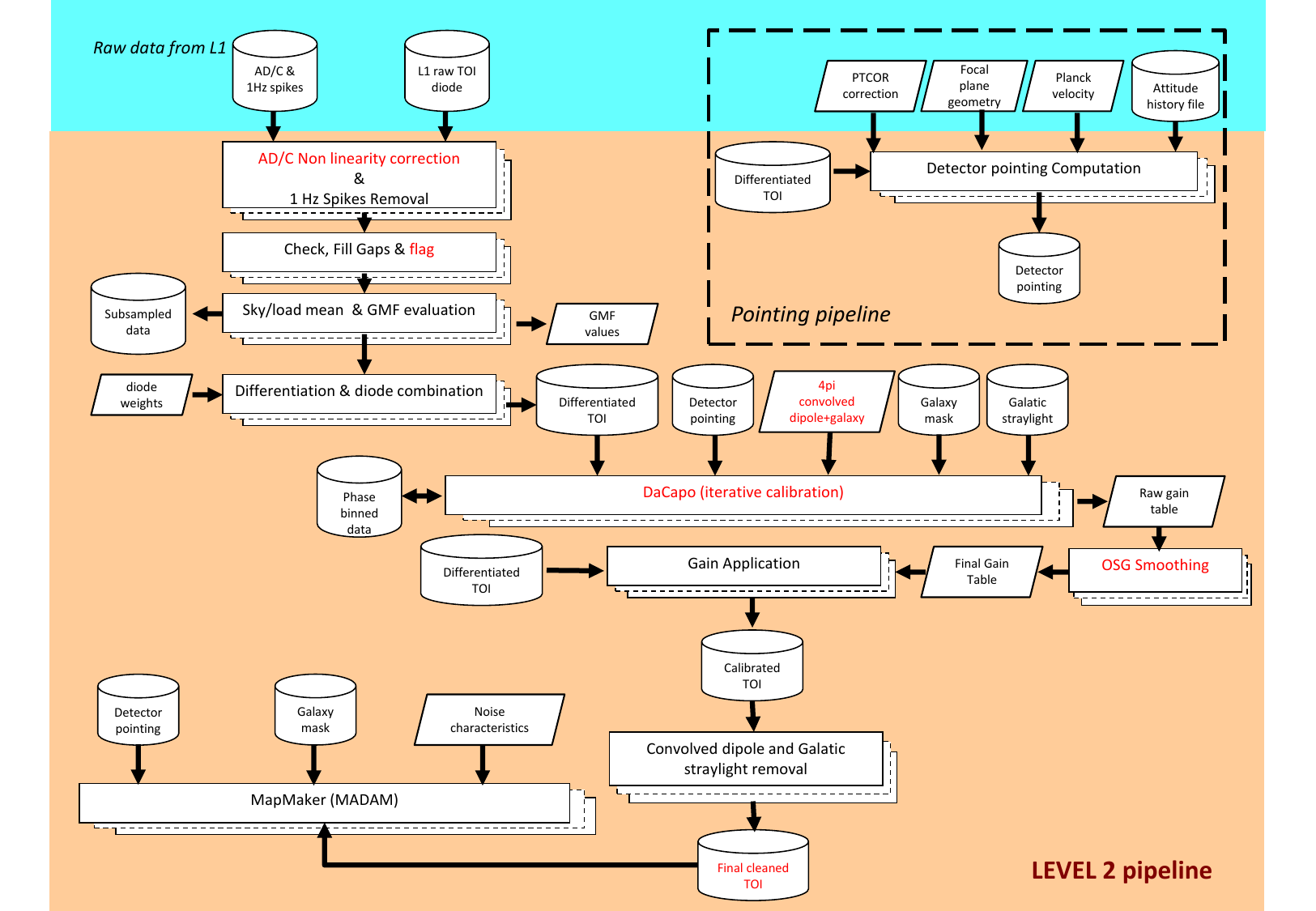}}
\caption{Schematic representation of the LFI data-processing pipeline from raw telemetry down to frequency maps. Elements
in red are those changed or improved with respect to \citet{planck2014-a03}.}
\label{figure_1}
\end{figure*}

Processing LFI data is divided into three main levels (see Fig.~\ref{figure_1}).
In Level~1, the process starts with the ingestion of the required information from
the telemetry data packets and auxiliary data received from the Mission Operation
Centre; both the science and housekeeping information is then transformed into a
format suitable for  Level~2 processing.
The goal of Level~2 is the creation of calibrated maps at all LFI frequencies in
both temperature and polarization, with known systematic and instrumental effects
removed. Finally, Level~3 requires the combination of both LFI and HFI data to
perform astrophysical component separation (both CMB and foregrounds), extraction
of CMB angular power spectra, and determination of cosmological parameters. This
last level is not described in this paper: we refer readers to 
\citet{planck2016-l01}, \cite{planck2016-l04}, \cite{planck2016-l05}, and \cite{planck2016-l06}.

Level~2 includes three main blocks of the analysis pipeline: TOI processing
or `preprocessing'; calibration; and mapmaking.  Preprocessing starts with
flagging data that were made unusable due to lost telemetry packets and spacecraft manoeuvres.
It continues with corrections for nonlinearity in the analogue-to-digital converters
(ADCs) and for small spurious electronic signals at 1\,Hz.  The ADCs convert analogue
output voltages from the detectors into digital form. Any departure from exact linearity
creates a distortion in the response curve of the radiometer.  The current implementation
of the algorithm to correct for ADC nonlinearity includes improvements made since
\citet{planck2014-a03}, which are described in  Sect.~\ref{sec_toiprocessing}.
The 1-Hz electronic spikes result from an unwanted, low-level interaction between
the electronic clock and the science data, and occur in the data-acquisition electronics
after the acquisition of raw data from the radiometer diodes, and before ADC conversion
\citep{meinhold2009,mennella2010,planck2011-1.4}.  They appear as a 1-Hz square wave,
synchronous with the on-board time signal. The procedure for correcting the data is the
same as described in \citet{planck2014-a03}, and consists of fitting and subtracting a
1-Hz square wave template from the time-domain data.

In the pseudo-correlation scheme adopted for the LFI radiometers \citep{bersanelli2010},
each radiometer diode produces an alternating sequence of sky and reference load signals
at the 4096-Hz phase-switch frequency. The $1/f$ noise of the sky and reference data streams
are highly correlated. Subtracting the optimally scaled reference data stream from the sky
data stream reduces the $1/f$ noise in the sky data by several orders of magnitude.
We calculate this optimal gain modulation factor (GMF) using the same method as for the
2015 release \citep{planck2014-a03}.

The last preprocessing step is diode combination.  This reduces the impact of imperfect
isolation between the two diodes of each LFI radiometer. The weighted combinations are
unchanged since the 2015 release, and may be found in Table~3 of \citet{planck2014-a03};
typical values range between $0.4$ and $0.6$ (a perfect diodes isolation
would yield $0.5$ equal weights).

The pointing pipeline runs in parallel with the preprocessing pipeline just described.
It uses the focal plane geometry, the spacecraft velocity and attitude, and `PTCOR' a
long-time-scale pointing correction (which takes account of both the distance from
the Sun and thermometry from the Radiometer Electronics Box Assembly). The pointing
pipeline reconstructs the pointing position and horn orientation for each sample in
the data stream. PTCOR is unchanged from the 2015 release, and is described in \citet{planck2014-a01}.

The next step in the pipeline is photometric calibration. In addition to the pointing
and sky-minus-load differences, auxiliary information is required to obtain accurate
calibration. This includes the $4\pi$ beam response, a model of the CMB dipole together with
the time-varying modulation of its amplitude induced by the motion of the spacecraft along its orbit
--- our primary calibration source ---
and a model of Galactic emission whose contribution through
the beam far sidelobes is modelled and subtracted from each time line.  The calibration
process converts raw voltages at the output of the radiometers into thermodynamic
temperatures. The basic calibration reference signal is the \Planck\, dipole convolved with the
full $4\pi$ beam response, properly weighted according to the bandpass of each radiometer.
The \Planck\, dipole used in this step is identical to the one employed in the earlier
2015 release \citep[][see further details in the following sections]{planck2014-a01}.
On the other hand, the calibration algorithm has been significantly improved
compared to the previous release by including Galactic emission along with
the CMB dipole in the calibration model.
This is particularly important during periods when the spacecraft spin axis is nearly
aligned with the CMB dipole, and the variation of the dipole signal along scan circles
is small (`dipole minima'). The final gain solution is obtained with an iterative
destriper, {\tt{DaCapo}}, which at each step determines radiometer gains, constraining
the data to fit the dipole + Galaxy beam-convolved model. The output gain solutions
are noisy during dipole minima (especially in Surveys~2 and 4). Therefore, as in the
previous release (but with further optimization), we employ an adaptive smoothing
algorithm that reduces scatter in gain solutions, but preserves real discontinuities
caused by abrupt changes in the radiometer operating conditions. Finally, these smoothed
gain solutions are applied to raw data streams, after subtraction of both the dipole and an estimated
signal contributed by Galactic emission into the beam sidelobes.

The final step of the Level~2 pipeline is mapmaking, in other words,
using the calibrated data and pointing information to create
Stokes $I$, $Q$, and $U$ maps of the sky at each frequency. The LFI mapmaking code is {\tt{Madam}}, 
fully described in \citet{keihanen2005} and \citet{planck2014-a07}, which removes
correlated $1/f$ noise with a destriping approach.  Correlated noise is modelled as a
single baseline \citep{maino2002}.  The algorithm makes use of the redundancy in the
observing strategy to constrain these baselines, which are then subtracted from the
time-ordered data in the creation of the sky maps. The algorithm allows a selection of baseline
lengths, which is always a compromise between optimal noise removal and computational cost.
As in the 2015 release, we adopt baselines of 1\,s at  44 and 70\GHz, and 0.25\,s at 30\GHz.
The shorter baseline at 30\GHz\ is appropriate for the higher $1/f$ noise of the radiometers
at this frequency (see Table~\ref{tab_summary_instperf}), which introduces a larger correlated
component in the noise.

\begin{table*}[htpb]
\begingroup
\newdimen\tblskip \tblskip=5pt
\caption{LFI performance parameters.}
\label{tab_summary_instperf}
\nointerlineskip
\vskip -3mm
\footnotesize
\setbox\tablebox=\vbox{
   \newdimen\digitwidth
   \setbox0=\hbox{\rm 0}
   \digitwidth=\wd0
   \catcode`*=\active
   \def*{\kern\digitwidth}
   \newdimen\signwidth
   \setbox0=\hbox{+}
   \signwidth=\wd0
   \catcode`!=\active
   \def!{\kern\signwidth}
\halign{\hbox to 2.7in{#\leaderfil}\tabskip=3em&
        \hfil#\hfil&
        \hfil#\hfil&
        \hfil#\hfil\tabskip=0pt\cr
\noalign{\doubleline}
\noalign{\vskip -3pt}
\omit\hfil Parameter\hfil&30\,GHz&44\,GHz&70\,GHz\cr
\noalign{\vskip 3pt\hrule\vskip 5pt}
Centre frequency [GHz]&28.4&44.1&70.4\cr
\noalign{\vskip 3pt}
Bandwidth [GHz]& \getsymbol{LFI:bandwidth:30GHz}& \getsymbol{LFI:bandwidth:44GHz}& \getsymbol{LFI:bandwidth:70GHz}\cr
\noalign{\vskip 3pt}
Scanning beam FWHM$^{\rm a}$ [arcmin]&33.10&27.94&13.08\cr
\noalign{\vskip 3pt}
Scanning beam ellipticity$^{\rm a}$&1.37&1.25&1.27\cr
\noalign{\vskip 3pt}
Effective beam FWHM$^{\rm b}$ [arcmin]& 32.29 & 26.99 & 13.22\cr
\noalign{\vskip 3pt}
White-noise level in timelines$^{\rm c}$ [$\mu\mathrm{K_{CMB}}s^{1/2}$]&147.9&174.0&151.9\cr
\noalign{\vskip 3pt}
$f_{\rm knee}$$^{\rm c}$ [mHz]&113.9&53.0&19.6\cr
\noalign{\vskip 3pt}
$1/f$ slope$^{\rm c}$&\llap{$-$}0.92&\llap{$-$}0.88&\llap{$-$}1.20\cr
\noalign{\vskip 3pt}
Overall calibration uncertainty$^{\rm d}$ [\%]&0.17&0.12&0.20\cr
\noalign{\vskip 5pt\hrule\vskip 3pt}}}
\endPlancktablewide
\tablenote a Determined by fitting Jupiter observations
directly in the timelines.\par
\tablenote b Calculated from the main-beam solid angle of the effective beam. These values
are used in the source extraction pipeline~\citep{planck2014-a35}.\par
\tablenote c Typical values derived from fitting noise spectra (see Sect.~\ref{sec_general_noise}).\par
\tablenote d Difference between first and last iteration of the iterative calibration (for 30 and 44\GHz) or E2E 2015 result (for 70\GHz). In 2015, the calibration uncertainty was 0.35\,\% and 0.26\,\% at 30 and 44\GHz, respectively.\par
\endgroup
\end{table*}

Table~\ref{tab_summary_instperf} gives typical values for the main instrument
performance parameters measured in flight;  similar tabulations were given in
previous releases.  Beam and optical properties are derived from Jupiter transits,
and are consistent with 2015 results. Major improvements in the calibration uncertainty
are reflected in more stable results for noise parameters. Values reported are averages
among the radiometers operating at a given frequency. At 44\,GHz the FWHM is not entirely
representative of the actual beamwidth, since one of the three 44-GHz horns is located
on the opposite site of the focal plane from the other two \citep{planck2014-a05}.

\section{Time-ordered information (TOI) processing}
\label{sec_toiprocessing}

The main changes in the Level~1 pipeline since the last release are related to data
flagging and to correcting the nonlinearity in the analogue-to-digital converter (ADC). 

We revised our flagging procedure to use more conservative and rigorously homogeneous
criteria. The new procedure results in a slightly higher flagging rate, particularly
during the first 200 operational days (ODs) of the mission; however, the fraction of
flagged data remains negligible.  Table~\ref{tab_data_flags_percentage} gives final
values for the missing and unusable data for the full mission; changes from the release
reported in \citet{planck2014-a03} are a fraction of one percent.  Since the fraction of
flagged data is negligible, so the effect on science is also negligible.  It is worth mentioning
that although the LFI radiometers are quite stable, there are occasional jumps in gain that
if not treated properly would impact the calibration procedure well beyond the single data
point in which the jump occurs.  These jumps are now properly identified and taken into account.

\begin{table}[htpb]
  \begingroup
  \newdimen\tblskip \tblskip=5pt
  \caption{Percentage of LFI observation time lost due to missing or unusable data, and to manoeuvres.}
  \label{tab_data_flags_percentage}
  \nointerlineskip
  \vskip -3mm
  \footnotesize
  \setbox\tablebox=\vbox{
    \newdimen\digitwidth
    \setbox0=\hbox{\rm 0}
    \digitwidth=\wd0
    \catcode`*=\active
    \def*{\kern\digitwidth}
    \newdimen\signwidth
    \setbox0=\hbox{+}
    \signwidth=\wd0
    \catcode`!=\active
    \def!{\kern\signwidth}
    \halign{\hbox to 1.31in{#\leaderfil}\tabskip=1em&
      \hfil#\hfil&
      \hfil#\hfil&
      \hfil#\hfil\tabskip=0pt\cr 
      \noalign{\doubleline}
      \omit\hfil Category\hfil& 30\,GHz& 44\,GHz& 70\,GHz\cr
      \noalign{\vskip 3pt\hrule\vskip 5pt}
        Missing [\%]& *0.15425& *0.15425& *0.15433\cr
      Anomalies [\%]& *0.82402& *0.50997& *0.84842\cr
      Manoeuvres [\%]& *8.03104& *8.03104& *8.03104\cr
\noalign{\vskip 3pt}
      Usable [\%]& 90.99069& 91.30474& 90.96621\cr
      \noalign{\vskip 5pt\hrule\vskip 3pt}
    }}
  \endPlancktable
\endgroup
\end{table}

Nonlinearity in the ADCs that convert analogue detector voltages into numbers
distorts the radiometer response, possibly mimicking a sky signal. For the present
release, we developed a new approach to the correction of this effect that produces
significantly better results at 30\,GHz. 

The first step in the correction is calculation of the white-noise amplitude, given
by the difference between the sum of the variances and twice the covariances of
adjacent samples in the time-stream.  Specifically,
$\sigma^2_{\mathrm{WN}} = {\mathrm{Var}}[X_{\rm o}] + {\mathrm{Var}}[X_{\rm e}] - 2 {\mathrm{Cov}}[X_{\rm o},X_{\rm e}]$,
where $X_{\rm o}$  and $X_{\rm e}$ are data points with time-stream odd and even
indices respectively. ADC nonlinearity produces a variation in the white-noise
amplitude as a function of the detector voltage. 

In the previous release, we fitted the white-noise amplitudes binned with respect
to detector voltage with a simple spline curve, and translated the results into a
correction curve as described in appendix~A of \citet{planck2013-p02a}. For this
release, we tried a more physically motivated fitting function based on the fact
that ADCs suffer from a linearity error $\varepsilon$ on each bit.  We modelled
the output voltage $V_0$ as
\begin{equation}
V_0 = V_{\mathrm{adu}}\sum_{i=0}^{\mathrm{nbit-1}}2^i b_i \left(1+\varepsilon_i/2^i\right) -
V_{\mathrm{off}}\, ,
\label{adceq}
\end{equation}
where $b_i$ is 1 if the $i$th bit is set and 0 otherwise, $\varepsilon_i$ is
the linearity error of the  $i$th bit (which is between $-0.5$ and +0.5), $V_{\mathrm{adu}}$
is the voltage step for one binary level change (one analogue-to-digital unit or adu),
and $V_{\mathrm{off}}$ allows for a possible offset (see figures~9 and 10 of \citet{planck2013-p02a}).
Due to complex degeneracies in Eq.~(\ref{adceq}), we adopted an annealed optimization
procedure to avoid local minima in the $\chi^2$ fit to this model.  

Even this improved model proved to be too simple, however, as it did not reproduce
some of the asymmetries present in the original ADC curve, which appeared to be due to
coupling between adjacent bits.  We therefore add to the previous expression an extra
summation for adjacent coupled bits:
\begin{equation}
V_0 = V_{\mathrm{adu}} \sum_{i=0}^{\mathrm{nbit-1}} 2^i b_i\left(1+\varepsilon_i/2^i\right) + 
\sum_{i=0}^{\mathrm{nbit-2}} b_i b_{i+1}\varepsilon_{i,i+1} -V_\mathrm{off}\, ,
\label{newadceq}
\end{equation}
where $\varepsilon_{i,i+1}$ is the coupled error between bits $i$ and $i+1$.

We compared the results between this method and the previous one by means of null maps,
checking the consistency of the resulting new gain solution with the new ADC correction
applied. This was done by computing the rms scatter from the eight different survey maps,
taking into account pixel hits and zero levels. A ``goodness" parameter can then be derived
from the mean level of the masked null map made between these survey scatter maps. 

The null maps showed substantial improvement at 30\,GHz, but little improvement at  44
and 70\,GHz.  Inspection of the ADC solutions revealed that the higher noise per radiometer
and low ADC nonlinearity at 70\GHz\ did not allow for a good fit. At 44\,GHz, on the other
hand, the ADC effect was so large that the new model could not reproduce some of the details,
and so led to some small residuals.  The 30-GHz system has much lower noise and less thermal
drift in the gain, meaning that more voltage levels were revisited more often,  yielding a
more consistent ADC model curve.  It was therefore decided to keep the new solution only for
the 30-GHz channels.  The other two frequencies thus have the same correction for ADC
nonlinearity as in the previous release.

\section{Photometric calibration}
\label{sec_calibration}

The raw output from an LFI radiometer is a voltage, $V$, which we can write \citep{planck2014-a03}  as 
\begin{equation}
V(t) = G(t) \times \left[ B \ast (D_{\mathrm{solar}} + D_{\mathrm{orbital}}
+ T_{\mathrm{sky}}) + T_0\right],
\label{eq:calib}
\end{equation}
where $G$ is the gain $B$ encodes both convolution with a $4\pi$ instrumental
beam and the observation scanning strategy, $D_{\mathrm{solar}}$ and $D_{\mathrm{orbital}}$
are the solar and orbital CMB dipoles,\footnotemark\footnotetext{The solar dipole is the dipole anisotropy in the CMB
  induced by the motion of the solar System barycentre with respect to the rest frame
  of the CMB itself. The orbital dipole is the modulation of the solar dipole due
  to the orbital motion of the spacecraft around the Sun.},  
$T_{\mathrm{sky}}$ represents the sum of the CMB and foreground fluctuations, and $T_0$
is the sum of the $2.7$-K CMB temperature, other astrophysical monopole
terms, and any internal instrumental offsets.  Photometric calibration is the process
of determining $G(t)$ accurately over time, which is critical for the quality of the final maps.

In the \Planck\ 2013 release, based on 15\,months of data, an accurate \Planck\ determination
of $D_{\mathrm{orbital}}$ was not possible, and $G(t)$ was estimated from $D_{\mathrm{solar}}$ alone.
We used the best-fit, 9-year \WMAP\ dipole estimate as the reference model against which
to compare the measured voltages \citep{planck2013-p02b,bennett2012}. 
Successive analyses \citep{planck2013-p01a,planck2014-a01} showed that this model resulted
in gain estimates that were offset by about $0.3\,$\% (due largely to foreground
contamination in the \WMAP\ dipole), within the originally estimated error uncertainty.

In the \Planck\ 2015 release, we implemented internal and self-consistent estimation of
the solar dipole by using the orbital dipole for absolute calibration (see Sect.~\ref{sec_lfidipole}
for further details).  The orbital dipole is much smaller than the solar dipole, but
is known absolutely with exquisite accuracy from the orbital motion of \Planck\ itself.
This resulted in relative calibration uncertainties $\lsim 0.3$\,\% \citep{planck2014-a03},
adequate to allow high-precision cosmology based on temperature measurements. However, for
polarization even a relative error of $10^{-3}$ is non-negligible, and a large fraction of
the LFI work on data quality since the 2015 release has revolved around reducing this error further.

As discussed extensively in \citet{planck2014-a03} and \citet{planck2014-a13}, one of the most notable
problems in the 2015 LFI processing was the failure of a specific internal null test, namely
that taken between Surveys~1, 3, 5, 6, 7, and 8, and Surveys~2 and 4. In particular,
Surveys~2 and 4 showed significantly larger uncertainties in their gain estimation than
the other Surveys (see figure~4 in \citealp{planck2014-a03}),  and, critically, they also
showed significant excess $B$-mode power on the very largest scales. Although it was
well-known that Surveys~2 and 4 happen to be aligned with the \Planck\ scanning strategy
in such a way that the dipole modulation reaches very low minima, thus exacerbating the
impact on calibration of any potential systematic effect, we could not identify the
specific source of the anomaly.  Nevertheless, because of the null-test failure, those
two surveys were removed from the final polarization maps and likelihood analysis.

Since that time, we have performed a series of detailed end-to-end simulations designed
specifically to identify the source of this null-test failure, and this work ultimately
led to a minor, but important, modification of the calibration scheme outlined above
and described in detail in \citet{planck2014-a03}. In short, the survey null-test
failure was due to not accounting for the polarized component of the sky signal in
Eq.~(\ref{eq:calib}). This has now been done, as described in detail below. Thus
the updated calibration scheme represents the logical conclusion of Eq.~(\ref{eq:calib}),
since we now account for all terms as far as we are able to model them.

\subsection{Joint gain estimation and component separation}

Before describing the updated calibration scheme, we first establish some useful intuition
regarding the physical effect in question.  We start with the raw gains, $G(t)$,
as measured in 2015 \citep{planck2014-a03}. Overall, this function may be crudely
modelled over the course of the mission as a sum of a linear term and a 1-year sinusoidal term:
\begin{equation}
G_{\mathrm{model}}(t) = (a+b\,t) +
c\,\sin\left(\frac{2\pi}{365\,\mathrm{days}}\,t + d\right),
\end{equation}
where the four free parameters, \{$a$,$b$,$c$,$d$\}, must be fitted
radiometer by radiometer. From this model, we compute
the ``normalized fractional gain'' as
\begin{equation}
\hat{G}(t) = 100 \frac{G(t)-G_{\mathrm{model}}(t)}{a}.
\end{equation}
This function is simply the fractional gain excess (or deficit) relative to a smoothly varying model, expressed as a percentage.

\begin{figure*}[htpb]
\centerline{
\includegraphics[width=18cm]{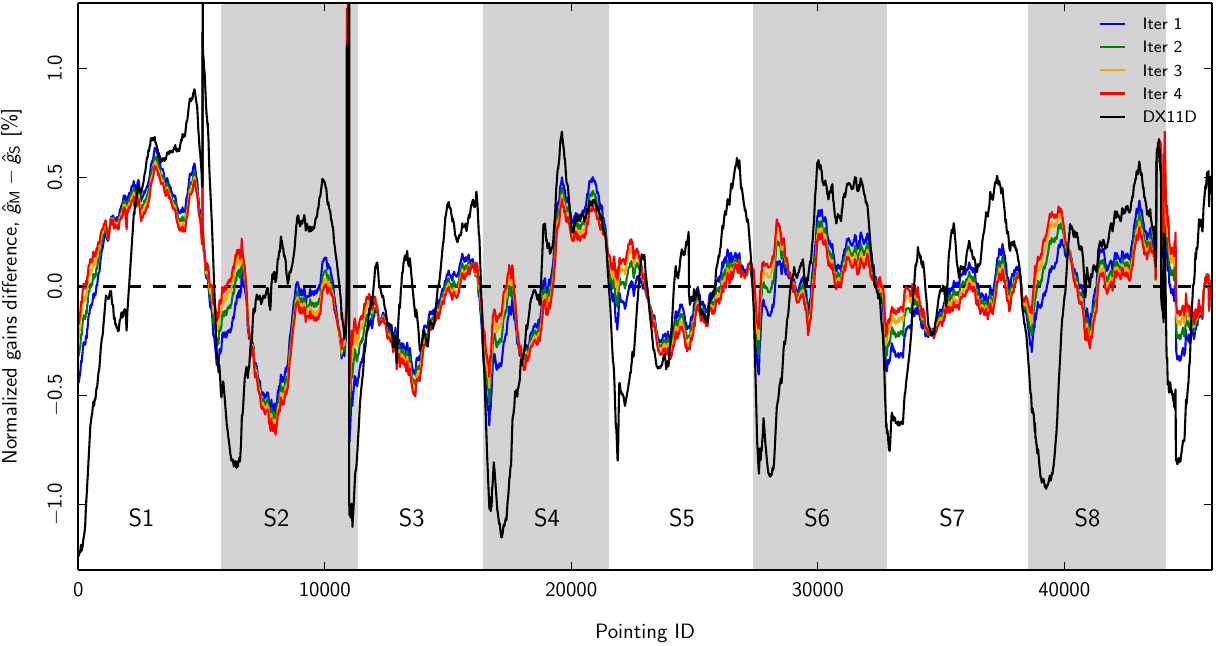}}
\caption{Normalized gain difference (see main text for precise definition) between
  two radiometers inside the same horn (28M and 28S here) as a function of pointing ID for
  both the 2015 (DX11D) and 2018 calibration schemes. The coloured lines show this function
  for the various iterations in the new \{gain-estimation + mapmaking + component separation\}
  calibration scheme. Sky surveys are indicated by alternating white and grey vertical bands.
  Other 30- and 44-GHz radiometers show qualitatively similar behaviour, whereas the 70-GHz
  radiometers exhibit too much noise, and corresponding iterations for these detectors do not
  converge within this scheme.}
\label{fig:gain_vs_iter}
\end{figure*}

Each LFI horn feeds two independent polarization-sensitive radiometers with
polarization angles rotated $90^{\circ}$ with respect to each other \citep{planck2013-p02};
these are called `M' (main) and `S' (side), respectively. Since two such radiometers are
often susceptible to the same instrumental effects (thermal, sidelobes, etc.), it is useful
to study differences between them to understand instrumental systematic effects.
For example, Fig.~\ref{fig:gain_vs_iter} shows the difference in normalized fractional
gain for two 30-GHz radiometers, namely $\hat{G}_{\mathrm{28M}}-\hat{G}_{\mathrm{28S}}$. Other
30- and 44-GHz radiometers show qualitatively similar behaviour, at the sub-percent
level, whereas the 70-GHz radiometers behave differently, for reasons explained below.
The following discussion therefore applies in detail only to the 30 and 44\,GHz
radiometers, while the 70-GHz radiometers will be treated separately.

\subsection{Calibration at 30 and 44\,GHz}
\label{sect:calresults}

The black curve in Fig.~\ref{fig:gain_vs_iter} shows the normalized gain difference
for the default 2015 orbital dipole-based calibration scheme (DX11D), and exhibits a striking
oscillatory pattern with a period equal to one survey. Such a pattern is very difficult
to explain instrumentally, since the two radiometers reside inside the same horn,
while it is consistent with the effect of polarized foregrounds. Since the polarization
angles of the two radiometers are rotated by $90^{\circ}$, any polarized signal on the sky
will at any given time be observed with opposite signs by the two radiometers. Strong
polarized foregrounds therefore lead to the kind of difference shown in
Fig.~\ref{fig:gain_vs_iter}, with a sign given by the relative orientation of the
satellite and the Galactic magnetic field. Furthermore, this difference will be
repeatable across surveys. This is confirmed by simulations --- inserting a polarized
foreground sky into end-to-end simulations induces precisely the same pattern as shown here.

The solution to this problem is to include the sky signal, $T_{\mathrm{sky}}$, in the
calibrator, on the same footing as the orbital and solar dipoles, including both
temperature and polarization fluctuations. This is non-trivial, since the purpose
of the experiment is precisely to measure the polarized emission from the sky.
A good approximation can be established, however, through an iterative process that
alternates between gain calibration, map making, and astrophysical component separation, using the following steps.
\begin{enumerate}
  \setcounter{enumi}{-1} 

\item Let $T_{\mathrm{sky}}$ be the full best-fit
  (\texttt{Commander}-based; \citealp{eriksen2008,planck2014-a12}) \Planck\ 2015
  astrophysical sky model, including CMB, synchrotron, free-free, thermal and
  spinning dust, and CO emission for temperature maps, plus CMB, synchrotron, and thermal dust in polarization.  

\item Estimate $G$ from Eq.~(\ref{eq:calib}), explicitly including the temperature
  and polarization component of $T_{\mathrm{sky}}$ in the calibration on the same
  footing as $D_{\mathrm{solar}}$ and $D_{\mathrm{orbital}}$.

\item Compute frequency maps with these new gains.

\item Determine a new astrophysical model from the updated frequency maps
  using \texttt{Commander} (at present the sky model is adjusted only for LFI frequencies.)

\item Iterate steps (1) to (3).
\end{enumerate}

Since the true sky signal is stationary on the sky, while the spurious gain
fluctuations are not, this process will converge, essentially corresponding to
a generalized mapmaker in which the $G(t)$ is estimated jointly with the sky maps.
Alternatively, this process may also be considered as a Gibbs sampler that in turn
iterates through all involved conditional distributions, and thereby converges to
the joint maximum likelihood point \citep{eriksen2008}.

The process is, computationally expensive, however; each iteration takes about
one week to complete. For practical purposes, the current process was therefore
limited to four full iterations (not counting the 2015 model used for initialization).
The normalized gain differences established in each iteration are shown as coloured
curves in Fig.~\ref{fig:gain_vs_iter} for the same radiometer pair as discussed above (for example 28M and 28S).
Here we see that most of the effect is accounted for simply by introducing a rough
model, as already the first iteration is significantly flatter than the initial
model (black versus blue curves). Subsequent iterations make relatively small
differences, and, critically, the differences between consecutive iterations become
smaller by almost a factor of 2 in each case, indicating that the algorithm indeed converges.

\begin{figure}[htpb]
\centerline{
\includegraphics[width=8.8cm]{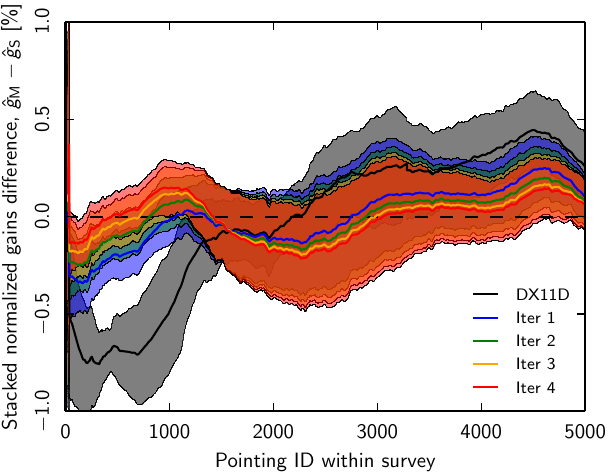}}
\caption{Same as Fig.~\ref{fig:gain_vs_iter}, but stacked over surveys. Each band
  corresponds to the mean and $1\,\sigma$ confidence region as evaluated from eight surveys. }
\label{fig:gain_vs_iter_stack}
\end{figure}

While the most obvious oscillatory pattern in the initial model has been eliminated
by the introduction of the astrophysical sky model, it is not clear whether a smaller
contribution remains.  In Fig.~\ref{fig:gain_vs_iter_stack} we therefore show the same
functions, but now with all surveys stacked together. Explicitly, the coloured regions
in Fig.~\ref{fig:gain_vs_iter_stack} represent the mean and standard deviations as
evaluated over the eight surveys. Since the surveys have slightly different lengths,
the stacking is done such that the starting pointing periods of the surveys are
aligned, and longer surveys are truncated at the end. 
These stacked functions will tend to suppress random signals across surveys,
but highlight those common to all eight surveys. 
As before (but now more clearly), we see that the 2015 model (grey band) exhibits a
highly significant Survey-dependent pattern. This pattern is greatly suppressed
simply by adding a rough model to the calibrator (blue band).  And the pattern
is additionally reduced by further iterations, with a convergence rate of
about a factor of 2 per iteration.

\begin{figure}[htpb]
\centerline{
\includegraphics[width=8.8cm]{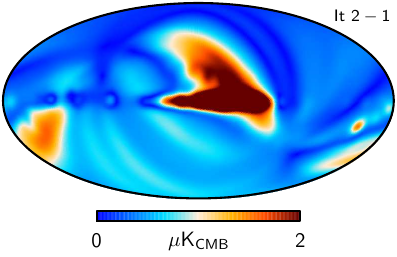}}
\vspace*{-13mm}
\centerline{
\includegraphics[width=8.8cm]{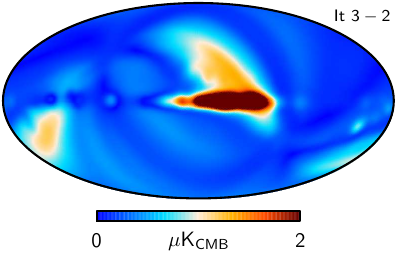}}
\vspace*{-13mm}
\centerline{
\includegraphics[width=8.8cm]{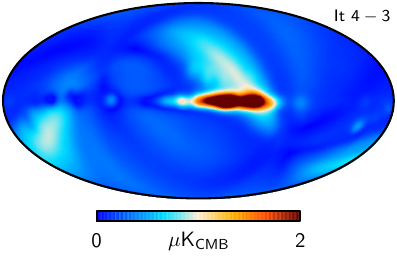}}
\caption{Polarization amplitude difference maps between consecutive
iterations of the internal foreground model evaluated at 30\,GHz, as
derived with \texttt{Commander}. The three panels show the differences
between: the second and first iterations (top); the
third and second iterations (middle); and the fourth
and third iterations (bottom). All maps are smoothed to an
effective angular resolution of $8^{\circ}$ FWHM.}
\label{fig:calib_conv}
\end{figure}

To understand the convergence rate in more detail, we show in Fig.~\ref{fig:calib_conv}
the differences in polarization amplitude between two consecutive iterations of the
full 30-GHz map. The top panel shows the difference between the second and first
iterations; the middle panel shows the difference between the third and second
iterations; and, finally, the bottom panel shows the difference between the fourth
and third iterations. As anticipated, here we see that the magnitude of the updates
decreases by a factor of 1.5--2 at high latitudes.

In addition to the decreasing amplitude with iterations, it is also important to
note that the \textit{morphology} of the three difference maps is very similar,
and dominated by a few scans that align with Ecliptic meridians. In other words,
most of the gain uncertainty is dominated by a few strong modes on the sky, and
the iterations described above largely try to optimize the amplitude of these
few modes. Furthermore, as seen in Figs.~\ref{fig:gain_vs_iter}--\ref{fig:calib_conv},
it is clear that we have not converged to numerical precision with only
four iterations. Due to the heavy computational demand of the iterative process, we could not produce further steps.  As a consequence, we expect that low-level residuals are still present in the 2018 LFI maps, with a pattern similar to that of the 2015 maps, though with significantly lower amplitude.  For the 2018 release, we adopt the difference between the two last iterations as a spatial template of residual gain uncertainties projected onto the sky.  This template is used only at 70\,GHz.

\subsection{Calibration at 70\,GHz}

\begin{figure}[htpb]
\centerline{
\includegraphics[width=8.8cm]{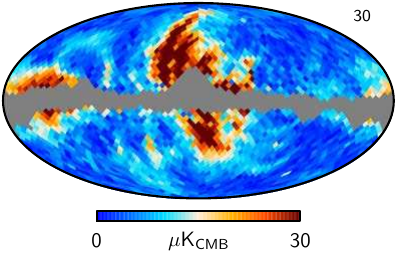}}
\centerline{
\includegraphics[width=8.8cm]{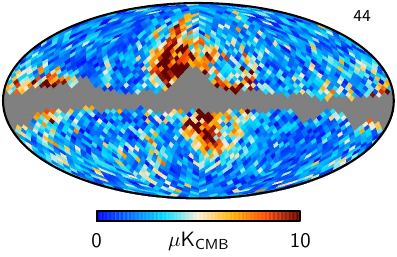}}
\centerline{
\includegraphics[width=8.8cm]{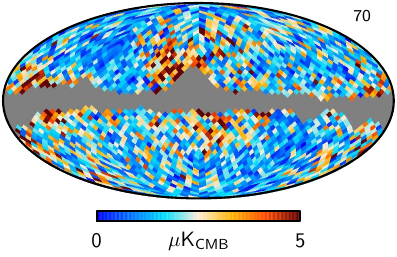}}
\caption{Final low-resolution LFI 2017 polarization amplitude sky maps.
  From \textit{top} to \textit{bottom}, the panels show the
  co-added 30-, 44-, and 70-GHz frequency maps. The grey regions indicate
  the mask used during gain calibration. Each pixel is
$3\pdeg8 \times 3\pdeg8$, corresponding to {\tt HEALPix} resolution $N_{\textrm{side}}=16$.}
\label{fig:calib_maps}
\end{figure}

As already mentioned, the above discussion applies only to the 30- and 44-GHz
radiometers, since the 70-GHz radiometers behave differently. The reason for this
may be seen in Fig.~\ref{fig:calib_maps}, which simply shows the final co-added 30-,
44-, and 70-GHz frequency maps, downgraded to {\tt HEALPix} \citep{gorski2005} $N_{\mathrm{side}}=16$
resolution ($3\pdeg8 \times 3\pdeg8$ pixels) to enhance the effective signal-to-noise
ratio per pixel. The grey regions show the Galactic calibration mask used in the analysis,
within which we do not trust the foreground model sufficiently precisely to use it in
gain calibration, primarily due to bandpass leakage effects \citep{planck2014-a03}.
Here we see a qualitative difference between the three frequency maps: while both
the 30- and 44-GHz polarization maps are signal-dominated, the 70-GHz channel is
noise-dominated. This has a detrimental effect on the iterative scheme described
above, ultimately resulting in a diverging process; essentially, the algorithm
attempts to calibrate on noise rather than actual signal.

\begin{figure}[htpb]
\centerline{
\includegraphics[width=8.8cm]{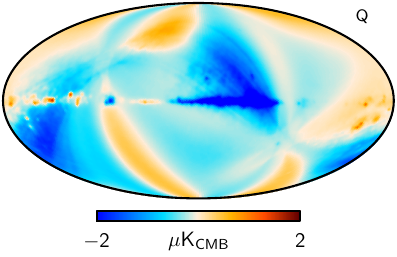}}
\vspace*{-13mm}
\centerline{
\includegraphics[width=8.8cm]{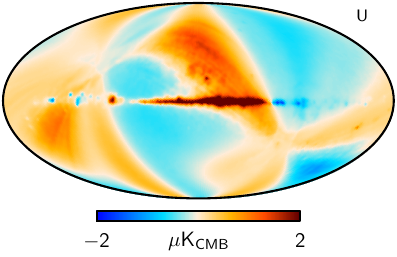}}
\centerline{
\includegraphics[width=8.8cm]{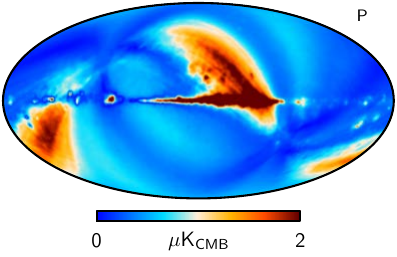}}
\caption{Gain correction template for the 70-GHz channel in terms of
Stokes $Q$ (top) and $U$ (middle), and the
polarization amplitude, $P$ (bottom). The template is
smoothed to $2^{\circ}$ FWHM, and its amplitude is normalized to the
best-fit value derived in a joint maximum likelihood analysis
of both the CMB power spectrum and template fit, as described in the text.}
\label{fig:gaintemp_70GHz}
\end{figure}

For this channel, we therefore retain the same calibration scheme used in 2015,
noting that the iterative algorithm fails because the foregrounds are weak at this
channel. While this is a problem when using foregrounds directly as a calibrator,
it also implies that foregrounds are much less of a problem than in the other channels.
In place of an iterative scheme, we adopt the corresponding internal differences
described above, obtained at 30\,GHz, as a tracer of gain residuals also for
the 70-GHz channel (shown in Fig.~\ref{fig:gaintemp_70GHz}), and marginalize
over this spatial template in a standard likelihood fit in pixel space.
Indeed, we provide this additive template as part of the LFI 2018 distribution,
with a normalization given by a best-fit likelihood accounting for both CMB and
astrophysical foregrounds. Thus, the best-fit amplitude of the provided template
is unity. For all the cosmological analysis involving the 70~GHz channel and 
presented in this final \Planck\, data release we accounted for gain residuals by
subtracting this correction. Therefore  we strongly recommend to do the same
for any other cosmological investigation involving the 70-GHz frequency channel.

\begin{figure}[htpb]
\centerline{
\includegraphics[width=8.8cm]{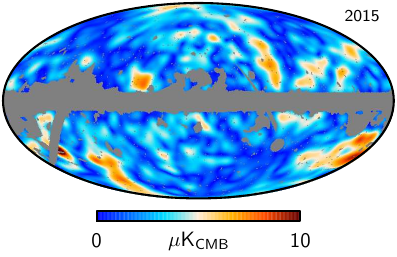}}
\vspace*{-13mm}
\centerline{
\includegraphics[width=8.8cm]{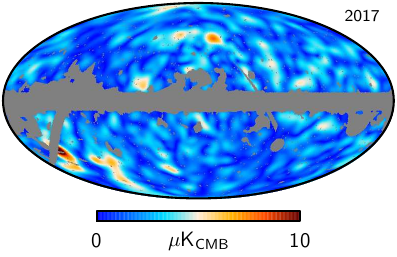}}
\caption{30\,GHz polarization amplitude null maps evaluated between survey
combinations \{S1,S3, S5--8\} and \{S2, S4\} for both the 2015
(top) and 2018 (bottom) calibration schemes. This
particular survey split is maximally sensitive to residual gain
uncertainties from polarized Galactic foreground contamination because
of the orientation of the scanning strategy employed in Surveys~2 and
4; see text and \citet{planck2014-a03} for further details. Both
maps are smoothed to an effective angular resolution of $8^{\circ}$ FWHM.}
\label{fig:calib_30null}
\end{figure}

We started this discussion by recalling the failure reported in \citet{planck2014-a03}
of a null test between Survey sets \{1,3, 5--8\} and \{2, 4\}. In Fig.~\ref{fig:calib_30null}
we therefore compare this particular null map as derived with the old (top panel)
and new (bottom panel) calibration schemes. The improvement is obvious, with most
fluctuations in the latter appearing consistent with noise. The remaining excesses
appear on angular scales compatible with the smoothing scale of $8^\circ$ FHWM, and
are consistent with the position of known variable point sources. 
A direct comparison with \WMAP\ frequency maps suggests similar improvements;
see Appendix~\ref{app_LFIWMAP} for details.

\section{The LFI Dipole}
\label{sec_lfidipole}

\begin{table*}[htpb]
\begingroup
\newdimen\tblskip \tblskip=5pt
\caption{Dipole characterization.}
  \label{tab_dipoleParameters}
  \nointerlineskip
  \vskip -2mm
  \footnotesize
  \setbox\tablebox=\vbox{
  \newdimen\digitwidth
  \setbox0=\hbox{\rm 0}
  \digitwidth=\wd0
  \catcode`*=\active
  \def*{\kern\digitwidth}
  \newdimen\signwidth
  \setbox0=\hbox{+}
  \signwidth=\wd0
  \catcode`!=\active
  \def!{\kern\signwidth}
\halign{\hbox to 1.3in{#\leaderfil}\tabskip=2em&
    \hfil$#$\hfil \tabskip 2em&
    \hfil$#$\hfil \tabskip 2em&
    \hfil$#$\hfil \tabskip 0em\cr
\noalign{\doubleline}
\omit&&\multispan2\hfil\sc Galactic Coordinates \hfil\cr
\noalign{\vskip -3pt}
\omit&\omit\hfil Amplitude\hfil&\multispan2\hrulefill\cr
\noalign{\vskip 3pt} 
\omit\hfil Radiometer\hfil&[\,\mu\mathrm{K}_{\rm CMB}]& l& b\cr
\noalign{\vskip 3pt\hrule\vskip 5pt}
\noalign{\vskip 2pt}
\omit{\bf 30\,GHz}\hfil\cr
\noalign{\vskip 4pt}
\hglue 2em 28M& 3357.82\pm1.09& 264\pdeg242\pm0\pdeg030& 48\pdeg174\pm0\pdeg009\cr
\hglue 2em 28S& 3363.98\pm1.37& 264\pdeg252\pm0\pdeg198& 48\pdeg192\pm0\pdeg062\cr
\hglue 2em 27M& 3366.99\pm1.46& 264\pdeg323\pm0\pdeg181& 48\pdeg195\pm0\pdeg042\cr
\hglue 2em 27S& 3368.23\pm1.15& 264\pdeg226\pm0\pdeg091& 48\pdeg160\pm0\pdeg043\cr 
\noalign{\vskip 5pt}
\omit{\bf 44\,GHz}\hfil\cr
\noalign{\vskip 4pt}
\hglue 2em 26M& 3363.83\pm0.34& 264\pdeg045\pm0\pdeg016& 48\pdeg253\pm0\pdeg007\cr 
\hglue 2em 26S& 3363.64\pm0.35& 264\pdeg026\pm0\pdeg015& 48\pdeg255\pm0\pdeg006\cr
\hglue 2em 25M& 3360.22\pm0.35& 264\pdeg021\pm0\pdeg014& 48\pdeg252\pm0\pdeg006\cr
\hglue 2em 25S& 3360.96\pm0.36& 264\pdeg009\pm0\pdeg014& 48\pdeg245\pm0\pdeg006\cr
\hglue 2em 24M& 3363.94\pm0.32& 264\pdeg012\pm0\pdeg014& 48\pdeg253\pm0\pdeg007\cr
\hglue 2em 24S& 3363.13\pm0.37& 264\pdeg034\pm0\pdeg015& 48\pdeg257\pm0\pdeg006\cr
\noalign{\vskip 5pt}
\omit{\bf 70\,GHz}\hfil\cr
\noalign{\vskip 4pt}
\hglue 2em 23M& 3364.18\pm0.30& 264\pdeg005\pm0\pdeg011& 48\pdeg267\pm0\pdeg005\cr
\hglue 2em 23S& 3364.69\pm0.30& 264\pdeg003\pm0\pdeg012& 48\pdeg271\pm0\pdeg005\cr
\hglue 2em 22M& 3364.15\pm0.28& 263\pdeg988\pm0\pdeg011& 48\pdeg269\pm0\pdeg005\cr
\hglue 2em 22S& 3364.63\pm0.28& 264\pdeg013\pm0\pdeg012& 48\pdeg263\pm0\pdeg005\cr
\hglue 2em 21M& 3364.16\pm0.30& 263\pdeg991\pm0\pdeg011& 48\pdeg267\pm0\pdeg005\cr
\hglue 2em 21S& 3364.20\pm0.27& 264\pdeg007\pm0\pdeg012& 48\pdeg262\pm0\pdeg006\cr
\hglue 2em 20M& 3364.74\pm0.27& 263\pdeg983\pm0\pdeg012& 48\pdeg266\pm0\pdeg005\cr
\hglue 2em 20S& 3364.38\pm0.27& 263\pdeg991\pm0\pdeg012& 48\pdeg261\pm0\pdeg006\cr
\hglue 2em 19M& 3363.94\pm0.27& 263\pdeg984\pm0\pdeg012& 48\pdeg265\pm0\pdeg005\cr
\hglue 2em 19S& 3364.58\pm0.28& 264\pdeg019\pm0\pdeg013& 48\pdeg264\pm0\pdeg006\cr
\hglue 2em 18M& 3364.42\pm0.28& 264\pdeg005\pm0\pdeg011& 48\pdeg267\pm0\pdeg005\cr
\hglue 2em 18S& 3364.23\pm0.29& 263\pdeg991\pm0\pdeg010& 48\pdeg263\pm0\pdeg005\cr
\noalign{\vskip 0.5em}
Combined$^a$& 3364.4\pm3.1& 263\pdeg998\pm0\pdeg051& 48\pdeg265\pm0\pdeg015\cr
\noalign{\vskip 5pt\hrule\vskip 10pt}}}
\endPlancktable 
\tablenote a This estimate is based on the collective sum of all the Markov chain Monte Carlo
(MCMC) samples for all 70-GHz channels.  Final amplitude error
bars include 0.07--0.11\,\% calibration uncertainty. \par

\endgroup
\end{table*}

\medskip
\noindent

The calibration signal for LFI is the dipole anisotropy due the motion of the
solar system relative to the CMB. Precise knowledge of the amplitude and
direction of the 3.3-mK solar dipole, however, requires another absolutely
defined signal. This is given by the orbital dipole,
the time-varying 200-$\mu$K modulation of
the dipole amplitude induced by the motion of the spacecraft in its yearly orbit
around the Sun (including the small velocity component due to the spacecraft
orbit around L2). As the amplitude and orientation of the orbital dipole can
be determined with exquisite accuracy from the satellite telemetry and orbital
ephemeris, it is the best absolute calibration signal in all of microwave space astrophysics.
It should be emphasized that the dipole determination is primarily a velocity
measurement, and that the actual dipole amplitude is derived from the velocity
assuming a value for the absolute temperature
of the CMB; together with HFI \citep{planck2016-l03} we use
the value $T_0 = 2.72548 $K \citep{fixsen2009}.

\subsection{Initial calibration to determine the amplitude and direction of
the solar dipole}

These two dipoles are merged into a single signal at any given time, but they can
be separated over the course of the mission, since the solar dipole is fixed on
the sky while the orbital dipole varies in amplitude and direction with the
satellite velocity as it orbits the Sun. It is therefore possible to base the
calibration entirely on the orbital dipole alone.  As in the previous release \citep{planck2014-a03},
we omit the solar dipole from the fit but retain the far-sidelobe-convolved orbital
dipole and the fiducial dipole convolved again with far sidelobes, and also
remove the restriction of having no dipole signal in the residual map.
In this way, the solar dipole is extracted as a residual, and its
amplitude and position can be determined.  In this section we discuss
the LFI 2018 measurements of the solar dipole and compare them to other measurements.

For accurate calibration,  we have to take into account two effects
that behave like the orbital dipole, in the sense that they are linked
to the satellite and not to the sky: polarized foregrounds; and pick-up in
the far sidelobes. While the orbital dipole calibration is robust against
unpolarized foregrounds, the polarized part of the foregrounds depends on
the orientation of the satellite. Similarly, the far sidelobes are also
locked to the direction in which the satellite is pointing.

The corrections for polarized foregrounds are made directly in the
timelines by unrolling the $Q$ and $U$ frequency maps from the previous
internal data release, in other words, projecting them into timelines according to
the scanning strategy and beam orientation,  and also taking into account
the gain calibration factor derived from an initial calibration run.  We find
that only one iteration is required to remove the polarized signal, with further
iterations in this cleaning process making no difference. The amplitude of the
polarized signal removed is about 40\muK\ at 30\GHz, 15\muK\ at 44, and 4\muK\
at 70\GHz, mainly due to the North Polar Spur and the Fan regions.

In the previous dipole analysis, the far sidelobes were removed using
the {\tt{GRASP}} beam model but reduced to the lowest multipoles to obtain
the expected, properly convolved dipole signal for both the
orbital and solar dipoles.
In the calibration code now used ({\tt{DaCapo}}; \citealt{planck2014-a03}),
we fit for the orbital dipole convolved with far sidelobes, as well as
for the convolution of the solar dipole with the far sidelobes. In such a
way, we force a pure dipole (without far sidelobes) into the residual map.
However, we found that far-sidelobe pick-up was not completely removed, which
resulted in a trend in the dipole amplitude with horn position on the focal
plane, as well as small differences between the orbital and solar dipole
calibration factors.  By adjusting the direction of the large-scale compoment
of the far sidelobes, we find a correction between $1$ and $10\muK$, depending
on horn focal plane position, which brings both calibration factors together
and, simultaneously, removes the asymmetry in the focal plane.

To calibrate on the orbital dipole, we need to mask the strong emission from
the Galactic plane.  The mask is generated using a 5\deg-smoothed 30-GHz
intensity map with different threshold cut-offs, which result in different
sky fractions. The orbital calibration is carried out using a sky fraction
of 94\,\%, since this calibration is robust to the intensity of unpolarized
foregrounds. For the analysis of the resultant dipole maps, we use instead
an 80\,\% sky fraction, accounting for the presence of unpolarized foregrounds
in the residual maps.  An even more conservative mask with a sky fraction
of 60\,\% gives about 15\,\% more scatter in the dipole position between channels,
but not in any systematic way.  This is consistent with a lower signal-to-noise
ratio due to the poorer sky coverage.  The actual fit is performed with a
Markov chain Monte Carlo (MCMC) approach, where we search for dipole position and
amplitude as well as the amplitudes of synchrotron, dust, and free-free
templates derived from {\tt{Commander}} \citep{planck2014-a12}.  We performed
several tests, varying the amplitude of the mask (ranging from 60\,\% to 80\,\% sky
fraction), with and without point source masks, in the derivation of the
foreground templates. We found that use of a mask with 80\,\% sky fraction
and masked foregrounds reduces the scatter in the dipole estimation at 70\GHz\ by about 15\,\%.

\subsection{The solar dipole}

\begin{figure}[htpb]
\includegraphics[width=88mm]{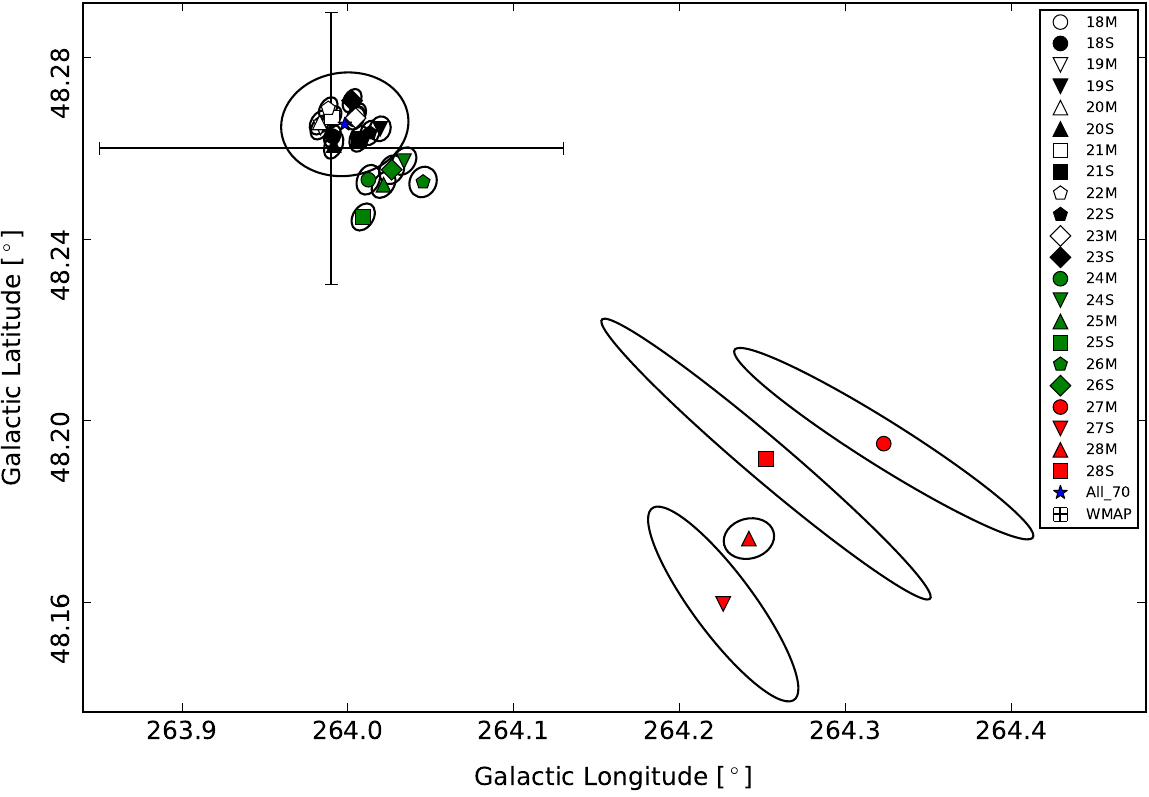}
\vskip 5pt
\includegraphics[width=88mm]{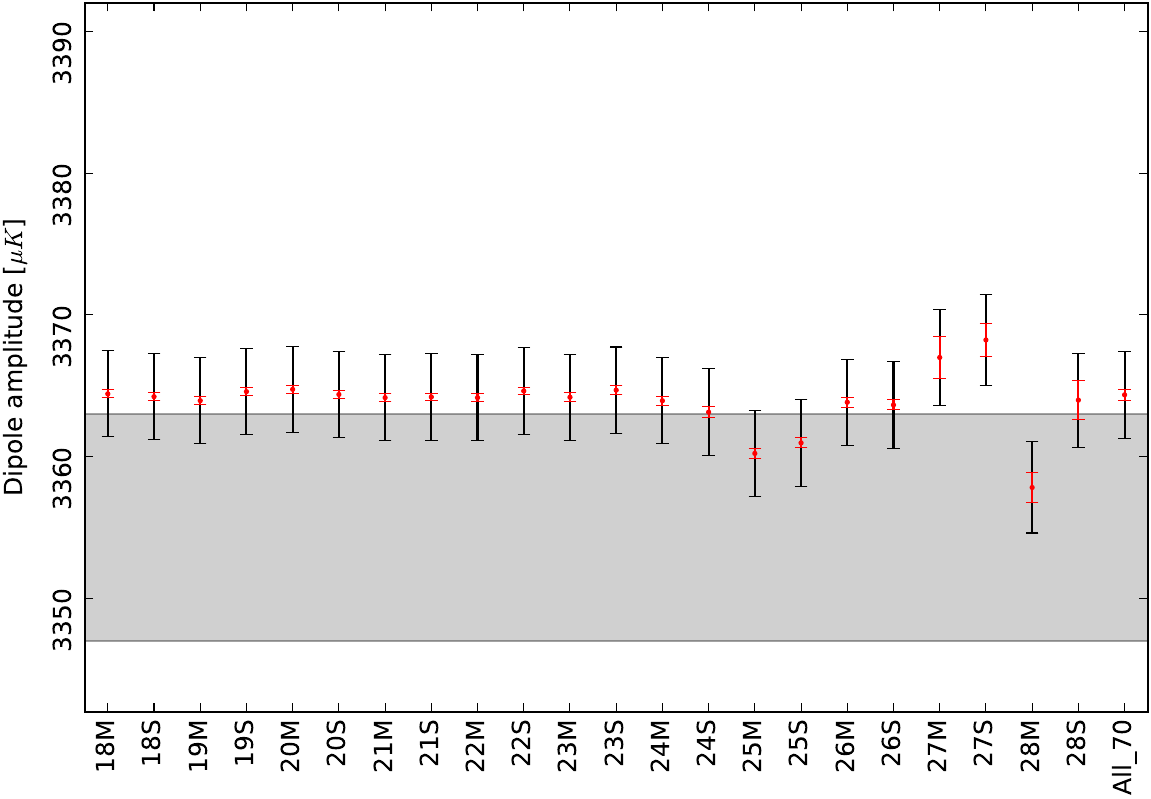}
\caption{Direction (top) and amplitude (bottom) of the solar dipole
  determined from each of the LFI detectors.  Uncertainties in direction are
  given by 95\,\% ellipses around symbols, colour-coded for frequency (70\,GHz
  black or unfilled, 44\,GHz green, and 30\,GHz red). The four discrepant points
  are at 30\GHz, where we expect foregrounds affect dipole estimation.
  Amplitude uncertainties
  are dominated by the systematic effects of gain uncertainties. The grey band in
  the bottom panel shows the \WMAP\ dipole amplitude for comparison.}
\label{fig_dipoleParametersPlot}
\end{figure}

From the MCMC samples, the 1\,\% and 99\,\% values are used to set the limits on the
dipole amplitude and position. Figure~\ref{fig_dipoleParametersPlot} shows results
for the three LFI frequencies for both dipole direction (upper panel) and 
amplitude (lower panel).  There is a clear trend with frequency in dipole
direction, due to foreground contamination.
As expected, the 70-GHz channel has the lowest foreground signal, and it
is used to derive the final LFI dipole. With respect to the previous release,
the use of the small far-sidelobe correction has removed the systematic amplitude
variation with focal plane position.  Thus the cross-plane null pairs that were
used in the previous release are not needed.  This results in a smaller scatter
of both dipole positions and derived amplitude, as shown in the bottom panel.
For each LFI data point we report two error bars: the small (red) one is the actual error
in the fit (also reported in Table~\ref{tab_dipoleParameters});
and the large (black) one is obtained by summing the calibration
error in quadrature.
The grey band represents the \WMAP\ derived dipole amplitude, for comparison.
Numerical results are summarized in Table~\ref{tab_dipoleParameters}, where single
radiometer errors are derived from the MCMC samples. The final uncertainty in
the LFI dipole derived from only the 70\GHz\, measurements, however, also takes
also into account gain errors, estimated through the use of
dedicated simulations with {\tt{DaCapo}},
in the range 0.07--0.11\,\%.  This yields a final dipole amplitude $D=3364\pm3\muK$ and
direction in Galactic coordinates 
$(l,b)=(263\pdeg998\pm0\pdeg051, 48\pdeg265\pm0\pdeg015)$.

\section{Noise estimation}
\label{sec_general_noise}

\medskip
\noindent

We estimated the basic noise properties of the receivers (for example, knee-frequency
and white-noise variance) throughout the mission lifetime.  This is a simple way
to track variations and possible instrument anomalies during operations.
Furthermore, a detailed knowledge of noise properties is required for other
steps of the analysis pipeline, such as optimal detector combination in the
mapmaking process or Monte Carlo simulations used for error evaluation at the power spectrum level.

\begin{figure}[htpb]
\centerline{
  \includegraphics[width=88mm]{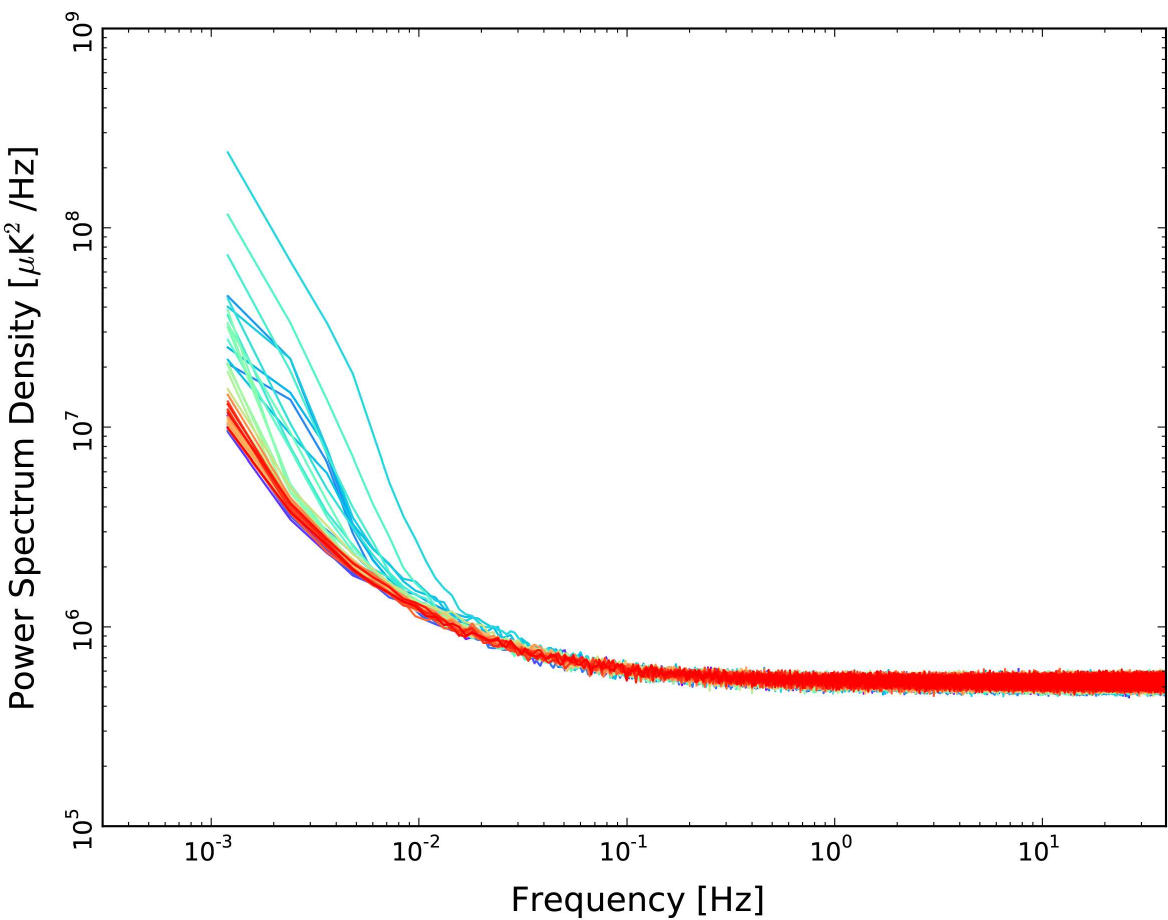}}
\centerline{
  \includegraphics[width=88mm]{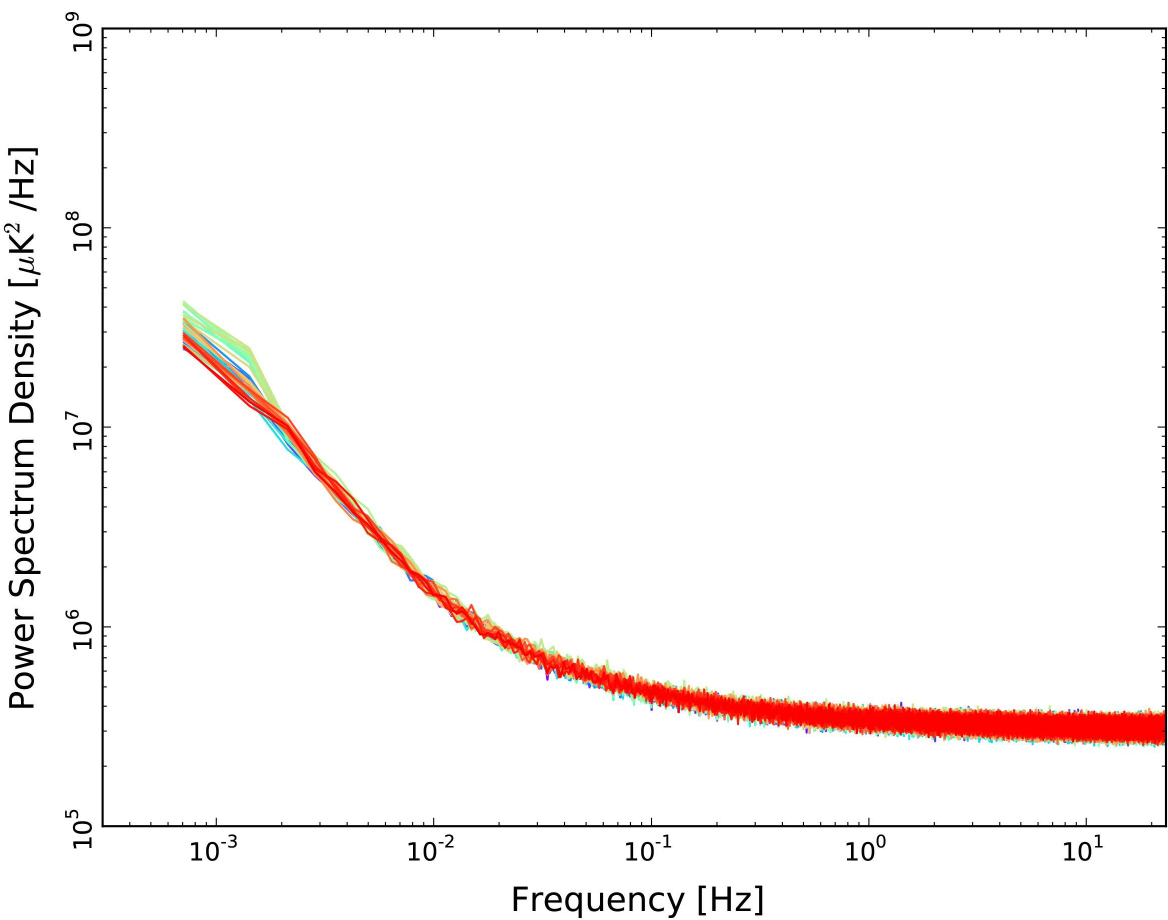}}
\centerline{
\includegraphics[width=88mm]{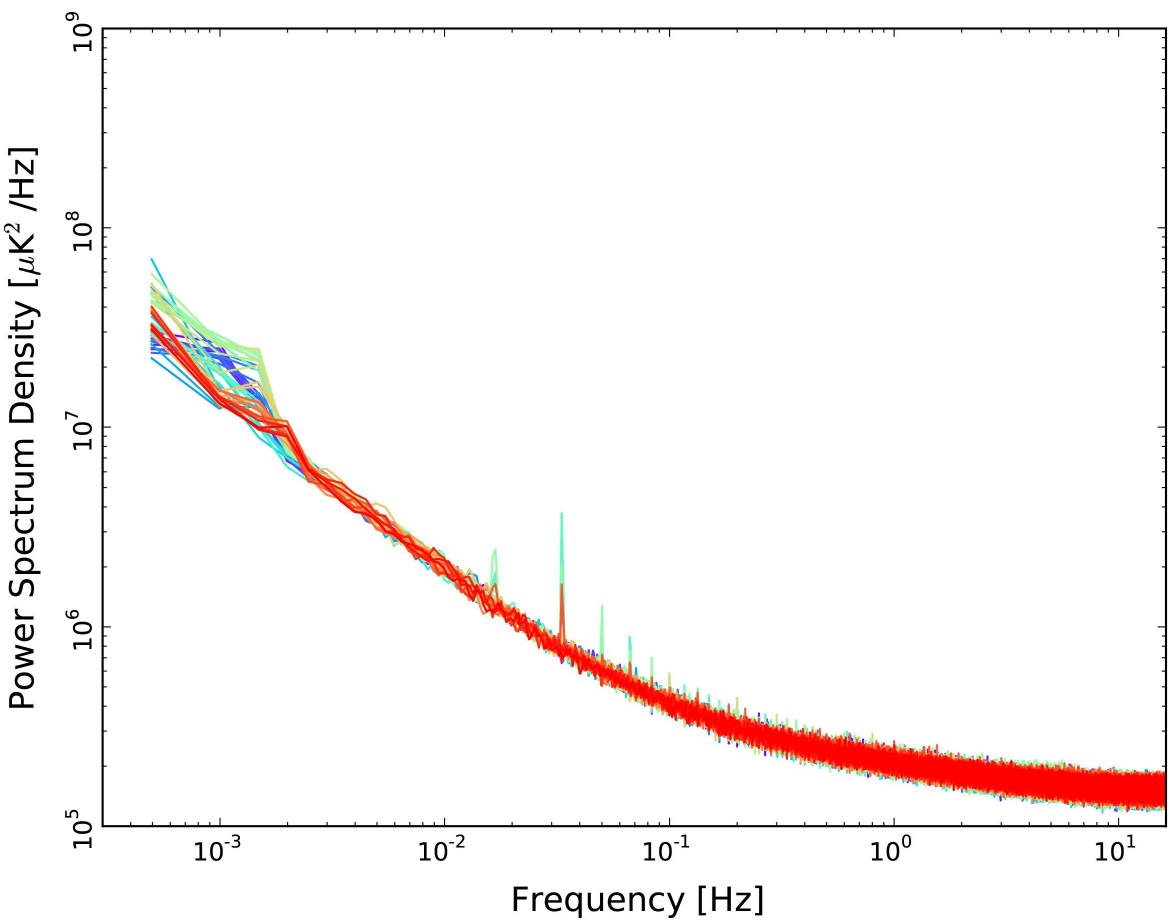}}
\caption{Evolution of noise spectra over the mission lifetime for radiometer 18M (70\GHz\, top),
  25S (44\GHz\, middle), and 27M (30\GHz\, bottom). Spectra are colour-coded, ranging
  from OD 100 (blue) to OD 1526 (red), with intervals of about 20 ODs. White-noise levels and
  slope stability are considerably better than in the 2015 release, being at the 0.1\,\% level
  for noise, while knee-frequencies show variations at the 1.5\,\% level.} 
\label{noisespectra}
\end{figure}

The noise model and the approach for noise estimation is the same as
described in \citet{planck2014-a03} and \citet{planck2013-p02}. We employed a noise model of the form
\begin{equation}
P(f) = P_0^2\left[ 1+\left(\frac{f}{\fknee}\right)^\beta\right]\, ,
\label{noisemodel}
\end{equation}
\noindent
%\DS{I think that if this is a power spectrum expression, then $\sigma$ isn't
%the right quantity to put here.  Similarly, I don't think that the thing
%quoted in Table~4 is actually $\sigma$ either (and I changed the wording there
%to ``white-noise level'', rather than ``sensitivity'').
%Maybe we should call this term something like
%$P_0^2$?  And then would the thing in Table~4 be $P_0$?}
where $P_0^2$ is the white-noise power spectrum level, and \fknee\ and $\beta$ encode
the non-white ($1/f$-like) low-frequency noise component. We estimate $P_0^2$ by taking
the mean of the noise spectrum in the last few bins at the highest values
of $f$ (typically 10\,\% at 44 and 70\,GHz, and 5\,\% at 30\,GHz due to the higher
knee-frequency).  For the knee frequency and slope, we exploited the same MCMC
engine as in the previous release.  Tables~\ref{tab_white_noise_per_radiometer}
and \ref{tab_one_over_f_noise_per_radiometer} give white noise and low-frequency
noise parameters, respectively. Comparing these results with those from the 2015
release \citep{planck2014-a03}, we see that both the white-noise level and
slope $\beta$  show variations well below 0.1\,\%, while \fknee \ varies by less
than 1.5\,\%. However, error bars (rms fitted values
over the mission lifetime) are in some cases larger than before.  This results
from both improved TOI processing (mainly flagging) and an improved calibration
pipeline that allows us to detect a sort of bimodal distribution (at the $\simeq 1\,\%$ level)
in \fknee\, for three of the LFI radiometers. Figure~\ref{noisespectra} shows typical
noise spectra at several times during the mission lifetime for three representative
radiometers, one for each LFI frequency. Spikes at the spin-frequency (and its harmonics) are visible
in the 30-GHz spectra due to residual signal left over in the noise estimation procedure
\citep[see][]{planck2014-a03}. These are due to the combined effect of the large signal (mainly
from the Galaxy) and the large value of the knee frequency, together with the limited time
window (only 5 ODs) on which spectra are computed.

  \begin{table}[htpb]
  \begingroup
  \newdimen\tblskip \tblskip=5pt
  \caption{White-noise levels for the LFI radiometers.}
  \label{tab_white_noise_per_radiometer}
  \nointerlineskip
  \vskip -5mm
  \footnotesize
  \setbox\tablebox=\vbox{
  \newdimen\digitwidth
  \setbox0=\hbox{\rm 0}
  \digitwidth=\wd0
  \catcode`*=\active
  \def*{\kern\digitwidth}
  \newdimen\signwidth
  \setbox0=\hbox{+}
  \signwidth=\wd0
  \catcode`!=\active
  \def!{\kern\signwidth}
  \halign{\hbox to 1.3in{#\leaderfil}\tabskip=2em&
      \hfil#\hfil&
      \hfil#\hfil\tabskip=0pt\cr
  \noalign{\doubleline}
  \omit&\multispan2\hfil W{\sc hite}-N{\sc oise} L{\sc evel}\hfil\cr
  \noalign{\vskip -4pt}
  \omit&\multispan2\hrulefill\cr
  \omit&Radiometer M&Radiometer S\cr
  \omit\hfil\sc Radiometer\hfil&[$\,\mu\mathrm{K}_{\rm CMB}\, \mathrm{s}^{1/2}$]&[$\,\mu\mathrm{K}_{\rm CMB}\, \mathrm{s}^{1/2}$]\cr
  \noalign{\vskip 3pt\hrule\vskip 5pt}
  \omit{\bf 70\,GHz}\hfil\cr
  \noalign{\vskip 4pt}
  \hglue 2em LFI-18&\getsymbol{LFI:white:noise:sensitivity:LFI18:Rad:M}$\,\pm\,$\getsymbol{LFI:white:noise:sensitivity:uncertainty:LFI18:Rad:M}&\getsymbol{LFI:white:noise:sensitivity:LFI18:Rad:S}$\,\pm\,$\getsymbol{LFI:white:noise:sensitivity:uncertainty:LFI18:Rad:S}\cr
  \hglue 2em LFI-19&\getsymbol{LFI:white:noise:sensitivity:LFI19:Rad:M}$\,\pm\,$\getsymbol{LFI:white:noise:sensitivity:uncertainty:LFI19:Rad:M}&\getsymbol{LFI:white:noise:sensitivity:LFI19:Rad:S}$\,\pm\,$\getsymbol{LFI:white:noise:sensitivity:uncertainty:LFI19:Rad:S}\cr
  \hglue 2em LFI-20&\getsymbol{LFI:white:noise:sensitivity:LFI20:Rad:M}$\,\pm\,$\getsymbol{LFI:white:noise:sensitivity:uncertainty:LFI20:Rad:M}&\getsymbol{LFI:white:noise:sensitivity:LFI20:Rad:S}$\,\pm\,$\getsymbol{LFI:white:noise:sensitivity:uncertainty:LFI20:Rad:S}\cr
  \hglue 2em LFI-21&\getsymbol{LFI:white:noise:sensitivity:LFI21:Rad:M}$\,\pm\,$\getsymbol{LFI:white:noise:sensitivity:uncertainty:LFI21:Rad:M}&\getsymbol{LFI:white:noise:sensitivity:LFI21:Rad:S}$\,\pm\,$\getsymbol{LFI:white:noise:sensitivity:uncertainty:LFI21:Rad:S}\cr
  \hglue 2em LFI-22&\getsymbol{LFI:white:noise:sensitivity:LFI22:Rad:M}$\,\pm\,$\getsymbol{LFI:white:noise:sensitivity:uncertainty:LFI22:Rad:M}&\getsymbol{LFI:white:noise:sensitivity:LFI22:Rad:S}$\,\pm\,$\getsymbol{LFI:white:noise:sensitivity:uncertainty:LFI22:Rad:S}\cr
  \hglue 2em LFI-23&\getsymbol{LFI:white:noise:sensitivity:LFI23:Rad:M}$\,\pm\,$\getsymbol{LFI:white:noise:sensitivity:uncertainty:LFI23:Rad:M}&\getsymbol{LFI:white:noise:sensitivity:LFI23:Rad:S}$\,\pm\,$\getsymbol{LFI:white:noise:sensitivity:uncertainty:LFI23:Rad:S}\cr
  \noalign{\vskip 5pt}
  \omit{\bf 44\,GHz}\hfil\cr
  \noalign{\vskip 4pt}
  \hglue 2em LFI-24&\getsymbol{LFI:white:noise:sensitivity:LFI24:Rad:M}$\,\pm\,$\getsymbol{LFI:white:noise:sensitivity:uncertainty:LFI24:Rad:M}&\getsymbol{LFI:white:noise:sensitivity:LFI24:Rad:S}$\,\pm\,$\getsymbol{LFI:white:noise:sensitivity:uncertainty:LFI24:Rad:S}\cr
  \hglue 2em LFI-25&\getsymbol{LFI:white:noise:sensitivity:LFI25:Rad:M}$\,\pm\,$\getsymbol{LFI:white:noise:sensitivity:uncertainty:LFI25:Rad:M}&\getsymbol{LFI:white:noise:sensitivity:LFI25:Rad:S}$\,\pm\,$\getsymbol{LFI:white:noise:sensitivity:uncertainty:LFI25:Rad:S}\cr
  \hglue 2em LFI-26&\getsymbol{LFI:white:noise:sensitivity:LFI26:Rad:M}$\,\pm\,$\getsymbol{LFI:white:noise:sensitivity:uncertainty:LFI26:Rad:M}&\getsymbol{LFI:white:noise:sensitivity:LFI26:Rad:S}$\,\pm\,$\getsymbol{LFI:white:noise:sensitivity:uncertainty:LFI26:Rad:S}\cr
  \noalign{\vskip 5pt}
  \omit{\bf 30\,GHz}\hfil\cr
  \noalign{\vskip 4pt}
  \hglue 2em LFI-27&\getsymbol{LFI:white:noise:sensitivity:LFI27:Rad:M}$\,\pm\,$\getsymbol{LFI:white:noise:sensitivity:uncertainty:LFI27:Rad:M}&\getsymbol{LFI:white:noise:sensitivity:LFI27:Rad:S}$\,\pm\,$\getsymbol{LFI:white:noise:sensitivity:uncertainty:LFI27:Rad:S}\cr
  \hglue 2em LFI-28&\getsymbol{LFI:white:noise:sensitivity:LFI28:Rad:M}$\,\pm\,$\getsymbol{LFI:white:noise:sensitivity:uncertainty:LFI28:Rad:M}&\getsymbol{LFI:white:noise:sensitivity:LFI28:Rad:S}$\,\pm\,$\getsymbol{LFI:white:noise:sensitivity:uncertainty:LFI28:Rad:S}\cr
  \noalign{\vskip 5pt\hrule\vskip 3pt}}}
  \endPlancktable
  \endgroup
  \end{table}

  % Table with 1/f noise properties
  \begin{table*}[htpb]
  \begingroup
  \newdimen\tblskip \tblskip=5pt
  \caption{Knee frequencies and slopes for each of the LFI radiometers.}
  \label{tab_one_over_f_noise_per_radiometer}
  \nointerlineskip
  \vskip -3mm
  \footnotesize
  \setbox\tablebox=\vbox{
  \newdimen\digitwidth
  \setbox0=\hbox{\rm 0}
  \digitwidth=\wd0
  \catcode`*=\active
  \def*{\kern\digitwidth}
  \newdimen\signwidth
  \setbox0=\hbox{+}
  \signwidth=\wd0
  \catcode`!=\active
  \def!{\kern\signwidth}
  \halign{\tabskip=0pt\hbox to 1.3in{#\leaderfil}\tabskip=2em&
      \hfil#\hfil&
      \hfil#\hfil&
      \hfil#\hfil&
      \hfil#\hfil\tabskip=0pt\cr
  \noalign{\doubleline}
  \omit&\multispan2\hfil K{\sc nee} F{\sc requency} $f_{\rm knee}$ [mHz]\hfil&\multispan2\hfil S{\sc lope} $\beta$\hfil\cr
  \noalign{\vskip -4pt}
  \omit&\multispan2\hrulefill&\multispan2\hrulefill\cr
  \omit& Radiometer M&Radiometer S&Radiometer M&Radiometer S\cr
  \noalign{\vskip 3pt\hrule\vskip 5pt}
  \omit{\bf 70\,GHz}\hfil\cr
  \noalign{\vskip 4pt}
  \hglue 2em LFI-18&*\getsymbol{LFI:knee:frequency:LFI18:Rad:M}$\,\pm\,$\getsymbol{LFI:knee:frequency:uncertainty:LFI18:Rad:M}&
              *\getsymbol{LFI:knee:frequency:LFI18:Rad:S}$\,\pm\,$\getsymbol{LFI:knee:frequency:uncertainty:LFI18:Rad:S}&
              \getsymbol{LFI:slope:LFI18:Rad:M}$\,\pm\,$\getsymbol{LFI:slope:uncertainty:LFI18:Rad:M}&
              \getsymbol{LFI:slope:LFI18:Rad:S}$\,\pm\,$\getsymbol{LFI:slope:uncertainty:LFI18:Rad:S}\cr
  \hglue 2em LFI-19&*\getsymbol{LFI:knee:frequency:LFI19:Rad:M}$\,\pm\,$\getsymbol{LFI:knee:frequency:uncertainty:LFI19:Rad:M}&
              *\getsymbol{LFI:knee:frequency:LFI19:Rad:S}$\,\pm\,$\getsymbol{LFI:knee:frequency:uncertainty:LFI19:Rad:S}&
              \getsymbol{LFI:slope:LFI19:Rad:M}$\,\pm\,$\getsymbol{LFI:slope:uncertainty:LFI19:Rad:M}&
              \getsymbol{LFI:slope:LFI19:Rad:S}$\,\pm\,$\getsymbol{LFI:slope:uncertainty:LFI19:Rad:S}\cr
  \hglue 2em LFI-20&**\getsymbol{LFI:knee:frequency:LFI20:Rad:M}$\,\pm\,$\getsymbol{LFI:knee:frequency:uncertainty:LFI20:Rad:M}&
              **\getsymbol{LFI:knee:frequency:LFI20:Rad:S}$\,\pm\,$\getsymbol{LFI:knee:frequency:uncertainty:LFI20:Rad:S}&
              \getsymbol{LFI:slope:LFI20:Rad:M}$\,\pm\,$\getsymbol{LFI:slope:uncertainty:LFI20:Rad:M}&
              \getsymbol{LFI:slope:LFI20:Rad:S}$\,\pm\,$\getsymbol{LFI:slope:uncertainty:LFI20:Rad:S}\cr
  \hglue 2em LFI-21&*\getsymbol{LFI:knee:frequency:LFI21:Rad:M}$\,\pm\,$\getsymbol{LFI:knee:frequency:uncertainty:LFI21:Rad:M}&
              *\getsymbol{LFI:knee:frequency:LFI21:Rad:S}$\,\pm\,$\getsymbol{LFI:knee:frequency:uncertainty:LFI21:Rad:S}&
              \getsymbol{LFI:slope:LFI21:Rad:M}$\,\pm\,$\getsymbol{LFI:slope:uncertainty:LFI21:Rad:M}&
              \getsymbol{LFI:slope:LFI21:Rad:S}$\,\pm\,$\getsymbol{LFI:slope:uncertainty:LFI21:Rad:S}\cr
  \hglue 2em LFI-22&**\getsymbol{LFI:knee:frequency:LFI22:Rad:M}$\,\pm\,$\getsymbol{LFI:knee:frequency:uncertainty:LFI22:Rad:M}&
              *\getsymbol{LFI:knee:frequency:LFI22:Rad:S}$\,\pm\,$\getsymbol{LFI:knee:frequency:uncertainty:LFI22:Rad:S}&
              \getsymbol{LFI:slope:LFI22:Rad:M}$\,\pm\,$\getsymbol{LFI:slope:uncertainty:LFI22:Rad:M}&
              \getsymbol{LFI:slope:LFI22:Rad:S}$\,\pm\,$\getsymbol{LFI:slope:uncertainty:LFI22:Rad:S}\cr
  \hglue 2em LFI-23&*\getsymbol{LFI:knee:frequency:LFI23:Rad:M}$\,\pm\,$\getsymbol{LFI:knee:frequency:uncertainty:LFI23:Rad:M}&
              *\getsymbol{LFI:knee:frequency:LFI23:Rad:S}$\,\pm\,$\getsymbol{LFI:knee:frequency:uncertainty:LFI23:Rad:S}&
              \getsymbol{LFI:slope:LFI23:Rad:M}$\,\pm\,$\getsymbol{LFI:slope:uncertainty:LFI23:Rad:M}&
              \getsymbol{LFI:slope:LFI23:Rad:S}$\,\pm\,$\getsymbol{LFI:slope:uncertainty:LFI23:Rad:S}\cr
  \noalign{\vskip 5pt}
  \omit{\bf 44\,GHz}\hfil\cr
  \noalign{\vskip 4pt}
  \hglue 2em LFI-24&*\getsymbol{LFI:knee:frequency:LFI24:Rad:M}$\,\pm\,$\getsymbol{LFI:knee:frequency:uncertainty:LFI24:Rad:M}&
              *\getsymbol{LFI:knee:frequency:LFI24:Rad:S}$\,\pm\,$\getsymbol{LFI:knee:frequency:uncertainty:LFI24:Rad:S}&
              \getsymbol{LFI:slope:LFI24:Rad:M}$\,\pm\,$\getsymbol{LFI:slope:uncertainty:LFI24:Rad:M}&
              \getsymbol{LFI:slope:LFI24:Rad:S}$\,\pm\,$\getsymbol{LFI:slope:uncertainty:LFI24:Rad:S}\cr
  \hglue 2em LFI-25&*\getsymbol{LFI:knee:frequency:LFI25:Rad:M}$\,\pm\,$\getsymbol{LFI:knee:frequency:uncertainty:LFI25:Rad:M}&
              \getsymbol{LFI:knee:frequency:LFI25:Rad:S}$\,\pm\,$\getsymbol{LFI:knee:frequency:uncertainty:LFI25:Rad:S}&
              \getsymbol{LFI:slope:LFI25:Rad:M}$\,\pm\,$\getsymbol{LFI:slope:uncertainty:LFI25:Rad:M}&
              \getsymbol{LFI:slope:LFI25:Rad:S}$\,\pm\,$\getsymbol{LFI:slope:uncertainty:LFI25:Rad:S}\cr
  \hglue 2em LFI-26&*\getsymbol{LFI:knee:frequency:LFI26:Rad:M}$\,\pm\,$\getsymbol{LFI:knee:frequency:uncertainty:LFI26:Rad:M}&
              *\getsymbol{LFI:knee:frequency:LFI26:Rad:S}$\,\pm\,$\getsymbol{LFI:knee:frequency:uncertainty:LFI26:Rad:S}&
              \getsymbol{LFI:slope:LFI26:Rad:M}$\,\pm\,$\getsymbol{LFI:slope:uncertainty:LFI26:Rad:M}&
              \getsymbol{LFI:slope:LFI26:Rad:S}$\,\pm\,$\getsymbol{LFI:slope:uncertainty:LFI26:Rad:S}\cr
  \noalign{\vskip 5pt}
  \omit{\bf 30\,GHz}\hfil\cr
  \noalign{\vskip 4pt}
  \hglue 2em LFI-27&\getsymbol{LFI:knee:frequency:LFI27:Rad:M}$\,\pm\,$\getsymbol{LFI:knee:frequency:uncertainty:LFI27:Rad:M}&
              \getsymbol{LFI:knee:frequency:LFI27:Rad:S}$\,\pm\,$\getsymbol{LFI:knee:frequency:uncertainty:LFI27:Rad:S}&
              \getsymbol{LFI:slope:LFI27:Rad:M}$\,\pm\,$\getsymbol{LFI:slope:uncertainty:LFI27:Rad:M}&
              \getsymbol{LFI:slope:LFI27:Rad:S}$\,\pm\,$\getsymbol{LFI:slope:uncertainty:LFI27:Rad:S}\cr
  \hglue 2em LFI-28&*\getsymbol{LFI:knee:frequency:LFI28:Rad:M}$\,\pm\,$\getsymbol{LFI:knee:frequency:uncertainty:LFI28:Rad:M}&
              *\getsymbol{LFI:knee:frequency:LFI28:Rad:S}$\,\pm\,$\getsymbol{LFI:knee:frequency:uncertainty:LFI28:Rad:S}&
              \getsymbol{LFI:slope:LFI28:Rad:M}$\,\pm\,$\getsymbol{LFI:slope:uncertainty:LFI28:Rad:M}&
              \getsymbol{LFI:slope:LFI28:Rad:S}$\,\pm\,$\getsymbol{LFI:slope:uncertainty:LFI28:Rad:S}\cr
  \noalign{\vskip 5pt\hrule\vskip 3pt}}}
  \endPlancktable
  \endgroup
  \end{table*}

\section{Mapmaking}
\label{sec_map_making}

The methods and implementation of the LFI mapmaking pipeline are described in detail
in \citet{planck2014-a03}, \citet{planck2014-a07}, and \citet{keihanen2010}.  Here
we report only the changes introduced into the code with respect to the previous release.

Our pipeline still uses the {\tt{Madam}} destriping code, in which the correlated noise
component is modelled as a sequence of short baselines (offsets) that are determined
via a maximum-likelihood approach. In the current release, the most important change
is in the definition of the noise filter in connection with horn-uniform detector weighting.
When combining data from several detectors, we assign each detector a weight that is
proportional to
\begin{equation}
C_{\rm w}^{-1}=\frac{2}{\sigma_\mathrm{M}^2+\sigma_\mathrm{S}^2}\,,
\end{equation}
where $\sigma_\mathrm{M}^2$ and $\sigma_\mathrm{S}^2$ are the white-noise variances of
the two radiometers (``Main" and ``Side") of the same horn. The same weight is
applied to both radiometers.  In the 2015 release the weighting was allowed to
affect the noise filter as well: for the noise variance $\sigma^2$ in Eq.~(\ref{noisemodel})
we used the average value $C_{\rm w}$.  For the current release we have
completely separated detector weighting from noise filtering.  We use the
individual variances $\sigma_\mathrm{M}^2$ and $\sigma_\mathrm{S}^2$ for each
radiometer when building the noise filter.
In principle, this makes maximum use of the information we have about the
noise of each radiometer.

We compared the previous and the current versions of {\tt{Madam}}, using a
single noise filter, and found excellent agreement. Differences were at the
0.01$\muK$ level, and the code took the same number of iterations.  We then
compared the results obtained with the combined noise filter with those
obtained with the separated noise filters for each radiometer.  We found
that using separate filters for the two radiometers of the same horn has the
effect of reducing the total number of iterations required for convergence by
almost a factor of 2. The net effect of using both the new version of {\tt Madam}
and the separate noise filters is that we obtain the same maps as before,
but considerably faster.

\begin{figure*}[htpb]
\centerline{
\includegraphics[width=12.2cm]{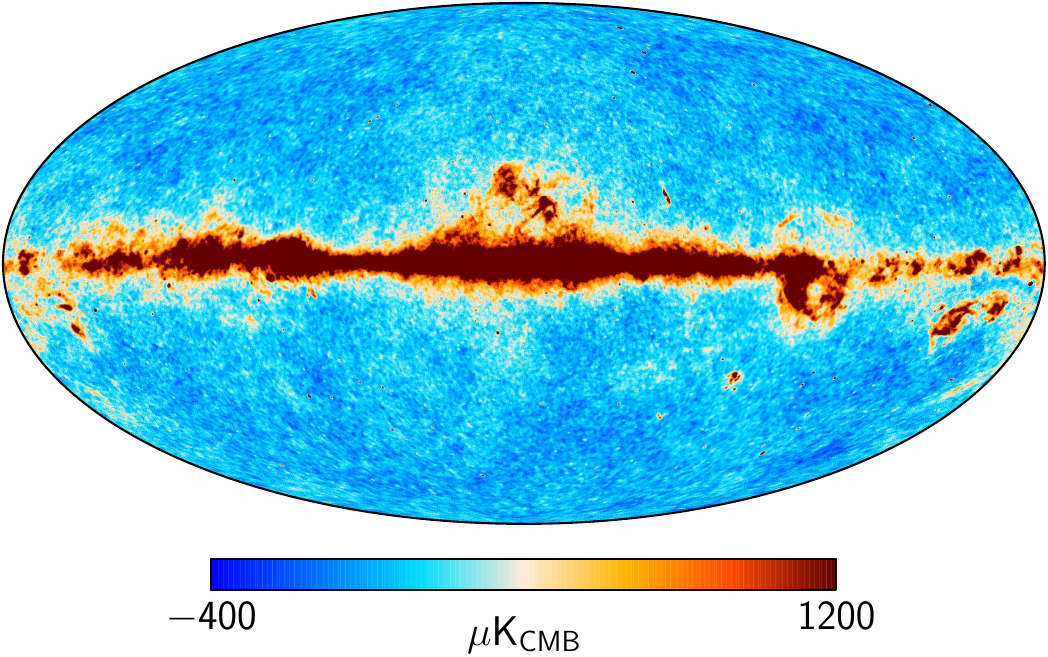}}
\centerline{
\includegraphics[width=12.2cm]{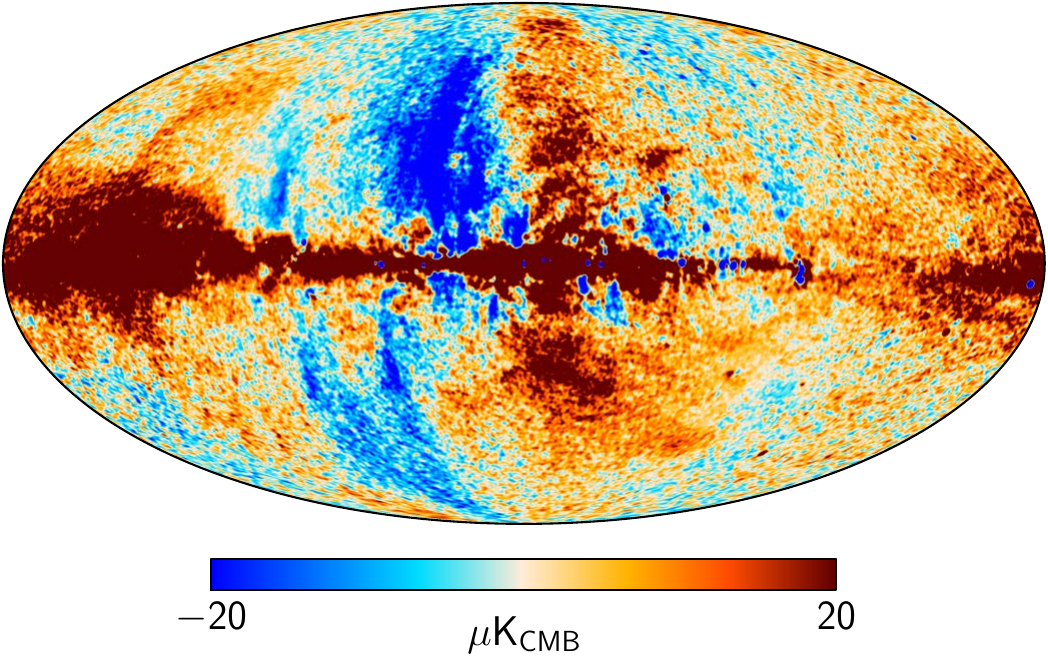}}
\centerline{
\includegraphics[width=12.2cm]{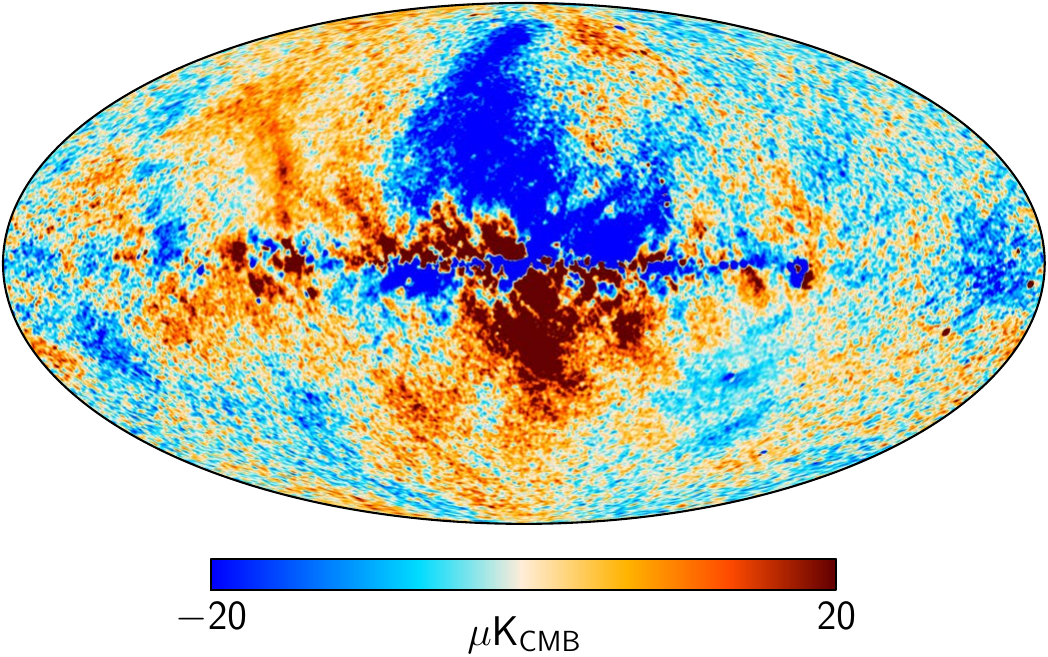}
}
\caption{LFI maps at 30\GHz: top, total intensity $I$;
  middle: $Q$ polarization component; bottom, $U$
  polarization component. Stokes $I$ is shown at instrument resolution and
  at $N_\mathrm{side}=1024$, while $Q$ and $U$ are smoothed to $1^\circ$
  resolution and at $N_\mathrm{side}=256$. Units are $\muK_\mathrm{CMB}$.
  The polarization components have been corrected for bandpass
  leakage (Sect.~\ref{sec_polarization})}
\label{maps30dx12}
\end{figure*}

\begin{figure*}[htpb]
\centerline{
\includegraphics[width=12.2cm]{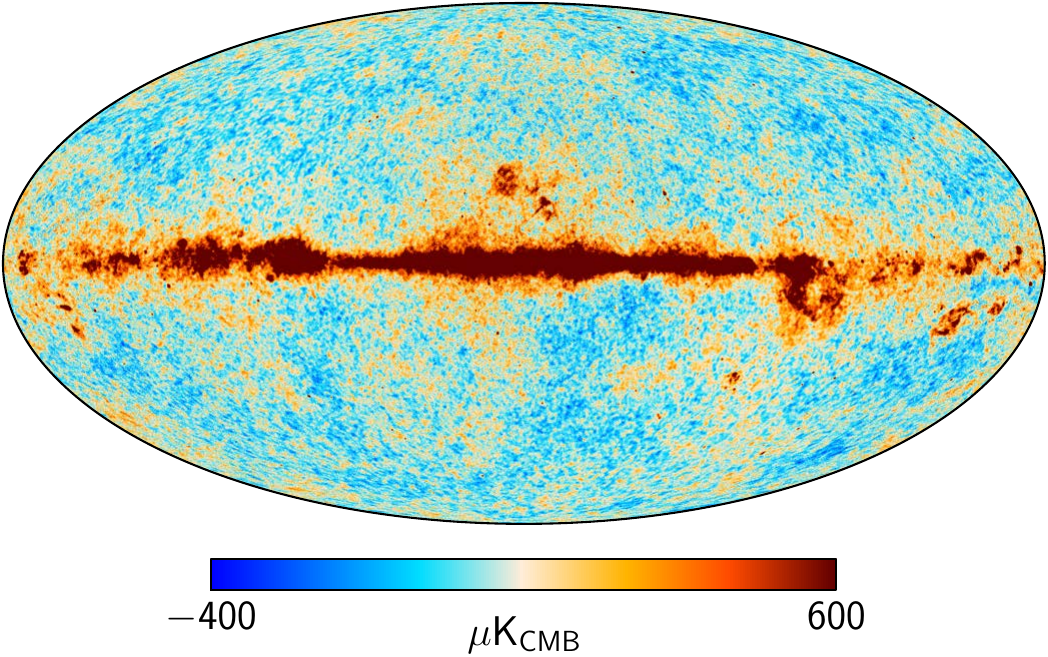}}
\centerline{
\includegraphics[width=12.2cm]{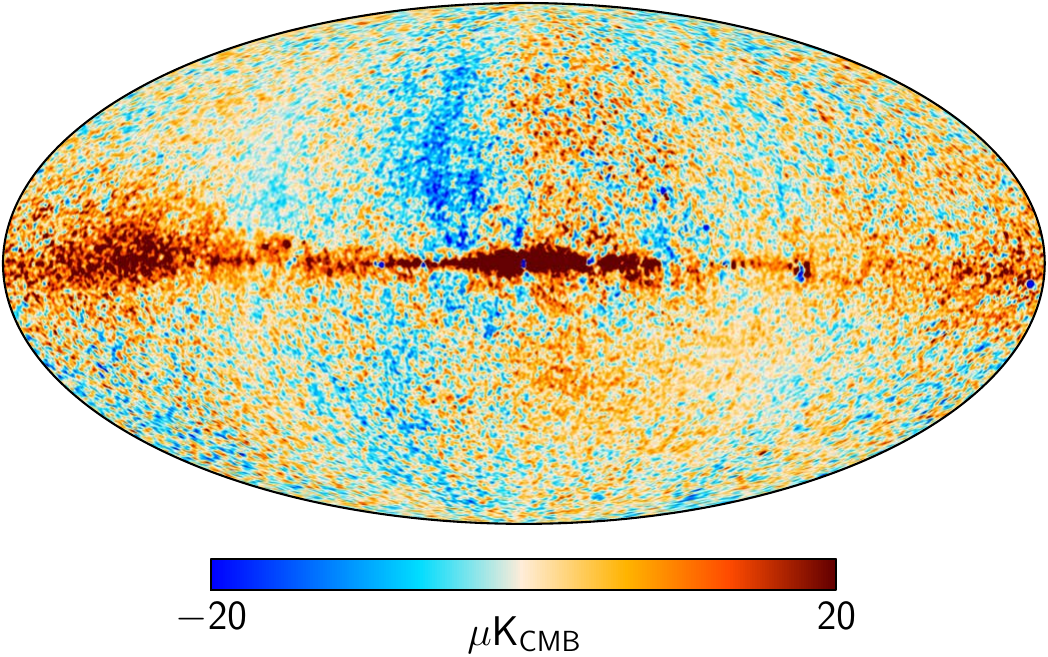}}
\centerline{
\includegraphics[width=12.2cm]{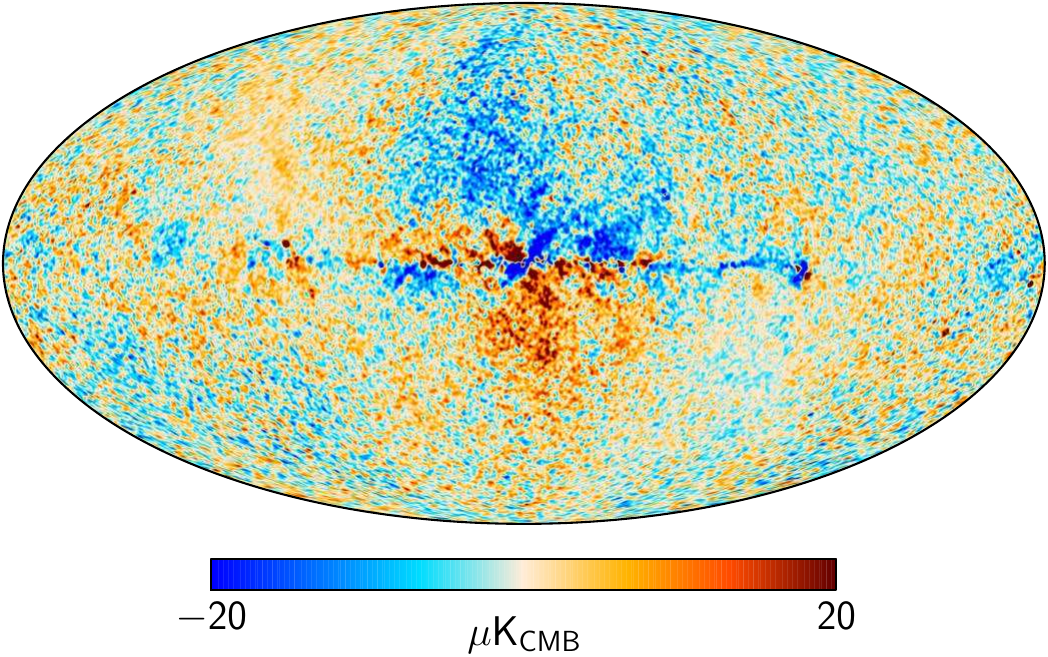}
}
\caption{Same as Fig.~\ref{maps30dx12}, for the 44-GHz channel.}
\label{maps44dx12}
\end{figure*}

\begin{figure*}[htpb]
\centerline{
\includegraphics[width=12.2cm]{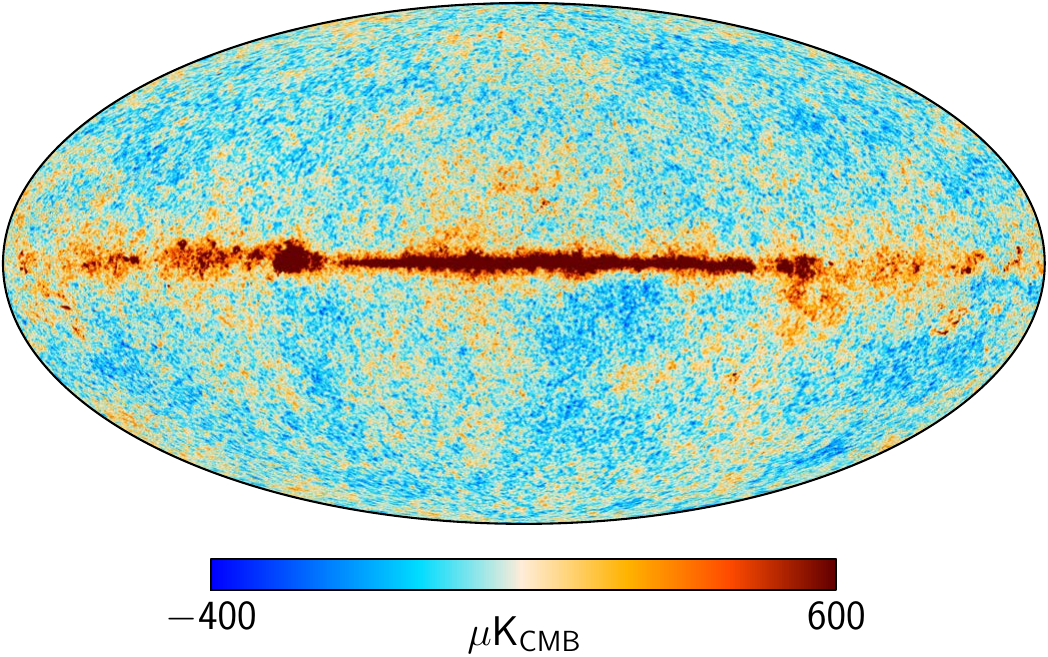}}
\centerline{
\includegraphics[width=12.2cm]{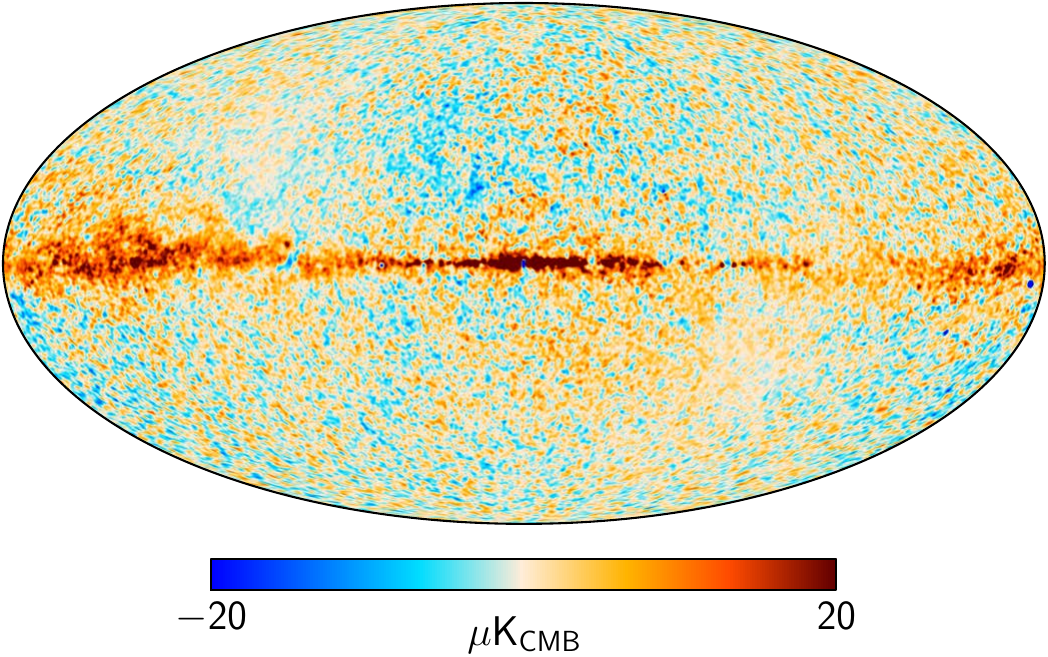}}
\centerline{
\includegraphics[width=12.2cm]{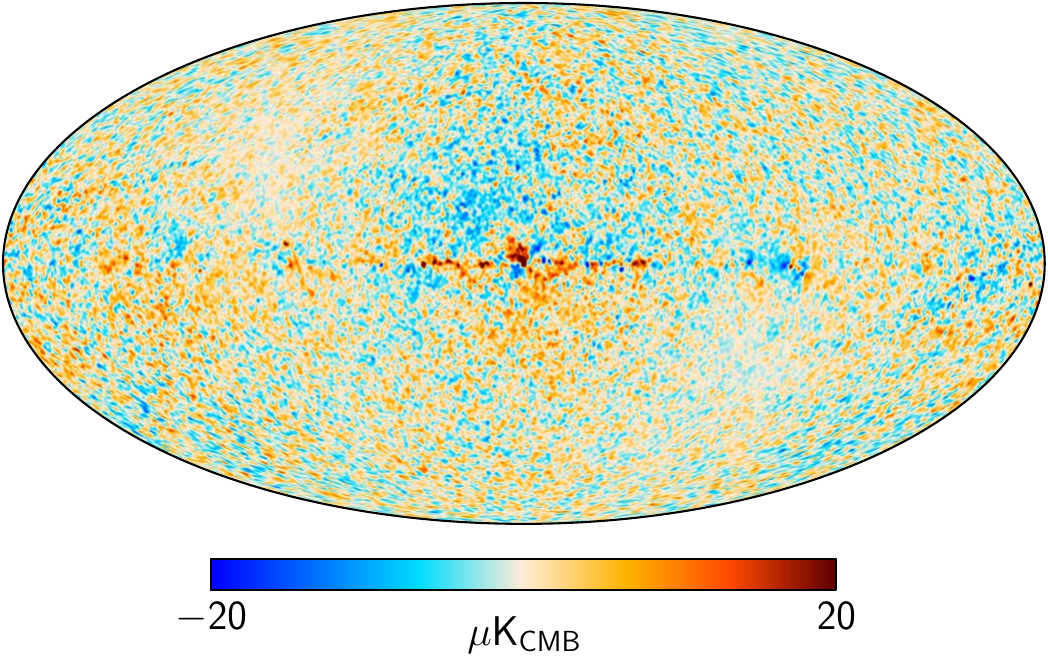}
}
\caption{Same as Fig.~\ref{maps30dx12}, for the 70-GHz channel.}
\label{maps70dx12}
\end{figure*}

In Figs.~\ref{maps30dx12} to \ref{maps70dx12}, we show the 30-, 44-, and
70-GHz frequency maps.  The top panels are the temperature ($I$) maps
based on the full observation period, and presented at the original native
instrument resolution, {\tt HEALPix} $N_\mathrm{side}=1024$. The middle and bottom
panels show the $Q$ and $U$ polarization components, respectively; these are
smoothed to $1\deg$ angular resolution and downgraded to $N_\mathrm{side}=256$.
Polarization components have been corrected for bandpass leakage
(see Sect.~\ref{sec_polarization}). Table~\ref{tab_map_properties} gives
the main mapmaking parameters used in map production. All values are the same as
for the previous data release except the monopole term; although we used the
same plane-parallel model for the Galactic emission as for the 2015 data
release, the derived monopole terms are slightly changed at 30 and 44\,GHz for
the adopted calibration procedure.


\begin{table*}[htpb]
  \begingroup
  \newdimen\tblskip \tblskip=5pt
  \caption{Mapmaking parameters used in the production of maps.
    Details are reported in \citet{planck2014-a07}.}
  \label{tab_map_properties}
  \nointerlineskip
  \vskip -3mm
  \footnotesize
  \setbox\tablebox=\vbox{
 % \vbox{
  \newdimen\digitwidth
  \setbox0=\hbox{\rm 0}
  \digitwidth=\wd0
  \catcode`*=\active
  \def*{\kern\digitwidth}
  \newdimen\signwidth
  \setbox0=\hbox{+}
  \signwidth=\wd0
  \catcode`!=\active
  \def!{\kern\signwidth}
  \halign{\tabskip=0pt\hbox to 1.3in{#\leaderfil}\tabskip=2em&
      \hfil#\hfil&
      \hfil#\hfil&
      \hfil#\hfil&
      \hfil#\hfil&
      \hfil#\hfil&
      \hfil#\hfil\tabskip=0pt\cr
  \noalign{\doubleline}
  \omit&\omit 
&\multispan2\hfil Baseline length$^{\rm b}$\hfil&\multispan2\hfil Resolution$^{\rm c}$\hfil&\hfil Monopole, B$^{\rm d}$\hfil\cr
  \noalign{\vskip -4pt}
  \omit&\omit&\multispan2\hrulefill&\multispan2\hrulefill&\hrulefill\cr
\omit Channel\hfil& $f_{\rm samp}$ [Hz]$^{\rm a}$& [s]& Samples& $N_{\rm side}$& [arcmin]&$\,[\mu\mathrm{K}_{\rm CMB}$]\cr
  \noalign{\vskip 3pt\hrule\vskip 5pt}
30\,GHz& 32.508&  0.246& *8& 1024& 3.44&\llap{$+$}11.9$\,\pm\,$0.7\cr
44\,GHz& 46.545&  0.988& 46&  1024& 3.44&\llap{$-$}15.4$\,\pm\,$0.7\cr
70\,GHz& 78.769& 1.000& 79&  1024/2048& 3.44/1.72&\llap{$-$}35.7$\,\pm\,$0.6\cr
  \noalign{\vskip 5pt\hrule\vskip 3pt}}}
  \endPlancktablewide
  \tablenote a Sampling frequency.\par 
  \tablenote b Baseline length in seconds and in number of samples.\par
  \tablenote c {\tt HEALPix} $N_{\rm side}$ resolution parameter and averaged pixel size.\par
  \tablenote d Monopole removed from the maps and reported in the {\tt{FITS}} header.\par
    \endgroup
  \end{table*}

\section{Polarization: Leakage maps and bandpass correction}
\label{sec_polarization}

The small amplitude of the CMB polarized signal requires careful handling because of systematic
effects capable of biasing polarization results.  The dominant one is the leakage
of unpolarized emission into polarization; any difference in bandpass between the
two arms of an LFI radiometer will result in such leakage. In the case of the CMB,
this is not a problem.  That is because calibration of each radiometer uses the CMB dipole,
which has the same frequency spectrum as the CMB itself, and so exact gain calibration
perfectly cancels out in polarization. However, unpolarized foreground-emission
components
with spectra different from the CMB will appear with different amplitudes in the
two arms, producing a leakage into polarizaion.

In order to derive a correction for this bandpass mismatch, we exploit the $IQUSS$
approach \citep{page2007} used in the 2015 release.  The main ingredients in the
bandpass mismatch recipe are the leakage maps $L$, the spurious maps $S_k$ (see below),
the $a$-factors, and the $A_{Q[U]}$ maps \citep[see section~11 of][for definitions]{planck2014-a03}.
With respect to the treatment of bandpass mismatch in the previous release, we
introduce three main improvements in the computation of the $L$ and $A$ maps.
Leakage maps, $L$, are the astrophysical leakage term encoding our knowledge of
foreground amplitude and spectral index. These maps are derived from the output
of the {\tt{Commander}} component-separation code. In the present analysis this
is done using only \Planck\ data from the current data release at their full
instrumental resolution.  In contrast, in the earlier approach we also used \WMAP\ 9-year
data and applied a $1\deg$ smoothing prior to the component-separation process. We
also exclude \Planck\, channels at 100 and 217\GHz\,, since these could be
contaminated by CO line emission.  Spurious maps $S_k$ (one for each radiometer)
are computed from {\tt{Madam}} mapmaking outputs (for the full frequency map
creation run). Basically, spurious maps are proportional to the bandpass mismatch
of each radiometer, and can be computed directly from single radiometer timelines.
As described in \citet{planck2014-a03}, the output of the Main and Side arms of a
radiometer can be redefined, including bandpass mismatch spurious terms, as
\begin{eqnarray}
\text{LFI 27} \left\{
          \begin{array}{l l}
          d_{s1} = I + Q{\mathrm{cos}}(2\psi_{\mathrm{s1}}) +
                 U{\mathrm{sin}}(2\psi_{\mathrm{s1}}) + S_1\, ,\nonumber \\
          d_{m1} = I + Q{\mathrm{cos}}(2\psi_{\mathrm{m1}}) +
                 U{\mathrm{sin}}(2\psi_{\mathrm{m1}}) - S_1\, ,\
          \end{array}\right. \\
\text{LFI 28} \left\{
          \begin{array}{l l}
          d_{s2} = I + Q{\mathrm{cos}}(2\psi_{\mathrm{s2}}) +
                 U{\mathrm{sin}}(2\psi_{\mathrm{s2}}) + S_2\, ,\nonumber \\
          d_{m2} = I + Q{\mathrm{cos}}(2\psi_{\mathrm{m2}}) +
                 U{\mathrm{sin}}(2\psi_{\mathrm{m2}}) - S_2\, ,\
          \end{array}\right. \\
\end{eqnarray}
or in the more compact form
\begin{equation}
d_i = I + Q{\mathrm{cos}}(2\psi_i) + U{\mathrm{sin}}(2\psi_i) + \alpha_1 S_1 + \alpha_2 S_2\, .
\end{equation}
Here $\alpha_1$ and $\alpha_2$ can take the values $-1$, $0$, and $1$. The problem of estimating $m=[I,Q,U,S_1,S_2]$ is similar
to a mapmaking problem, with two extra maps. The pixel-noise covariance matrix is therefore given by the
already available {\tt{Madam}}-derived covariance matrix, with two additional rows and columns, as
\begin{eqnarray}
\tens{M}_p&=\sum_{i\in p} w_i \times \nonumber
\end{eqnarray}
\vspace{-0.75cm}
\begin{eqnarray}
\quad \left( \begin{array}{ccccc}
\dots& \dots& \dots& \alpha_1& \alpha_2 \\
\dots& \dots& \dots& \alpha_1\cos(2\psi_i)& \alpha_2\cos(2\psi_i) \\
\dots& \dots& \dots& \alpha_1\sin(2\psi_i)& \alpha_2\sin(2\psi_i) \\
\alpha_1& \alpha_1\cos(2\psi_i)& \alpha_1\sin(2\psi_i)& \alpha_1^2& 0 \\
\alpha_2& \alpha_2\cos(2\psi_i)& \alpha_2\sin(2\psi_i)& 0& \alpha_2^2
\end{array}\right).
\end{eqnarray}

The \Planck\ scanning strategy allows only a limited range of radiometer orientations.
We therefore compute a joint solution with all radiometers at each frequency
that helps also to reduce the noise in the final solutions.  Once spurious
maps are derived, we compute the $a$-factors from a $\chi^2$ fit between the leakage
map $L$ and the spurious maps $S_k$ on those
pixels close to the Galactic plane, $|b|<15^\circ$ (at higher latitudes both
foregrounds and spurious signals are weak and
do not add useful information).

The last improvement involves the final step in the creation of the correction
maps.  Recall that polarization data from a single radiometer probe only one
Stokes parameter in the reference frame tied to that specific feedhorn. This
reference frame is then projected onto the sky according to the actual orientation
of the spacecraft, which modulates the spurious signal of each radiometer
into $Q$ and $U$. This modulation can be obtained by scanning the estimated
spurious maps $\hat{S} = a L$, to create a timeline that is finally reprojected
into a map. In the previous release, instead of creating timelines and then
maps (a time- and resource-consuming operation) we built projection maps $A_{Q[U]}$
that accounted exactly for horn and radiometer orientation. The final correction
maps were
\begin{equation}
\Delta Q[U] = L\times \sum_k a_k A_{k,Q[U]}\, ,
\label{correction_map_eq}
\end{equation}
where $a_k$ and $A_{k,Q[U]}$ are the $a$-factors and the projection map for the
radiometer $k$ of a given frequency. In using this approach, however, there
were two drawbacks. The first and more important one is related to a monopole
term present in the leakage map $L$ that directly impacted the correction maps.
The synchrotron component, in fact, has a significant quasi-isotropic component
(perhaps related to the ARCADE2-measured excess; \citealt{arcade2}) and this
contributed exactly to a monopole term in the correction map.  $Q$ and $U$ maps,
however, are produced by {\tt{Madam}}, which tends to remove any possible monopole term.
Therefore, with the simple approach of Eq.~(\ref{correction_map_eq}), we transfered the
monopole term into the correction, and hence into the final bandpass-corrected map,
resulting in an overestimation of the actual real effect. This was negligible at
70\GHz, but important in the 30-GHz map, which is used to correct foregrounds in
the 70-GHz likelihood power spectrum estimation.  The second drawback was that
the resulting correction maps displayed sharp features, especially around the
ecliptic poles. These were intrinsic to the projection maps, and caused problems
with nearby point sources. 

To resolve these issues, given new computing resources available, we exploit
the scanning, timeline creation, and mapmaking approach. This is done using
the \Planck\ {\tt LevelS} simulation package \citep{reinecke2006}, which takes the
harmonic coefficients $a_{\ell m}$ of the leakage maps, multiplied by the
derived $a$-factors, and the actual scanning strategy, and then creates
timelines accounting for proper beam convolution as well. The resulting TOD
are used to create maps with the {\tt{Madam}} mapmaking code. It is clear that
in this way the final correction map is processed by the same mapmaking used
for official map production and hence removes the presence of unwanted monopoles.
In addition, accounting for beam convolution significantly alleviates the presence
of sharp features in the correction maps.

Table~\ref{tab_afactors} gives the estimated $a$-factors for the current release.
They are very close to the 2015 values at 30 and 44\GHz, with larger
(but within $1\,\sigma$) variations at 70\GHz.  We investigated the origin of
these variations by computing the $a$-factors with the present data, but using the
old version of the leakage map.  We find results in agreement with those obtained
before. We also performed the same analysis with old data but with the new version of
the leakage maps. In this case, we find results in line with the current estimates.
These very simple tests clearly indicate that the observed variations in
the $a$-factors are not due to the adopted calibration pipeline, but are mainly
due to the changes in the leakage maps derived using \Planck\ data only and
excluding CO-dominated HFI channels.
\begin{figure*}[htpb]
  \centerline{
    \includegraphics[width=5cm]{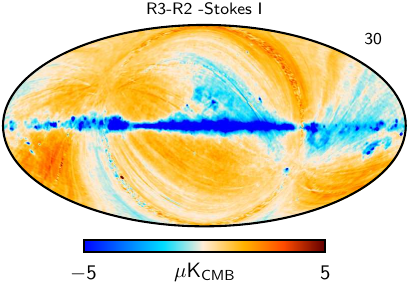}
    \includegraphics[width=5cm]{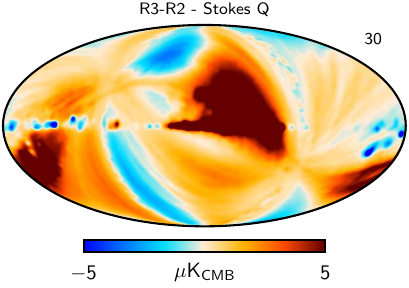}
    \includegraphics[width=5cm]{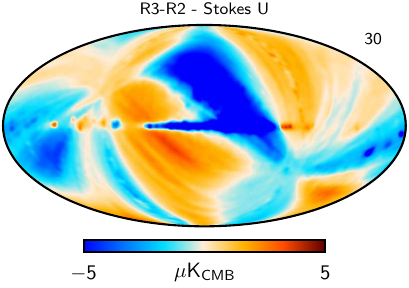}
  }
  \vspace{-0.75cm}
  \centerline{
    \includegraphics[width=5cm]{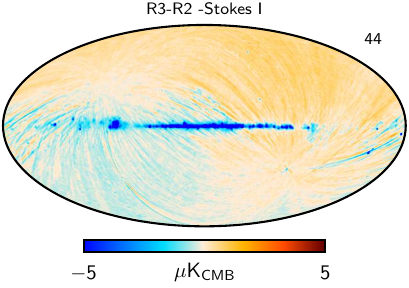}
    \includegraphics[width=5cm]{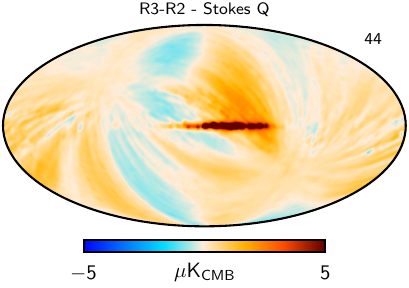}
    \includegraphics[width=5cm]{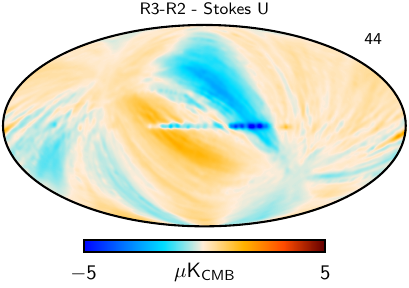}
  }    
  \vspace{-0.75cm}
  \centerline{
    \includegraphics[width=5cm]{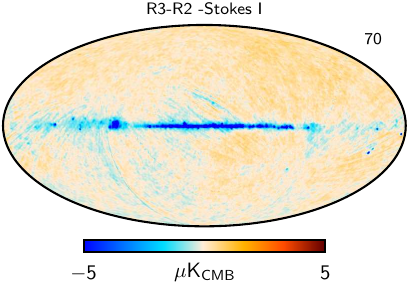}
    \includegraphics[width=5cm]{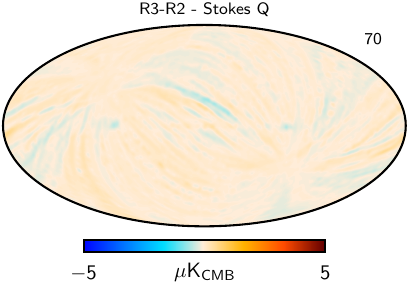}
    \includegraphics[width=5cm]{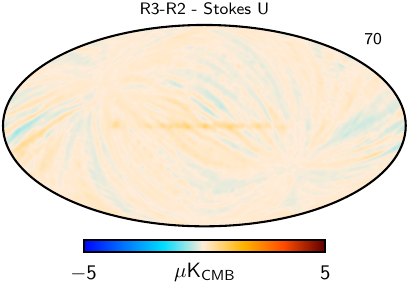}
  }      
  \caption{Differences between 2018 (PR3) and 2015 (PR2) frequency maps
    in $I$, $Q$, and $U$. Maps are smoothed to $1^\circ$ angular resolution
    for $I$ and to $3^\circ$ for $Q$ and $U$, in order to highlight large-scale
    features. Differences are clearly evident at 30 and 44\GHz, and are mainly
    due to changes in the calibration procedure.}
  \label{mapsdiffs}
\end{figure*}
\begin{table}[htpb]
\begingroup
\newdimen\tblskip \tblskip=5pt
\caption{Bandpass mismatch $a$-factors from a fit to $S_k = a_k L$.}
\label{tab_afactors}
\nointerlineskip
\vskip -3mm
\footnotesize
\setbox\tablebox=\vbox{
\newdimen\digitwidth
\setbox0=\hbox{\rm 0}
\signwidth=\wd0
\catcode`*=\active
\def*{\kern\digitwidth}
\newdimen\signwidth
\setbox0=\hbox{+}
\signwidth=\wd0
\catcode`!=\active
\def!{\kern\signwidth}
\halign{\hbox to 1.3in{#\leaderfil}\tabskip=2em&
	\hfil#\hfil\tabskip=0pt\cr
\noalign{\doubleline}
\noalign{\vskip -3pt}
\omit\hfil Horn\hfil& $a$-factor\cr
\noalign{\vskip 3pt\hrule\vskip 5pt}
\omit{\bf 70\,GHz}\hfil\cr
\noalign{\vskip 4pt}
\hglue 2em LFI 18& $-0.0030\pm0.0029$\cr
\hglue 2em LFI 19& $!0.0197\pm0.0030$\cr
\hglue 2em LFI 20& $!0.0051\pm0.0032$\cr
\hglue 2em LFI 21& $-0.0189\pm0.0031$\cr
\hglue 2em LFI 22& $!0.0063\pm0.0031$\cr
\hglue 2em LFI 23& $!0.0095\pm0.0031$\cr
\noalign{\vskip 5pt}
\omit{\bf 44\,GHz}\hfil\cr
\noalign{\vskip 4pt}
\hglue 2em LFI 24& $!0.0038\pm0.0004$\cr
\hglue 2em LFI 25& $!0.0006\pm0.0004$\cr
\hglue 2em LFI 26& $!0.0014\pm0.0004$\cr
\noalign{\vskip 5pt}
\omit{\bf 30\,GHz}\hfil\cr
\noalign{\vskip 4pt}
\hglue 2em LFI 27& $!0.0058\pm0.0001$\cr
\hglue 2em LFI 28& $-0.0101\pm0.0001$\cr
\noalign{\vskip 5pt\hrule\vskip 3pt}}}
\endPlancktable
\endgroup
\end{table}

\section{Data validation}
\label{sec_dataval_intro}

We verify the LFI data quality with the same suite of null tests used in
previous releases and described in \citet{planck2014-a03}. As before, null
tests cover different timescales (pointing periods, surveys, survey
combinations, and years) and data (radiometers, horns, horn-pairs, and
frequencies) for both total intensity and polarization. These allow us to
highlight possible residuals of different systematic effects still present in 
the final data products.

\subsection{Comparison between 2015 and 2018 frequency maps}

\noindent
Before presenting the null-test results, we compare the 2015 and 2018 maps.
We expect improvements especially at 30 and 44\GHz, where the calibration
procedure is significantly changed.  Figure~\ref{mapsdiffs} shows differences
between 2018 and 2015 frequency maps in $I$,$Q$, and $U$. Large scale differences
between the two set of maps are mainly due to changes in the calibration procedure,
but the exact origin of the differences is not revealed by these overall frequency maps. 

\begin{figure*}[htpb]
  \centerline{
    \includegraphics[width=6cm]{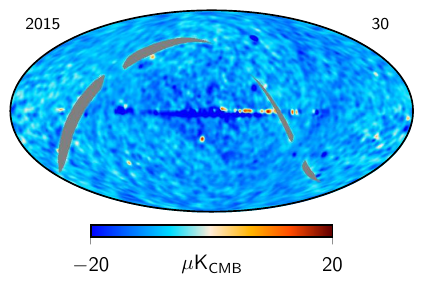}
    \includegraphics[width=6cm]{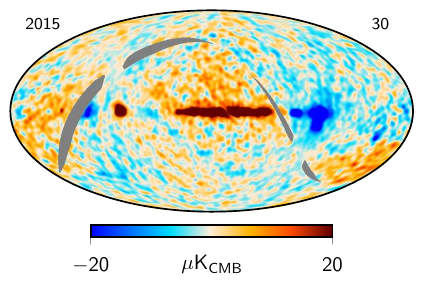}
    \includegraphics[width=6cm]{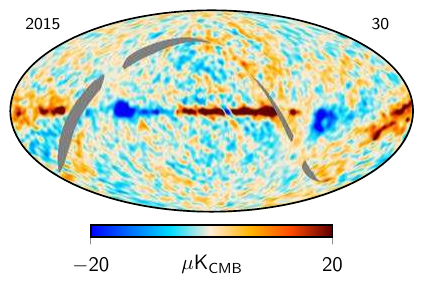}
  }
  \vspace{-1.cm}
  \centerline{
    \includegraphics[width=6cm]{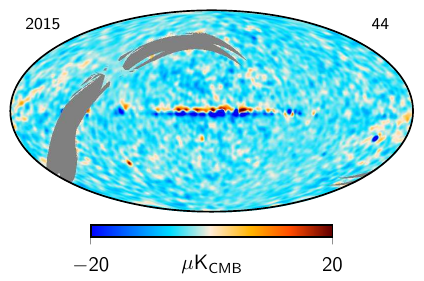}
    \includegraphics[width=6cm]{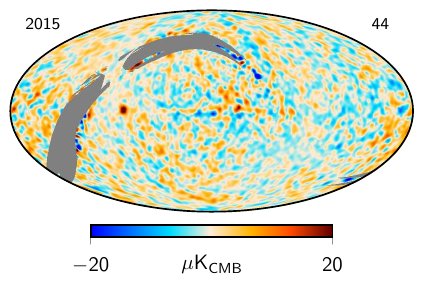}
    \includegraphics[width=6cm]{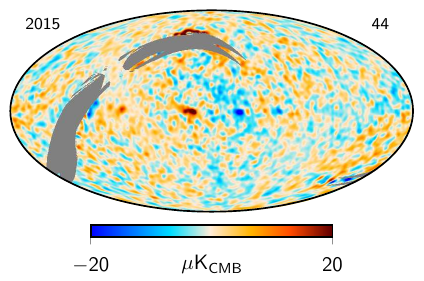}
  }
  \vspace{-1.cm}
  \centerline{
    \includegraphics[width=6cm]{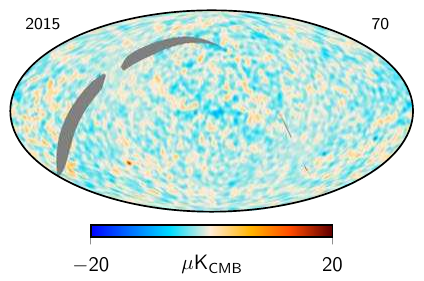}
    \includegraphics[width=6cm]{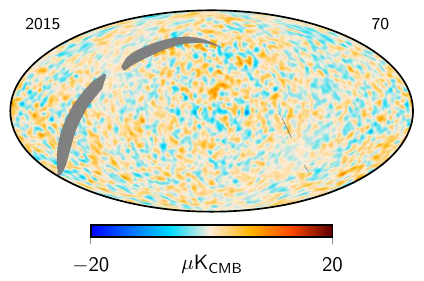}
    \includegraphics[width=6cm]{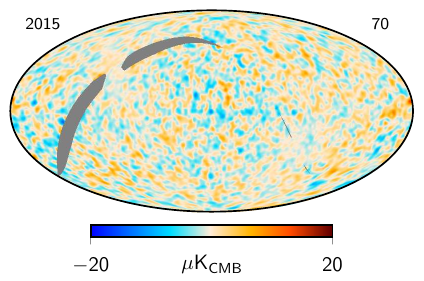}
  }  
  \vspace{-1.cm}
  \centerline{
    \includegraphics[width=6cm]{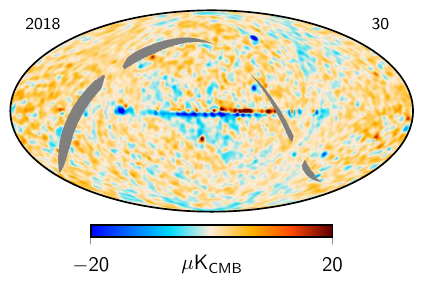}
    \includegraphics[width=6cm]{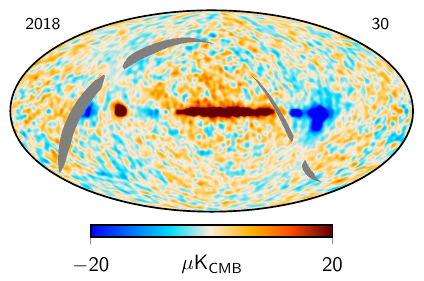}
    \includegraphics[width=6cm]{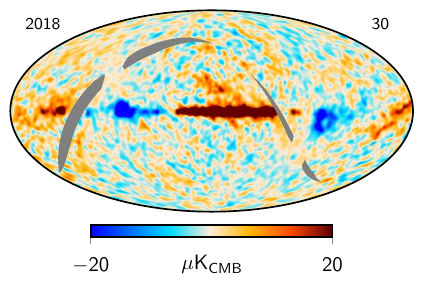}
  }
  \vspace{-1.cm}
  \centerline{
     \includegraphics[width=6cm]{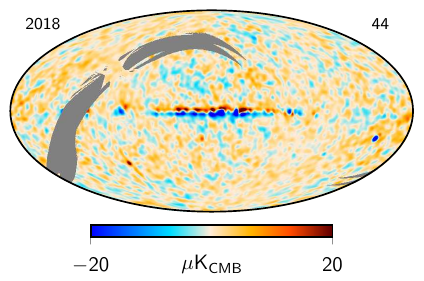}
     \includegraphics[width=6cm]{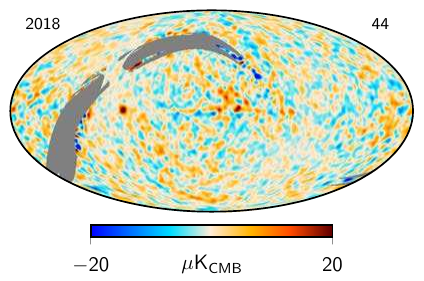}
     \includegraphics[width=6cm]{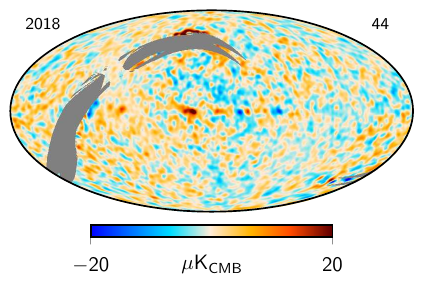}
  }
  \vspace{-1.cm}
  \centerline{
    \includegraphics[width=6cm]{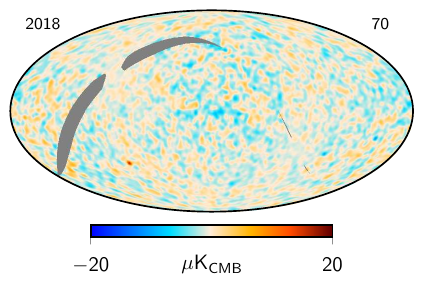}
    \includegraphics[width=6cm]{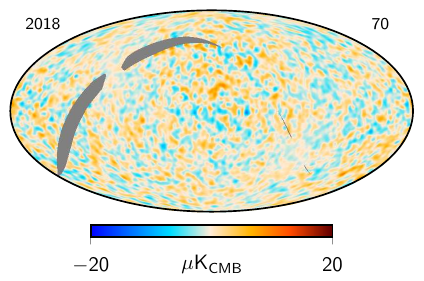}
    \includegraphics[width=6cm]{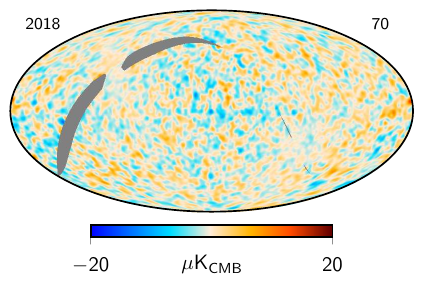}
  }
  \caption{Differences between odd (i.e., Surveys~1, 3, 5, and 7) and even
    (Surveys~2, 4, 6, and 8) surveys in $I, Q$, and $U$ (from left to right)
    for the 2015 (upper nine maps) and 2018 (lower nine maps) data releases.
    These maps are smoothed to 3\deg\ to reveal large-scale structures.}
 \label{oddeven20152017}
\end{figure*}

A clearer indication of the origin of improvements in 2018 is given by
survey differences at the frequency map level in temperature and polarization.
From results for the previous release, we know that odd minus even surveys are
the most problematic because of the low dipole signal in even numbered Surveys
(especially in Surveys~2 and 4), which increases calibration uncertainty. This
indeed was the motivation for the changes made in the calibration pipeline.
In addition, since the optical coupling of the satellite with the sky is
reversed every 6 months, such survey differences are the most sensitive to
residual contamination from far sidelobes not properly accounted for and
subtracted during the calibration process. We therefore consider the set
of odd-even survey differences combining all eight sky surveys covered by LFI.
These survey combinations optimize the signal-to-noise ratio, and are shown in
Fig.~\ref{oddeven20152017} with a low-pass filter to highlight large-scale structures.
The nine maps at the top show odd-even survey differences for the 2015 release,
while the nine maps at the bottom show the same for the 2018 release.

The 2015 data show large residuals in $I$ at 30 and 44\GHz\ that bias the
difference away from zero. This effect is considerably reduced in the 2018 release,
as expected from the improvements in the calibration process. 
The $I$ map at 70\GHz\ also shows a significant improvement.
In the polarization maps, there is a general reduction in the amplitude
of structures close to the Galactic plane: the Galactic centre region and 
the bottom-right structure in $Q$ at 30\GHz, and the rightmost region on the Galactic plane in $U$.

\begin{figure}[htpb]
  \centerline{
    \includegraphics[width=9cm]{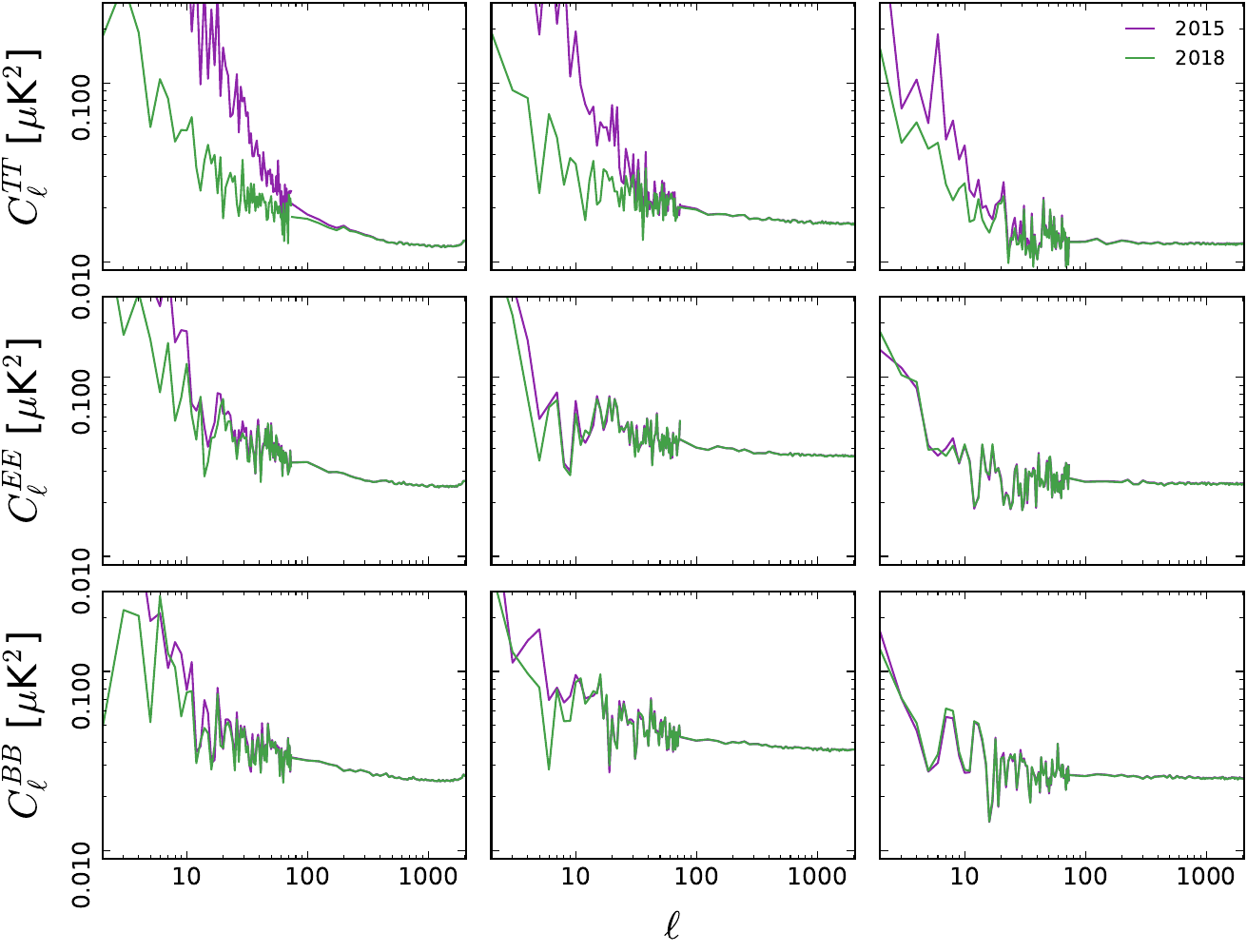}
  }
  \caption{Angular pseudo-power spectra of the odd-even survey difference maps for 30 (left column),
    44 (middle column), and 70\GHz\ (right column), with the 2015 data in purple and 2018 in green.}
\label{oddevenspectra}
\end{figure}

Figure~\ref{oddevenspectra} shows pseudo-angular power spectra from the odd-even
survey differences, using the same sky mask as for the null-test spectra
in Sect.~\ref{nulltest}, namely the union of all the single survey masks.  
There is great improvement in 2018 in removing large-scale structures
at 30\GHz\ in $TT$, $EE$, and somewhat in $BB$, and also in $TT$ at 44\GHz.
These improvements are again expected and the reason is two-fold.
First, the improved calibration now allows better tracing of the actual 
instrument gain, even in the low-dipole-signal period, both for 30 and 44\GHz.
Second, the improved calibration enhances our ability to remove
far-sidelobe contamination, resulting in significantly cleaner
30-GHz maps.  At 70\,GHz, even though the calibration procedure 
is almost unchanged from the 2015 release, we are able to reduce large-scale residuals in 
$TT$, thanks to the combined effect of data selection and the gain smoothing algorithm.

\subsection{Null-test results}
\label{nulltest}

\begin{figure}[htpb]
\centerline{
\hspace{-.5cm}
\includegraphics[width=4.5cm]{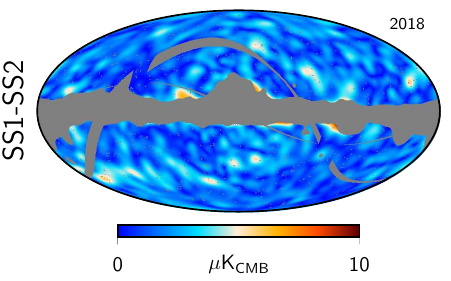}
\hspace{-8.9cm}
\includegraphics[width=4.5cm]{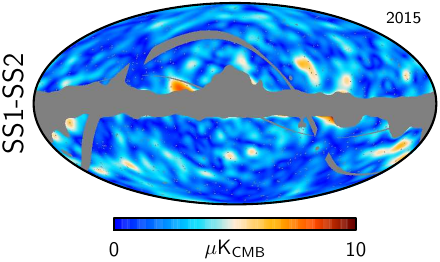}
}
\vspace{-0.7cm}
\centerline{
  \hspace{-.5cm}
\includegraphics[width=4.5cm]{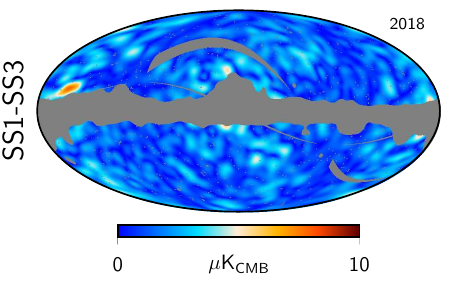}
\hspace{-8.9cm}
\includegraphics[width=4.5cm]{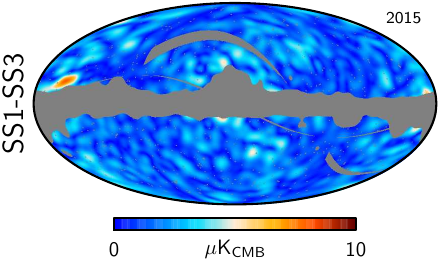}
}
\vspace{-0.7cm}
\centerline{
\hspace{-.5cm}
  \includegraphics[width=4.5cm]{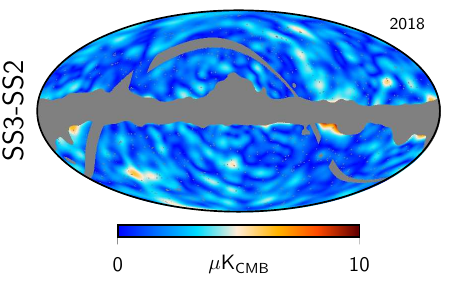}
\hspace{-8.9cm}
  \includegraphics[width=4.5cm]{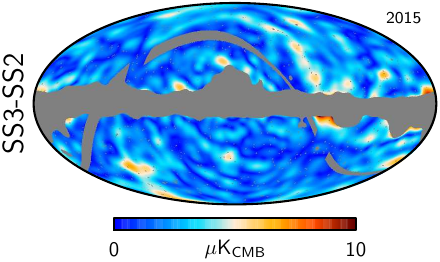}
}
\caption{Survey difference maps of polarization amplitude at 30\GHz\ for the current
  2018 (right)  and 2015 (left) releases. The improvement is evident,
  especially in odd-even difference maps, showing lower residuals due to the new
  calibration approach.  Maps are smoothed with an 8$\deg$ Gaussian beam to show large-scale structures.}
\label{fig_Survey_diff_maps}
\end{figure}

These findings are confirmed by specific null tests, taking differences of frequency
maps for odd and even surveys.  As for the previous release, we present differences
among the first three sky surveys.  Figure~\ref{fig_Survey_diff_maps} shows the
total amplitude of the polarized signal at 30\GHz\, (the channel with the largest
expected differences), smoothed with an 8\deg\ Gaussian beam. Odd-even Survey
differences reveal clear structures on large angular scales that are significantly
reduced in the 2018 data set. In contrast, the Survey~1 versus Survey~3 difference
map shows no large-scale features.  This is expected, since for both Surveys~1
and 3 the dipole signal used for calibration is large.  Moreover, the far
sidelobes are oriented similarly with respect to the sky for these two surveys.

\begin{figure*}[htpb]
\hspace{-0.1cm}
\centerline{
\includegraphics[width=18cm]{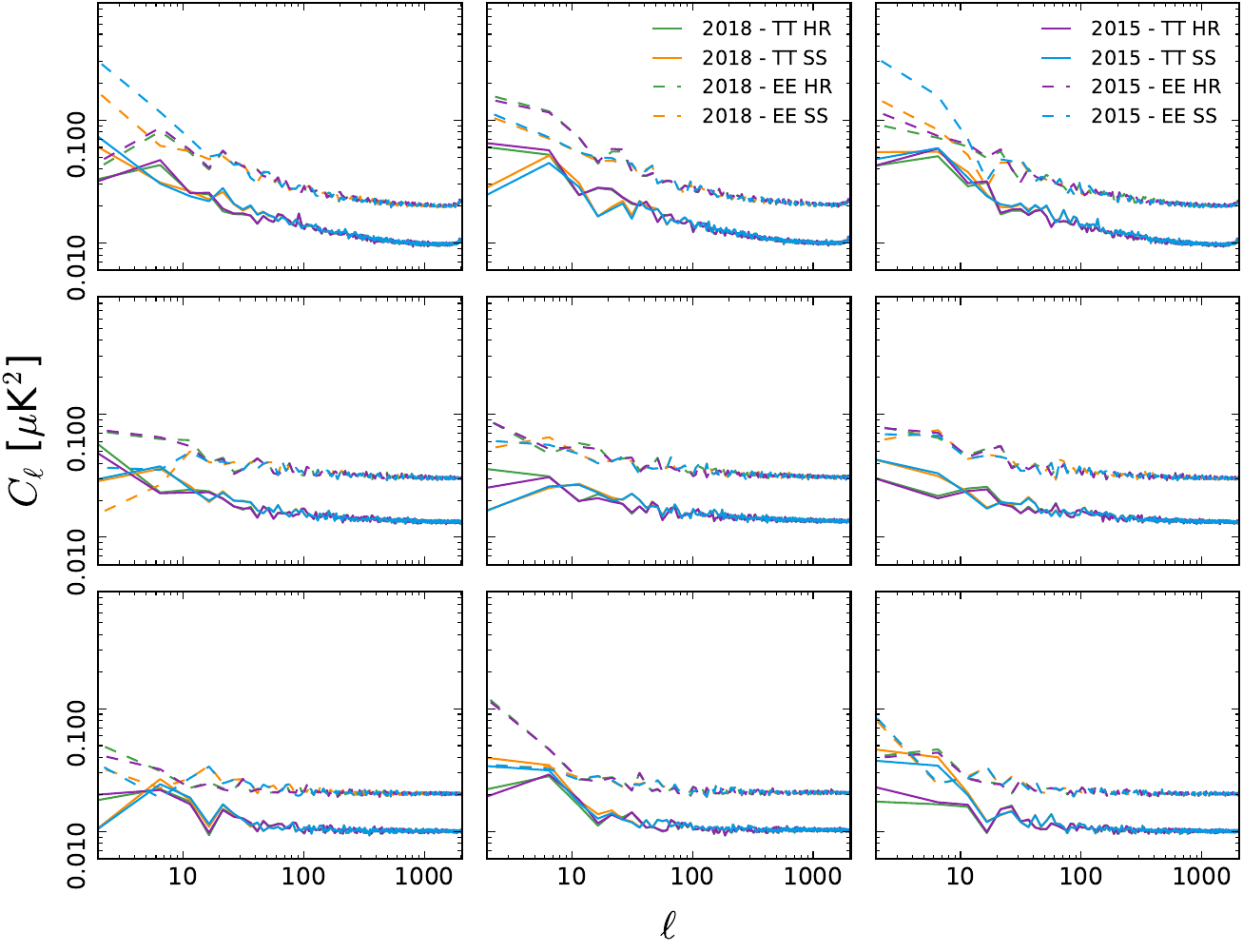}}
\caption{Null-test results comparing power spectra from survey differences
  to those from the half-ring maps. Differences are: left, Survey~1 $-$
  Survey~2; middle, Survey~1 $-$ Survey~3; and right, Survey~1 $-$ Survey~4.
  These are for 30\,GHz (top), 44\,GHz (middle), and 70\,GHz (bottom),
  for both $TT$ and $EE$ power spectra. There is a significant improvement
  in Survey~1 $-$ Survey~2 and Survey~1 $-$ Survey~4 at 30\GHz,
  especially in $EE$.}
\label{nullspectra}
\end{figure*}

We also inspect angular power spectra of odd-even survey differences, adopting
as a figure of merit the noise level derived from the `half-ring' difference
maps (made from the first and second half of each stable pointing period) weighted
by the hit count.  This quantity traces the instrument noise, but filters away
any component fluctuating on timescales longer than the pointing period.
To illustrate the general trend in null tests and the improvements in the 2018
release, Fig.~\ref{nullspectra} shows $TT$ and $EE$ Survey-difference power
spectra for the 2015 and 2018 data sets.  We compare these spectra with
noise levels derived from the corresponding half-ring maps.

Results at 30 and 44\GHz\ are in line with expectations. In particular we see
improvements at 30\GHz\ (survey differences are close to half-ring spectra) when
considering odd-even survey differences.  The better agreement results from the
improved treatment of residual polarization by iterating Galactic modelling during
calibration. The  44- and 70-GHz results are basically in line with the previous
release findings.  That is of course expected at 70\GHz, since the calibration procedure
is almost the same as in the previous release, except for the gain smoothing algorithm
and the foreground model adopted (now based on the {\tt Commander} solutions using only \Planck\ data). 

A more quantitative way to represent null-test results, especially at low multipoles,
is to compute deviations from the half-ring noise in terms of
\begin{equation}
\chi_\ell^2 = \frac{\sqrt{2\ell +1}}{2}\left(\frac{C_\ell^{\mathrm{SS}}-C_\ell^{\mathrm{hr}}}{C_\ell^{\mathrm{hr}}}\right)\, .
\end{equation}
We specifically sum each single $\chi_\ell^2$ in the range $\ell = 2$--50.
Then, from the total value of $\chi^2$ and $N_{\mathrm{dof}}$, we
derive $p$-values of the distribution. While a proper set of noise simulations
should in principle be considered, for this inspection it is adequate to use simple
half-ring noise. Nonetheless we should be aware of the fact that any result derived with this
approach is only indicative of possible issues and that a more detailed and refined
analysis is required.  Table~\ref{tab_chi2pvalue} reports both $\chi^2$ and $p$-values
from the three survey differences, as shown in Fig.~\ref{nullspectra} for polarization
spectra at the three LFI frequencies for the 2018 and 2015 data releases. A comment
is in order here.  On the one hand, we see that Survey~2 and Survey~4 seem to have some
problems at 70\GHz, as highlighted by the poor $\chi^2$ and $p$-values. However, this is expected,
since we made only relatively small changes in the calibration pipeline at 70\GHz, and
these surveys were known to be problematic in 2015. Nevertheless,  we can anticipate
that a power spectrum analysis of low-$\ell$ polarization at 70\GHz\ will find good
results even including Survey~2 and Survey~4, thanks to the use of the calibration
template described in Sect.~\ref{sec_calibration}. We note, however, that such a
template does not help to improve the $\chi^2$ and $p$-values, since these are derived
from survey differences; any global template applied to data from both surveys would
cancel out and leave $\chi^2$ results unaffected.
While at 44\GHz\ the picture is practically unchanged with respect to 2015, results at 30\GHz\ show 
in general a good trend of improvement for even survey differences, as indicated by $\chi^2$ values,
and underlining again the benefit of the new calibration scheme.
However, such values are far from being optimal and may indicate the presence of residuals
showing up in the difference maps. Moreover we stress that this kind of analysis is
only indicative and is used internally as an additional validation test.

%no problematic
%issues for even surveys. It is important to note here that with the current
%calibration scheme we have improved our estimate of the noise as encoded by the half rings.
%This has an impact, especially when dealing with Survey~4 at 30\GHz, as shown by its poor $\chi^2$ value.

\begin{table}[htpb]
  \begingroup
  \newdimen\tblskip \tblskip=5pt
  \caption{Odd-even surveys $\chi^2$ and $p$-values ($2\leq\ell\leq 50$).}
  \label{tab_chi2pvalue}
  \nointerlineskip
  \vskip -6mm
  \footnotesize
  \setbox\tablebox=\vbox{
    \newdimen\digitwidth
    \setbox0=\hbox{\rm 0}
    \digitwidth=\wd0
    \catcode`*=\active
    \def*{\kern\digitwidth}
    \newdimen\signwidth
    \setbox0=\hbox{+} 
    \signwidth=\wd0
    \catcode`!=\active
    \def!{\kern\signwidth}
    \halign{\hbox to 1.31in{#\leaderfil}\tabskip=1em&
      \hfil#\hfil\tabskip=0.4em&
      \hfil#\hfil\tabskip=1.2em&
      \hfil#\hfil\tabskip=0.4em&
      \hfil#\hfil\tabskip=0pt\cr 
      \noalign{\doubleline}
      \omit&\multispan2\hfil 2015\hfil&\multispan2\hfil 2018\hfil\cr
      \noalign{\vskip -4pt}
      \omit&\multispan2\hrulefill&\multispan2\hrulefill\cr      
      \omit\hfil Survey Differences\hfil& $\chi^2$& $p$-value& $\chi^2$& $p$-value\cr
      \noalign{\vskip 3pt\hrule\vskip 5pt}
      \omit{\bf 30\,GHz}\hfil\cr
      \noalign{\vskip 4pt}
\hglue 2em S1 $-$ S2&*89.91& 1.6$\times 10^{-4}$&*63.65& 0.0539\cr
\hglue 2em S1 $-$ S3&*40.00& 0.755*&*41.36& 0.705*\cr
\hglue 2em S1 $-$ S4&246.3*& $<1\times 10^{-10}$&146.1*& $4\times 10^{-12}$\cr
      \noalign{\vskip 5pt}
      \omit{\bf 44\,GHz}\hfil\cr
      \noalign{\vskip 4pt}
\hglue 2em S1 $-$ S2&*44.69& 0.568*&*44.28& 0.585*\cr
\hglue 2em S1 $-$ S3&100.4*& 9$\times 10^{-6}$&108.3*& 9$\times 10^{-7}$\cr
\hglue 2em S1 $-$ S4&*46.48& 0.494*&*59.5*& 0.105*\cr
      \noalign{\vskip 5pt}
      \omit{\bf 70\,GHz}\hfil\cr
      \noalign{\vskip 4pt}
\hglue 2em S1 $-$ S2& 82.01& 0.0012&*86.48&  3.97$\times 10^{-4}$\cr
\hglue 2em S1 $-$ S3& 63.11& 0.0582&*64.39& 0.0467\cr
\hglue 2em S1 $-$ S4& 74.06& 0.0071&*74.28& 0.0068\cr
      \noalign{\vskip 5pt\hrule\vskip 3pt}
    }}
  \endPlancktable

  \endgroup
\end{table}

\subsection{Half-ring test}

The actual noise in the LFI data is given directly by the half-ring difference
maps.  Detailed noise characterization is of paramount importance for the
creation of adequate noise-covariance matrices (NCVMs),
as well as for the noise MC realizations that are required in subsequent steps
of the data analysis. Any noise model has to be validated against such
half-ring difference maps.  In the current release, we follow the same
processing steps as in the previous releases. Specifically, we compute
{\tt anafast} auto-spectra in temperature and polarization of the half-ring
difference maps for the period covering the full mission. This is also done on
MC noise simulations produced using noise estimation at the TOI level taken
from FFP10.\footnotemark\footnotetext{This is the latest version of the Full
Focal Plane \Planck\ simulations similar to the FFP8 version used for the
2015 releases (see \citealt{planck2014-a12} for futher details)}
Half-ring noise power spectra are compared with the
distribution of noise spectra derived from the noise simulations and with the white-noise
level computed from the white-noise covariance matrices (WNCVM) produced during the mapmaking process.

\begin{figure*}[htpb]
\centerline{
\includegraphics[width=18cm]{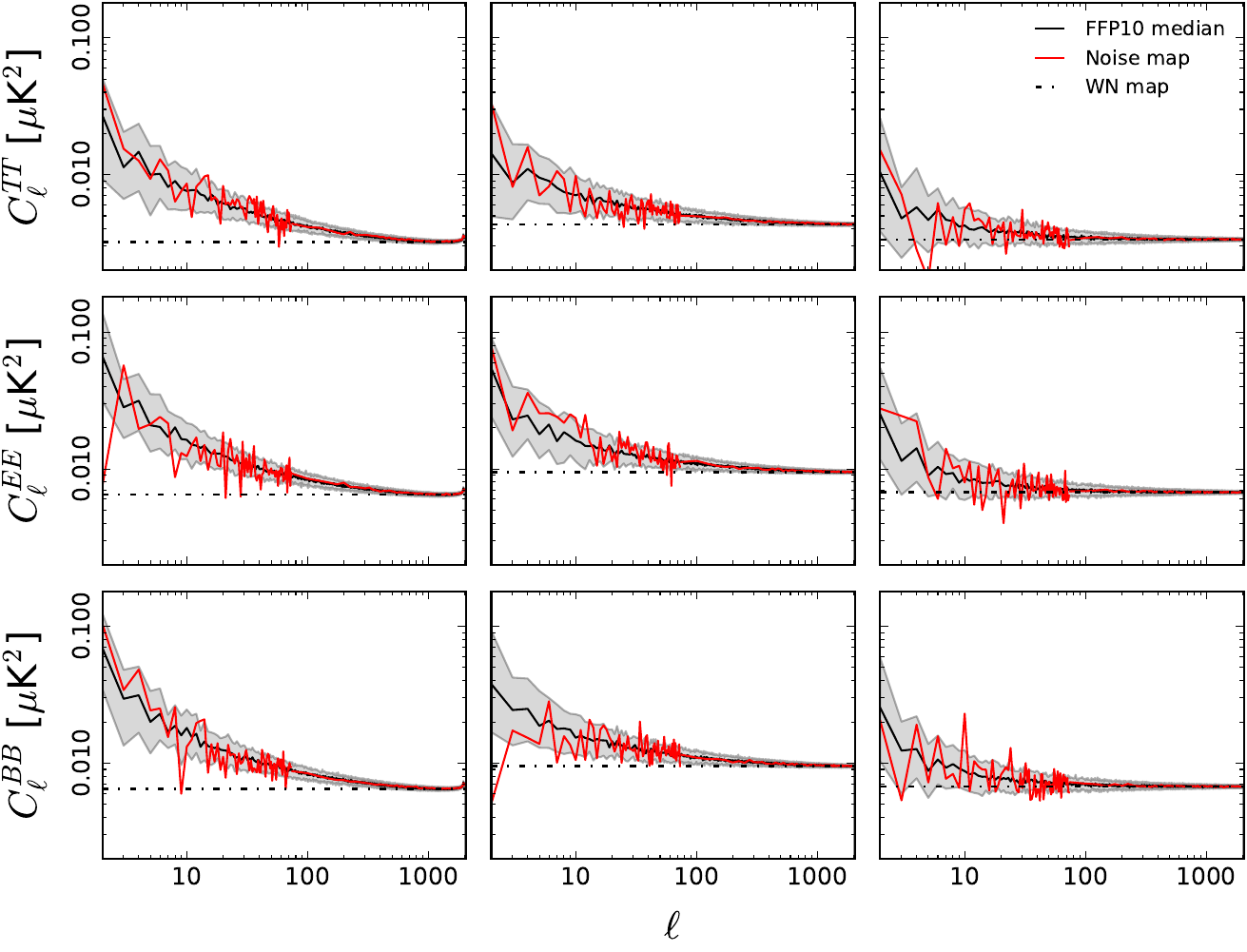}
}
\caption{Consistency check at the power spectrum level between half-ring difference maps (red),
  white-noise covariance matrices (black dash-dotted lines), and 100 full-noise MC simulations
  (grey band showing range for 16\,\% and 84\,\% quantiles of noise simulations,
and the black solid
  lines giving the median, i.e., 50\,\% quantile, of the distributions). From top to bottom we show
  $TT$, $EE$, and $BB$ power spectra for 30\,GHz (left), 44\,GHz (centre),
  and 70\GHz\ (right).  Half-ring spectra are binned with $\Delta \ell=25$ for $\ell \geq75$. }
\label{halfringspectra}
\end{figure*}

Figure~\ref{halfringspectra} shows such a comparison for $TT$, $EE$, and $BB$ spectra.
The grey bands represent the 16\,\% and 84\,\% quantiles of the noise MC, while the black
solid line is the median (50\,\% quantile) of these distributions. The half-ring spectra
are depicted in red, and for $\ell\geq 75$ are binned over a range of $\Delta \ell = 25$.
Even by eye the agreement is extremely good, and makes us confident about proper noise
characterization in LFI data. 

\begin{figure}[htpb]
\centerline{
\includegraphics[width=\columnwidth]{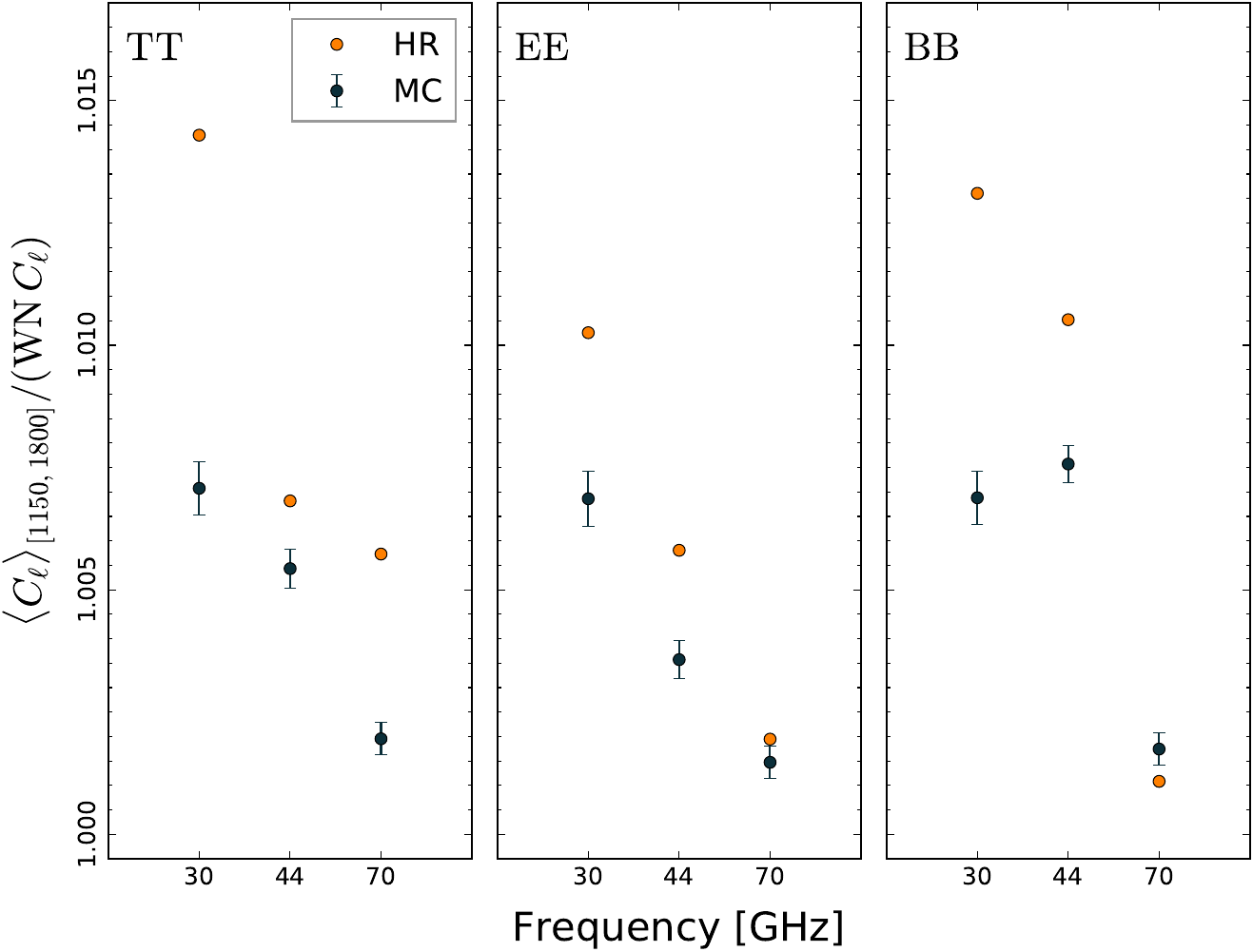}
}
\caption{Ratio of mean noise at high multipoles ($1150\leq \ell\leq 1800$) to
  white noise, derived from white-noise covariance matrices produced during the mapmaking process}
\label{ratioWNCVM}
\end{figure}

We futher investigate the noise properties in the high-$\ell$ regime, taking the
average of $\mathcal{C}_\ell$ in the range $1150\leq \ell \leq 1800$ for both temperature
and polarization, and then comparing with the WNCVM.  Figure~\ref{ratioWNCVM} displays the
result.  As already shown in previous releases, there is still an excess of $1/f$ noise
in this high-$\ell$ regime, meaning that both the real data and the noise MCs predict
slightly larger noise than the WNCVM. It is important to note that such noise excess
is reduced considerably with resepct to the 2015 release, thanks mainly to the new
and more accurate calibration procedure adopted. Residuals are $\la 1.4$\,\% at 30\,GHz,
$\la 1$\,\% at 44\,GHz, and $\la 0.6$\,\% at 70\GHz, for both temperature and polarization.
In addition, agreement between actual noise data and MC simulations is extremely good,
with deviations of only fractions of a percent.

\subsection{Intra-frequency consistency check}

Data consistency can also be checked by means of power spectra, as done in previous
releases \citep{planck2014-a03,planck2013-p02}. We consider frequency maps at 30,
44, and 70\,GHz, and take the cross-spectra between half-ring maps at each frequency
for the full mission time span. Taking cross-spectra has the advantage that we do
not need to consider noise bias at the power spectrum level.  We make use of the
{\tt{cROMAster}} code, a pseudo-$C_\ell$ cross-spectrum estimator
\citep{master,polenta_CrossSpectra}. Results obtained are sub-optimal with
respect to a maximum likelihood approach, but are less computationally
demanding and accurate enough for our purposes.

\begin{figure}[htpb]
\centerline{
\includegraphics[width=9cm]{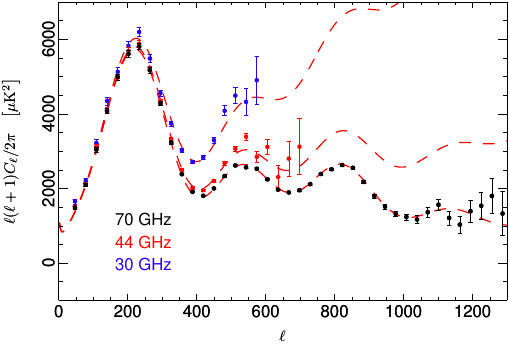}}
\caption{Binned $TT$ cross-spectra from half-ring maps at 30, 44,
  and 70\,GHz. Foregrounds are accounted for only through the use of the proper
  frequency-dependent Galactic masks. Red-dashed lines are the \Planck\
  best-fit $TT$ spectrum, to which a contribution from unresolved point
  sources (specific to each frequency) has been added.}
\label{combinedSpectra}
\end{figure}

The actual spectra are computed using a Galactic mask obtained with the
combination of the \Planck\ G040, G060, and G070 masks at 30, 44, and 70\GHz,
respectively, and accounting for the proper frequency-dependent masks for
resolved point sources. Figure~\ref{combinedSpectra} shows cross-spectra
from the half-ring maps. They agree well among the three frequencies, which
is remarkable since foregrounds have not been removed from the maps, except in
the masked regions.  The red-dashed lines are the 2015 \Planck\ best-fit $TT$
spectrum, augmented by the contribution from unresolved point sources.  The data
are in good agreement with this model at all three frequencies. More quantitatively,
we build a simple Gaussian likelihood (without any beam or foregrounds
modelling) and consider multipole bins up to where the beam function blows up.
We obtain the following $p-$values: 0.196 at 70\GHz\, ($50\lsim\ell\lsim 1300$);
0.262 at 44\GHz\, ($50\lsim\ell\lsim 500$); and 0.017 at 30\GHz\, ($50\lsim\ell\lsim 500$).
This shows that even with this simplified approach 70 and 44\GHz\, are consitent with
the model. The 30\GHz\, channel is marginally consistent with the null hypothesis
and clearly requires a more detailed treatment of foregrounds. 
   
\begin{figure*}[htpb]
\centerline{
\includegraphics[width=6cm]{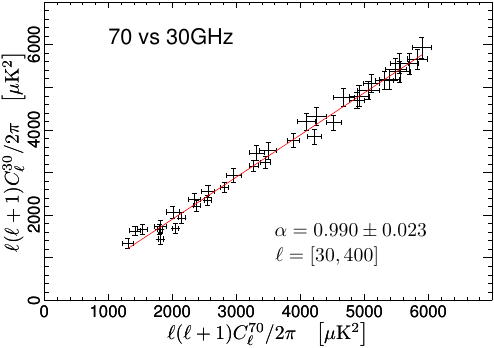}
\includegraphics[width=6cm]{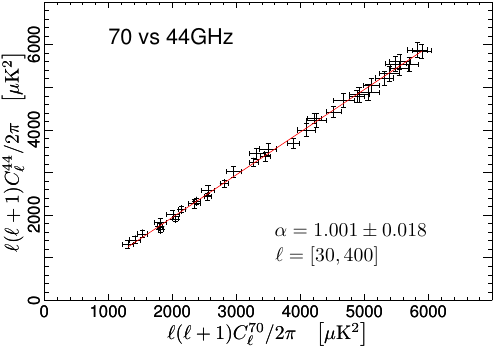}
\includegraphics[width=6cm]{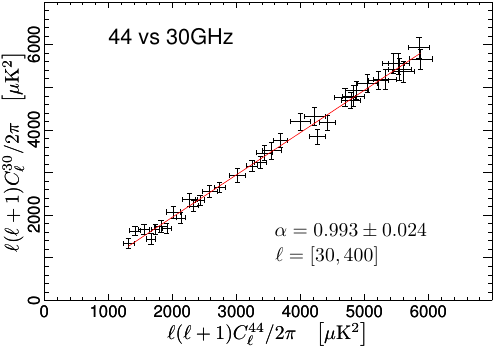}}
\caption{Scatter plots of LFI angular cross-spectra around the first acoustic peak for frequency pairs: left, 70\,GHz and 30\,GHz; middle, 70\,GHz and 44\,GHz; right, 44\,GHz and 30\,GHz. The solid line is the linear fit 
obtained accounting for error bars on both axes. Slopes, within expected uncertainty, show good consistency among
data sets}  
\label{ttplots}
\end{figure*}

As a more quantitative test of data consistency, we build the usual scatter
plots of angular power spectra for the three frequency pairs.  To ensure a proper comparison, we
remove the frequency-dependent contribution from unresolved sources, and
perform a linear fit, accounting for errors on both axes.  Figure~\ref{ttplots}
shows results in the multipole range around the first peak, where the effect of
different angular resolutions at the three frequencies is still manageable.
The agreement is extremely good: spectra are consistent with deviations
between 0.9\,\% and 0.1\,\%, a result in line with findings from the previous
release \citep{planck2014-a03}. This agreement is also an indication of the
calibration accuracy, which is at the sub-percent level. This is a sign of the
good health of our data, given that fact that in these tests we have not
accounted for any errors in the window function, calibration, or foreground removal.

\subsection{Internal consistency check}

\begin{figure}[htpb]
\centerline{
\includegraphics[width=\columnwidth]{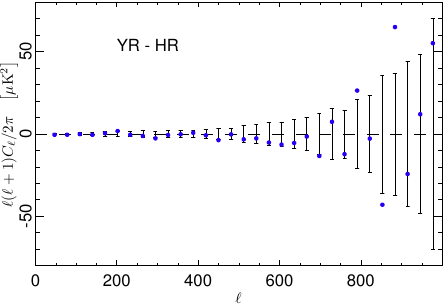}}
\caption{Residuals between cross-spectra at 70\GHz\ from half-ring (HR) and year
  maps (YR). Error bars come from FFP10 Monte Carlo simulations.}
  \label{cross_datasplit}
\end{figure}

\begin{figure*}[htpb]
\centerline{
    \includegraphics[width=6cm]{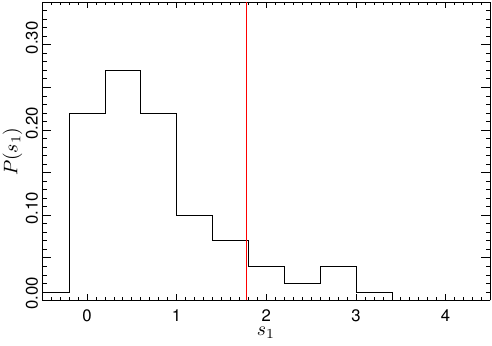}
    \includegraphics[width=6cm]{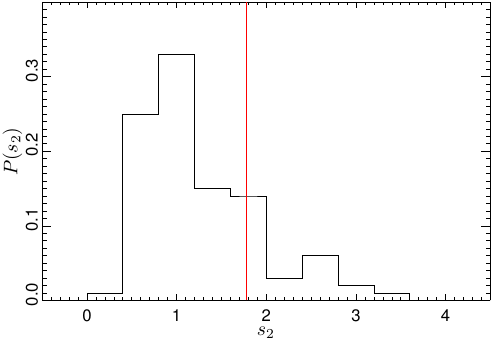}
    \includegraphics[width=6cm]{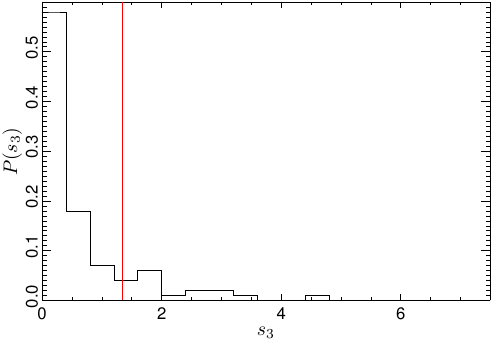}}
\caption{Empirical distributions derived from FFP10 simulations for the $s_1$, $s_2$,
  and $s_3$ statistics of the Hausman test on the 70-GHz data.  Vertical lines
  show values derived from half-ring and year maps at 70\GHz.}
\label{hauss}
\end{figure*}

We also check the internal consistency of the 70-GHz data used for the cosmological
analysis.  As done for the 2015 release, we create cross-spectra by taking different
data splits.  Specifically, we consider the half-ring maps and the year combination
maps Year~1 $-$ Year~3 and Year~2 $-$ Year~4 maps. Figure~\ref{cross_datasplit} shows the
residuals of the combination, compared with the expectations derived from FFP10
Monte Carlo simulations subjected to the same procedure. Residuals are clearly
compatible with the null hypothesis.
To quantify the agreement we apply the Hausman test \citep{polenta_CrossSpectra} to 70\GHz\ cross-spectra
from half-ring and year maps. As reported in \citet{planck2014-a03} we define the statistic
\begin{equation}
  H_\ell = \left(\hat{C}_\ell - \bar{C}_\ell\right)\bigg/\sqrt{{\mathrm{Var}}\left[\hat{C}_\ell - \bar{C}_\ell\right]},
\end{equation}
where $\hat{C}_\ell$ and $\bar{C}_\ell$ represent the two different cross-spectra.
We further compress the multipole information with
\begin{equation}
  B_L(r) = \frac{1}{\sqrt{L}}\sum_{\ell=2}^{\left[Lr\right]}H_\ell,\,\, r\in[0,1],
\end{equation}
where the operator $\left[.\right]$ returns the integer part.  It can be shown that
the distribution of $B_L(r)$ converges to Brownian motion, which can be simply
studied by means of three statistics: $s_1 = {\mathrm{sup}}\,B_L(r)$;  
$s_2 = {\mathrm{sup}}\left|B_L(r)\right|$; and $s_3 = \int_0^1B^2_L(r)dr$. Results are reported in Fig.~\ref{hauss}, where the three
statistics are compared to their expected distributions derived from FFP10
simulations. The corresponding $p$-values for $s_1$, $s_2$, and $s_3$
are, respectively, 0.11, 0.19 and 0.13, showing again that results are perfectly compatible with the
null hypothesis, and confirming the high level of internal consistency in the 70\GHz\ data. 

\subsection{Validation Summary}

At the end of this detailed validation process, the improved quality of the
2018 data release, especially at 30 and 44\GHz, is clear. This improvement is
mainly due to the new calibration procedure implemented. It is also evident that residuals at very
low multipoles (large angular scales) are still present at 30 and at 44\GHz.  One of the
reasons could be the fact that the initial foreground model used for the
iterative calibration, mapmaking and component-separation process was entirely based on
LFI data. We took this approach to avoid dependence on 
\WMAP\ and to avoid ingesting any HFI systematic effects. However, we realize that this results
in a foreground model with less power than expected. This choice should in principle be investigated further, but the actual level of residuals is small and does not prevent the use of the 30-GHz data as a synchrotron template in cosmological analyses
involving the 70-GHz LFI channel \citep[see][for a detailed
analysis of the low-$\ell$ likelihood]{planck2016-l05}

\section{Updated systematic effects assessment}
\label{sec_systematics}

\subsection{General approach}

Analysis of LFI systematic effects from the start of the mission to the
2015 release \citep{planck2013-p02a,planck2014-a03,planck2014-a04} identified
uncertainty in calibration as the dominant source of systematic error. This
source of error, and all other known systematic effects, are at least four
orders of magnitude below the measured CMB power spectrum in total intensity
for all relevant multipoles \citep[see figures~24--26 of][]{planck2014-a04}.
However, this is not the case for polarization.  Imperfect gain reconstruction,
obtained independently for the Main and Side radiometers of each horn,
translates directly into leakage of total intensity to polarization.
In particular, at the large angular scales relevant for probing the
reionization bump, the systematic error from calibration is
comparable with the expected signal, for reasonable values of the
optical depth parameter $\tau$. 

Our assessment of the overall systematic error budget remains essentially
unchanged from the 2015 release, as summarized in table~1 of \citet{planck2014-a04}
and references therein. For the present release, we have concentrated on
developing a detailed simulation programme to model all known instrumental
and astrophysical effects that produce uncertainty in the gain for
polarization data.  We first identify those parameters in the whole
calibration process that are affected by uncertainties, and then set up an ad
hoc Monte Carlo simulation strategy to judge the effect of varying
these parameters. This process has to be both accurate and realistic.
There are two possible sources of error that are quite different in
nature: {\emph{statistical}} uncertainties, related to instrumental
noise in conjunction with variations in the dipole amplitude on the
sky; and {\emph{systematic}} uncertainties due to effects and assumptions
in the calibration process that are not completely known.  The simulations
require a trade-off between a full, physically representative set of
simulations and a realistic, feasible number of simulations, given
the computation resources available.  Ideally, we would assume that all
the effects are indeed correlated with each other, and therefore the total
number of simulations would be the product of the number of simulations
required for each single effect. However, to reduce computer resources needed,
it is more convenient to assume that all the effects are un-correlated, and
the final error is obtained by summing in quadrature the error derived
from each single effect. This uncorrelated assumption is clearly not completely true:
it is impossible to separate the effect of different parameter values of the
gain smoothing algorithm from the actual noise realization. Therefore, we
follow a hybrid approach in the systematic effect simulations.

\subsection{Monte Carlo of systematic effects}

The FFP10 \citep[see][]{planck2014-a12} simulation
pipeline is the basis of our simulations pipeline, and the sky signal we
include makes use of the Planck Sky Model \citep{delabrouille2012}.  A detailed
description of the sky model and components is given in Appendix~\ref{app_psm}.
In the current implementation, the FFP10 pipeline is partially executed in full
timelines (data as a function of time at the original detector sampling
frequency), and partially in the ring-set domain (data binned into a
partial map for every pointing period covering full circles on the sky,
as identified by {\tt HEALPix} pixel indices).

We start by creating separate ring-sets for each signal of interest, using
the actual pointing information.  We have ring-sets for CMB, Galactic
foregrounds, extragalactic diffuse signals, point sources convolved with
the measured LFI beams, and finally the sum of the solar and orbital
dipoles, also convolved with the measured LFI beams.  In the second stage
of the pipeline, we perform the iterative calibration. First we divide
ring-sets by a fiducial set of gains (those estimated from actual data),
and then we process ring-sets to estimate the gain value for each pointing
period using the same pipeline as described (although using a different
implementation) in \citet{planck2014-a06}.
The reconstructed gains, however, are noisy
and biased when the orientation of the line of sight of the telescope scans
a region where the dipole is close to its minimum. Therefore we apply the
same smoothing algorithm developed by the DPC for the real data.  It is
important to note here that the smoothing algorithm is optimized through
a detailed study of real data, and we therefore do not expect that it
will perform optimally on each single simulation.  Observed differences
in the final maps could be used to estimated directly the impact of the
smoothing process on the reconstructed gains.

The final stage of simulations produces calibrated maps with a pipeline based on
{\tt{TOAST}}.\footnotemark\footnotetext{Time Ordered Astrophysics Scalable
  Tools, \url{http://hpc4cmb.github.io/toast/}, developed by the
Computational Cosmology Center C3 at Lawrence Berkeley National Laboratory
to handle MPI-based parallel data processing.}  We combine timelines of the
desired input signals and noise (exactly the same as already used to create
ring-sets for the calibration process), and divide by the fiducial gains,
converted into kelvin with the previously computed set of reconstructed gains.
From these calibrated timelines we subtract the simulated dipole signal, and
then create maps with {\tt{Madam}}. Any discrepancy between the fiducial and
reconstructed gains shows up as a residual dipole and a mis-calibration of the
Galactic signal and the noise. This gives an estimate of the calibration
error that we believe also affects the data.

We have also examined in more depth the impact of known systematic effects
on the calibration process, using targeted simulations. These simulations
include noiseless maps for a few different realizations of each individual
systematic effect: different solar dipole amplitudes and directions based on
the expected error on the \Planck\ dipole; different masks used inside the
calibration pipeline during the gain fitting process;  different beams used
to convolve foreground signal and dipoles, based on the expected error on
bandpasses and its impact on beams; ADC nonlinearities, based on the model
fitted to real data; and finally different sky signals for input and calibrator,
in order to simulate a discrepancy between input data and our sky model. We
create simulations of each single effect plus a set of 20 simulations that
include many different effects together by randomly sampling all the available
options from dipole, beam, and ADC systematics, plus realistic noise simulations.
It should be clear from this description that we are following a hybrid approach where,
although simulated independently, systematic effects can be combined together
to produce a new set of gains and maps.

In the following paragraph we describe results from these simulations of
systematic effects. There are two outcomes.  First, the results show how
well we are able to simulate instrumental and data-processing effects in a
direct comparison (mainly at power spectra level) with null tests on the data.
Second, they provide a quantitative estimation of the amplitude of systematic
effects.  We will also summarize the overall systematic error budget,
accounting for effects not directly simulated in 2018 by using
simulations performed for the 2015 data release. In the evaluation of the
various effects, we also used a perfect calibration simulation, where the
pipeline is exactly the same except that we use the same set of fiducial
gains in both the de-calibration and calibration processes.

\subsubsection{Gain smoothing error}

One of the most delicate steps in the calibration pipeline is the smoothing of the raw 
gains obtained from the dipole/sky fitting procedure. The smoothing algorithm is described 
in detail in \citet{planck2014-a06}, and has been tailored to the LFI data. Its performance 
is strictly linked to the actual noise in the data, the level of dipole signal with respect 
to the the sky signal, and the contribution of the far-sidelobe pickup.  An algorithm tuned 
to the specifics of the real data might not perform equally well on any single simulation, 
and one might be tempted to optimize the smoothing procedure for each one.  We choose not 
to do that, both because it would be a lengthly process, and also because it would introduce 
the gain smoothing algorithm itself as a new variable that would vary among simulations, making 
a direct comparison much more cumbersome.  Instead using exactly the same smoothing procedure 
applied to real data allows us to evaluate its impact on data when compared to a perfect calibration simulation. 

\begin{figure*}[htpb]
\centerline{
  \includegraphics[width=18cm]{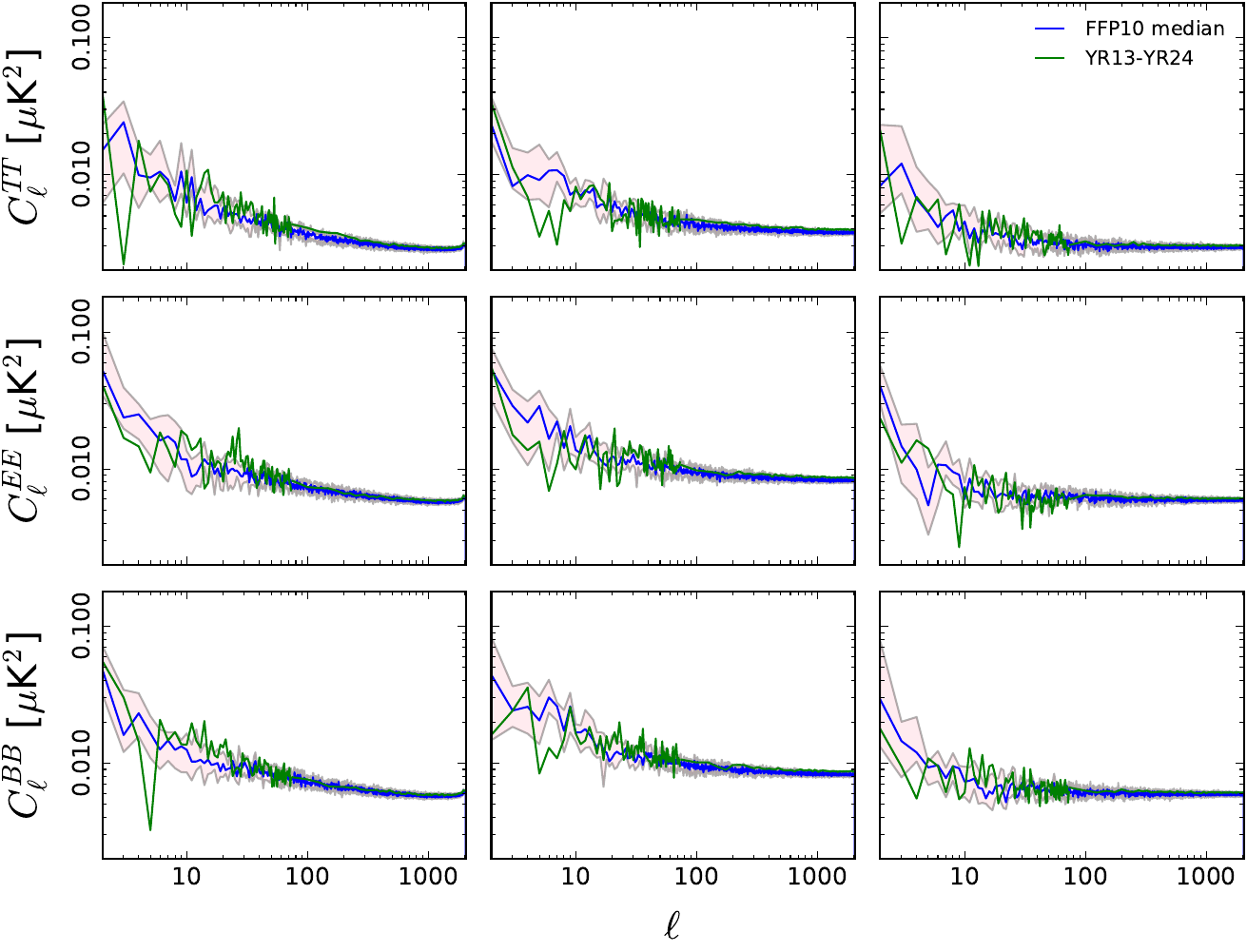}
   }
\vskip -10pt
\caption{Odd-even-year null spectra compared with simulations including the gain
  smoothing algorithm at 30 (left), 44 (middle), and 70\GHz\,(right). The pink
  band shows the 16\,\%--84\,\% quantile range of the simulations, with the median
  traced by the blue line.  On large angular scales the data show larger variations
  with multipole than the simulations.}
\label{fig_gainsmootherror_spectra}
\end{figure*}

Figure~\ref{fig_gainsmootherror_spectra} compares angular power spectra of odd-even-year null 
tests (in other words, the difference between the sum of Year~1 and Year~3 and the sum of Year~2 and Year~4) 
with the distribution (median plus 16\,\% and 84\,\% quantiles) derived from simulations.  
The simulations include the impact of the gain smoothing algorithm on the gain set that has 
been derived using the full set of systematic simulations, but applied to the noise-only timestream.  
We see that there is overall good agreement between data null tests and these pure-noise simulations. 
However, there are clearly some multipoles for which data null values are outside the range of simulations. 
This tells us that there are effects in the data other than the gain smoothing error.  Indeed, we made the 
same comparison for perfect calibration simulations, and found a very similar plot with very small 
differences in the distribution of simulated spectra.  In order to evaluate the impact of the gain 
smoothing error, we take the amplitude (for both perfect calibration and gain smoothing error) of 
the $1\,\sigma$ bands for some specific range of multipoles.  Considering $\ell=4$--5, the overall 
effect of the gain smoothing algorithm for the three frequency channels is to increase noise by 
$\la 3\times 10^{-3}\muK^2$ in total intensity and $\la 6\times 10^{-3}\muK^2$ in polarization, almost 
the same for all three frequencies.  Considering higher multipole ranges, $\ell\simeq100$ and $\ell\simeq500$, 
errors are reduced by at least one order of magnitude with respect to the $\ell=4$--5 range,
ranging between $10^{-4}$--$10^{-5}\mu\mathrm{K}^2$, 
both in temperature and polarization. 

\subsubsection{ADC nonlinearities}

The ADCs convert analogue detector outputs into numbers, and any nonlinearities
in their response could mimic a sky signal thereby affecting data calibration.  As
mentioned in Sect.~\ref{sec_toiprocessing}, for the current release we implemented
a new method to track and correct ADC nonlinearities that improves the quality
of the data at 30\GHz.

As well as the analysis done in the definition of the new correction method, we
also performed specific simulations of the effects of nonlinearities in the
ADCs. We created 10 noiseless simulations with the ADC effect, based on a model
fitted with real data.  In these simulations, we randomized the errors in the
voltage steps associated with each binary bit of the ADC in a way
that is consistent with what we found from the data.  Results are in line
with the findings in 2015, with improvements at 30\GHz\ where residuals decrease
by almost an order of magnitude.  Again considering the range $\ell=4$--5, the
ADC effects now contribute an increase in the noise of $\la 10^{-4}\muK^2$
at 30\GHz\ and around $10^{-5}\muK^2$
at  44 and 70\GHz. At higher multipoles the effect drops below $10^{-6}\muK^2$.
In polarization the effect is at least one order of magnitude smaller than in total intensity.

\subsubsection{Full systematic simulations}

The final set of simulations we consider includes all the effects that we expect
to directly impact calibration accuracy.  We now present the results obtained,
together with a final comprehensive table of the estimated impact of systematic
effects.  In addition, we create a summary plot like the ones in \citet{planck2014-a04},
but with updates for those effects being simulated for the current data release,
including plots of $EE$ and $BB$.

\begin{figure*}[htpb]
\centerline{
\includegraphics[width=18cm]{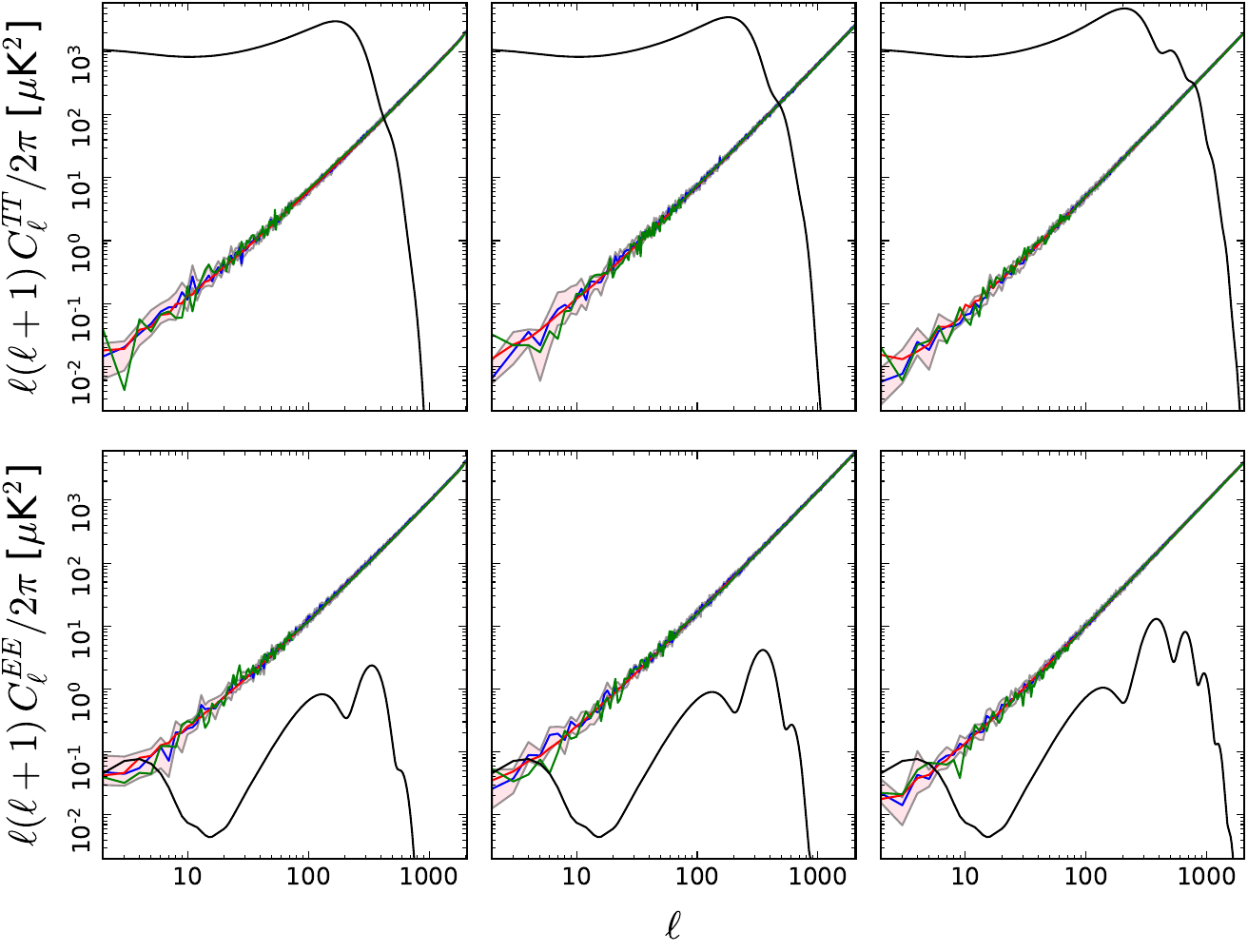}
}
\caption{Angular $TT$ (top) and $EE$ (bottom) power spectra at 30 (left),
  44 (middle), and 70\GHz\ (right) of a complete set of systematic effects for
  the 2015 data release (red) and the 2018 simulations
  (blue, 50\,\% percentile plus $1\,\sigma$ band from simulations).
  This shows the good agreement between the two releases. In
  green we show the null spectra from odd-even year difference.
  The CMB angular power spectra (black lines) are convolved with
  the frequency dependent window functions.}
\label{fig_sysTTspectra}
\end{figure*}

We begin by comparing the full systematic error budgets between the current
and the 2015 data release. Figure~\ref{fig_sysTTspectra} shows $TT$ and $EE$
angular power spectra at the three LFI frequencies.  Systematic simulations
made for the two data releases are in very good agreement; this justifies
our claim that the overall systematic error budget remains essentially unchanged
between 2015 and 2018.  Moreover, the null spectra from odd-even-year differences
are very close to the the systematic error expectations, both in temperature and
polarization.  There are a very few multipoles where null spectra lie outside
the $1\,\sigma$ band of the simulations.  Specifically, for $TT$ they are $\ell=2$
and 3 at 30\GHz, and $\ell=2$ for both 44 and 70\GHz. The extra variance in
the null tests is $\la 0.04\muK^2$ at 30 and 70\GHz, and around $0.065\muK^2$
at 44\GHz. At higher multipoles, the agreement with simulations, dominated by
instrumental noise, is extremely good.

For the $EE$ power spectra, we can see again very good agreement between
systematic simulations performed for the 2015 data release and those done
for the current release.  We see that the overall account of systematics
estimated in 2015 falls very close to the median of the present simulations
and, in any case, well within the $1\,\sigma$ limits. In addition, the odd-even-year
null difference is extremely well represented by our simulations, except
at a few multipoles.  This again tells us that we understand well the instrumental
and systematic effects actually present in the data. Evaluating the extra
variance in the data not accounted for by simulations, we see that
at 30\GHz\ the agreement is extremely good, at 44\GHz, there is an excess
at $\ell=4$--6 of $\la 0.05\muK^2$, while at 70\GHz, the very low multipoles
are in line with expectations (although for $\ell=10$ there is an excess of about $0.10\muK^2$).

Table~\ref{tbl_summary} summarizes systematic effects at the power spectrum
level for three multipole ranges: $\ell=4$--5; around the CMB first peak
$\ell\simeq100$; and the almost noise-dominated regime $\ell\simeq500$.

\begin{table*}[htpb]
  \begingroup
  \newdimen\tblskip \tblskip=5pt
  \caption{Additional noise in $\muK^2$ from systematic effects for three
    different multipole ranges ($\ell=4$--5, $\ell\simeq100$, and $\ell\simeq500$).}
  \label{tbl_summary}
  \nointerlineskip
  \vskip -5mm
  \footnotesize
  \setbox\tablebox=\vbox{
 % \vbox{
  \newdimen\digitwidth
  \setbox0=\hbox{\rm 0}
  \digitwidth=\wd0
  \catcode`*=\active
  \def*{\kern\digitwidth}
  \newdimen\signwidth
  \setbox0=\hbox{+}
  \signwidth=\wd0
  \catcode`!=\active
  \def!{\kern\signwidth}
 \halign{\hbox to 2.2in{#\leaderfil}\tabskip=2em&
      #\hfil&
      \hfil#\hfil&
      \hfil#\hfil&
      \hfil#\hfil\tabskip=0pt\cr
 \noalign{\doubleline}
\omit\hfil Effect\hfil&\omit\hfil Procedure\hfil& 30\GHz& 44\GHz& 70\GHz\cr
\noalign{\vskip 5pt\hrule\vskip 4pt}
\omit\boldmath$\ell=4$--5\hfil&&&&\cr
\noalign{\vskip 4pt}
\hglue 2em Gain smoothing error&                        Simulations + odd-even years&$\la 3\times10^{-3}$ ($TT$)& $\la 3\times10^{-3}$ $(TT)$& $\la 3\times 10^{-3}$ $(TT)$\cr
\omit&&                                                                                                                     $\la 6\times10^{-3}$ $(EE)$& $\la 6\times10^{-3}$ $(EE)$& $\la 6\times 10^{-3}$ $(EE)$\cr
\hglue 2em ADC&                                                  Simulations + half-rings&           $\la 1\times 10^{-4}$ $(TT)$& $\la 1\times10^{-5}$ $(TT)$& $\la 1\times10^{-5}$ $(EE)$\cr
\omit&&                                                                                                                     $\la 1\times 10^{-5}$ $(EE)$& $\la 1\times10^{-6}$ $(EE)$& $\la 1\times10^{-6}$ $(EE)$\cr
\hglue 2em Full ($4\pi$ beam + dipole params)& Simulations + odd-even years& $\la 7\times10^{-3}$ $(TT)$& $\la 4\times10^{-3}$ $(TT)$& $\la 1\times10^{-3}$ $(TT)$\cr
\omit&&                                                                                                                     $\la 1\times10^{-2}$ $(EE)$&  $\la 1\times10^{-2}$ $(EE)$& $\la 6\times10^{-3}$ $(EE)$\cr 
\noalign{\vskip 10pt}
\omit\boldmath$\ell\simeq100$\hfil&&&&\cr
\noalign{\vskip 4pt}
\hglue 2em Gain smoothing error&                         Simulations + odd-even years&$\la 1\times10^{-4}$ $(TT)$& $\la 1\times10^{-4}$ $(TT)$& $\la 1\times 10^{-5}$ $(TT)$\cr
\omit&&                                                                                                                     $\la 1\times10^{-4}$ $(EE)$&$\la 1\times10^{-4}$ $(EE)$& $\la 1\times10^{-4}$ $(EE)$\cr
\hglue 2em ADC&                                                   Simulations + half-rings&          $\la 1\times10^{-6}$ $(TT)$& $\la 1\times10^{-6}$ $(TT)$& $\la 1\times10^{-6}$ $(TT)$\cr
\omit&&                                                                                                                     $\la 1\times10^{-7}$ $(EE)$& $\la 1\times10^{-7}$ $(EE)$& $\la 1\times10^{-7}$ $(EE)$\cr
\hglue 2em Full ($4\pi$ beam + dipole params)&  Simulations + odd-even years& $\la 1\times10^{-4}$ $(TT)$& $\la 1\times10^{-3}$ $(TT)$& $\la 3\times 10^{-5}$ $(TT)$\cr
\omit&&                                                                                                                      $\la 1\times10^{-3}$ $(EE)$& $\la 6\times10^{-4}$ $(EE)$&    $\la 4\times10^{-4}$ $(EE)$\cr
\noalign{\vskip 10pt}
\omit\boldmath$\ell\simeq500$\hfil&&&&\cr
\noalign{\vskip 4pt}
\hglue 2em Gain smoothing error&                         Simulations + odd-even years&$\la 1\times10^{-5}$ $(TT)$& $\la 1\times10^{-5}$ $(TT)$& $\la 1\times 10^{-5}$ $(TT)$\cr
\omit&&                                                                                                                     $\la 1\times10^{-5}$ $(EE)$& $\la 1\times10^{-5}$ $(EE)$& $\la 1\times10^{-5}$ $(EE)$\cr
\hglue 2em ADC&                                                   Simulations + half-rings&          $\la 1\times10^{-6}$ $(TT)$& $\la 1\times10^{-6}$ $(TT)$& $\la 1\times10^{-6}$ $(TT)$\cr
\omit&&                                                                                                                     $\la 1\times10^{-7}$ $(EE)$& $\la 1\times10^{-7}$ $(EE)$& $\la 1\times10^{-7}$ $(EE)$\cr
\hglue 2em Full ($4\pi$ beam + dipole params)&  Simulations + odd-even years& $\sim 1\times10^{-5}$ $(TT)$& $\la 6\times10^{-5}$ $(TT)$& $\la 7\times10^{-6}$ $(TT)$\cr
\omit&&                                                                                                                     $\la 3\times10^{-4}$ $(EE)$& $\la 4\times10^{-4}$ $(EE)$& $\la 2\times10^{-4}$ $(EE)$\cr
\noalign{\vskip 5pt\hrule\vskip 3pt}}}
\endPlancktablewide
\endgroup
\end{table*}
  
A comment is in order here: for multipoles $\ell \ga 100$, it is clear
that the extra noise induced by systematic effects is well described by
simulations, and accounted for in the overall estimation of the noise budget.
At large angular scales, the improvement with the new calibration scheme is
evident from the comparison of null-test spectra on different data combinations
between the 2015 and 2018 data releases.  However, there are a few multipoles
that deserve particular attention, since specific null tests still show a noise
excess that is not completely traced by the simulations of systematics.  Nevertheless,
results presented in this section are intended to be a useful indication of
the overall amplitude of systematic effects.  More scientifically-oriented
analysis could in principle make use of the simulations to build a template
that could be fitted for in a maximum likelihood approach (along the same
lines as foreground emission templates),  and where errors in the simulations
could be properly propagated into the final low-$\ell$ angular power spectrum
and cosmological parameters.

\section{LFI data products available through the \Planck\ Legacy Archive}

Before concluding, we provide a list and short description of the \Planck\ LFI data
products available through the \Planck\ Legacy Archive,\footnotemark
\footnotetext{{\url{http://archive.esac.esa.int/pla2}}} based on the data
covering the operational lifetime of the instrument from 12 August 2009 to
23 October 2013 \citep[for further details on the data format refer to the
Explanatory Supplement,][]{planck2016-ES}.

\begin{itemize}

\item{Pointing timelines: identical to the 2015 release. One FITS file
  for each OD for each frequency. Each FITS file contains the OBT
  (on-board time) and the three angles, $\theta$, $\phi$, and $\psi$ that
  identify each sample on the sky.}

\item{Time timelines: identical to the 2015 release. One FITS file for
  each operational day (OD) for each frequency.  Each FITS file contains
  the OBT and its corresponding TAI (International Atomic Time)
  value (without leap second) in modified Julian-day format. The user can thus
  cross-correlate OBT with UTC.}

\item{Housekeeping timelines: identical to the 2015 release. All housekeeping
  parameters with their raw and calibrated values, separated by the housekeeping sources, for each OD.}

\item{Timelines in volts: raw scientific data in volts for each detector
  at 30,  44, and 70\GHz, and each OD, before any calibration procedure
  and with no instrumental systematic effects removed.}

\item{Cleaned and calibrated timelines: provided in ${\rm K}_{\rm CMB}$ units, for
  each detector at 30,  44, and 70\GHz, and each OD, after scientific
  calibration and with convolved dipoles and convolved Galactic straylight removed.}

\item{Scanning beam: $4\pi$ beam used in the calibration pipeline.}

\item{Effective beam: sky beam representation as a projection of scanning beam on the maps.}

\item{Full-sky maps at each frequency: maps of the observed sky at 30,
  44, and 70\GHz\ in temperature and polarization at $N_{\mathrm{side}}=1024$,
  and also at $N_{\mathrm{nside}}=2048$ for 70\GHz.  Maps are delivered for
  different data periods.  We note that the \Planck\ adopted polarization convention is
  not the one proposed by ``IAU'' but the one used more generally in CMB full-sky maps and
  referred to as ``COSMO'' \citep[see][]{planck2016-ES}.}
\item{Bandpass correction maps at each frequency and maps with the bandpass
  correction applied to delivered frequency sky maps (one specific example of each
  different data period) computed according to the prescription detailed in Sect.~\ref{sec_polarization}. }
\item{Gain Correction Template: template map to be subtracted from the
  delivered full sky map at 70\GHz\ in order to account for calibration uncertainties.
  This template has to be removed prior of any cosmological exploitation of the 70\GHz\ map.
  For completeness we also deliver the 70\GHz\ map with both bandpass and gain correction
  template applied.}
\item{RIMO (Reduced Instrument MOdel): model which includes parameters for
  the main instrument properties, including noise, bandpass, and beam function.}

\end{itemize}

\section{Discussion and conclusions}
\label{sec_conclusion}

\medskip
\noindent
We have presented a comprehensive description of the LFI data analysis
pipeline that produces final frequency maps to be used for scientific
exploitation.  The major improvements with respect to the 2015 data release
are in the new iterative calibration procedure at 30 and 44\GHz, which
uses sky estimation and component separation to create a sky model
to be fed into the calibration algorithm. Other minor improvements
are in the re-definition of the data flags that allow better selection of data.

The validation and improvements of the current data release are performed
with the usual battery of null tests on data with different observing
periods (half-rings, odd-even survey differences, odd-even year differences).
In addition, we performed an exhaustive comparison of the results on such
null tests obtained with the 2015 and 2018 data sets. These tests,
more than any others, clearly show the improvements in the data
quality, especially at 30 and 44\GHz, thanks to the new calibration
scheme. The better data selection in 2018 is also able to marginally improve
the quality of the 70\GHz.

For the analysis of systematic effects, we chose here not to consider
the whole set of effects simulated in 2015, but select only those expected to contribute to the final calibration accuracy. As a
result, the overall systematic error budget remains unchanged with respect
to 2015, and we have verified that this assumption is indeed true by comparing
present and past simulations with current null test spectra. Specifically, we consider
the ADC nonlinearity effect, the impact of the gain smoothing algorithm, and the
impact of parameters in both the $4\pi$ beams and the direction and amplitude
of the solar dipole.  The end result is that for multipoles $\ell \ga 100$,
the impact of systematic effects is well described by simulations, and is
well-accounted for in the overall error budget.  At large angular scales
there are still a very few multipoles that show a noise excess with
respect to simulations.  It is important to note that our power spectrum
analysis is performed on a masked sky with a pure pseudo-$C_\ell$ approach,
which is known to be sub-optimal at very low multipoles. Therefore our analysis
should be regarded as an overall indication of the amplitude of systematic
effects.  The 2018 release includes our end-to-end simulations, which allow those interested in systematic effects to create templates of the various systematics, similar to the gain-correction template described in Sect.~\ref{sect:calresults}.

In conclusion, we have demonstrated substantial improvement in the calibration of the LFI data over previous releases, achieving an overall calibration accuracy of 0.1--0.2\,\%.  We have provided a comprehensive description of the uncertainties, including systematic effects.  
Additional improvements are still possible, and can be anticipated in the future.

\begin{acknowledgements}

The Planck Collaboration acknowledges the support of: ESA; CNES and CNRS/INSU-IN2P3-INP (France); ASI, CNR, and INAF (Italy); NASA and DoE (USA); STFC and UKSA (UK); CSIC, MINECO, JA, and RES (Spain); Tekes, AoF, and CSC (Finland); DLR and MPG (Germany); CSA (Canada); DTU Space (Denmark); SER/SSO (Switzerland); RCN (Norway); SFI (Ireland); FCT/MCTES (Portugal); and ERC and PRACE (EU). A description of the Planck Collaboration and a list of its members, indicating which technical or scientific activities they have been involved  in, can be found at \url{http://www.cosmos.esa.int/web/planck/}. 
The simulations for systematics assessment used the Extreme Science and Engineering Discovery Environment 
(XSEDE, \citet{xsede}), supported by National Science Foundation grant number ACI-1548562, 
in particular the Comet Supercomputer at the San Diego Supercomputer Center through allocation 
AST160021: "Monte Carlo simulations for calibration uncertainty of the Planck mission", 
PI A. Zonca, CoPI P. Meinhold.

\end{acknowledgements}

\allearlypapers
\alltwentyfifteenresultspapers
\alltwentythirteenresultspapers

\bibliographystyle{aat}

\bibliography{Planck_bib,LFI_DPC_bib}

\begin{appendix}
\section{Comparison of LFI 30\,GHz with WMAP K and Ka bands}
\label{app_LFIWMAP}

\begin{figure*}[!ht]
  \centerline{
    \includegraphics[width=8.8cm]{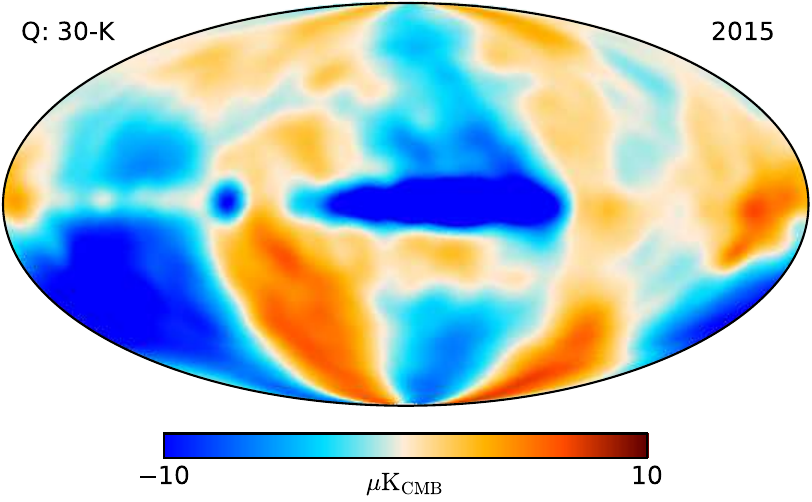}
    \includegraphics[width=8.8cm]{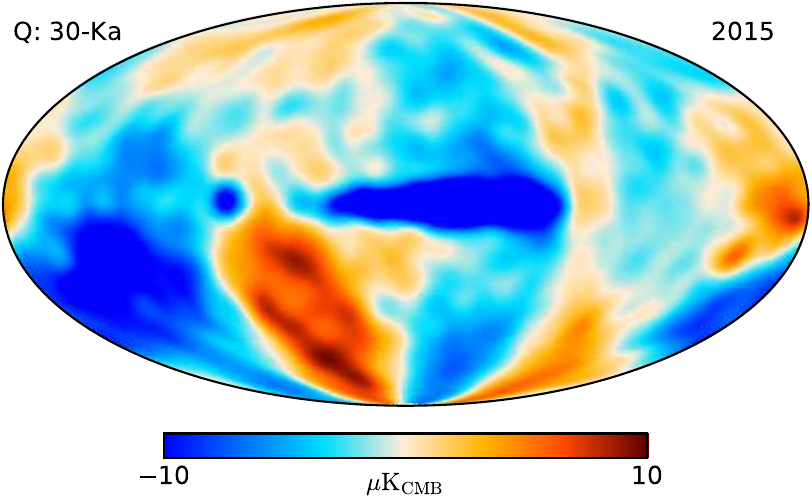}}
  \vspace{-1.cm}
  \centerline{
    \includegraphics[width=8.8cm]{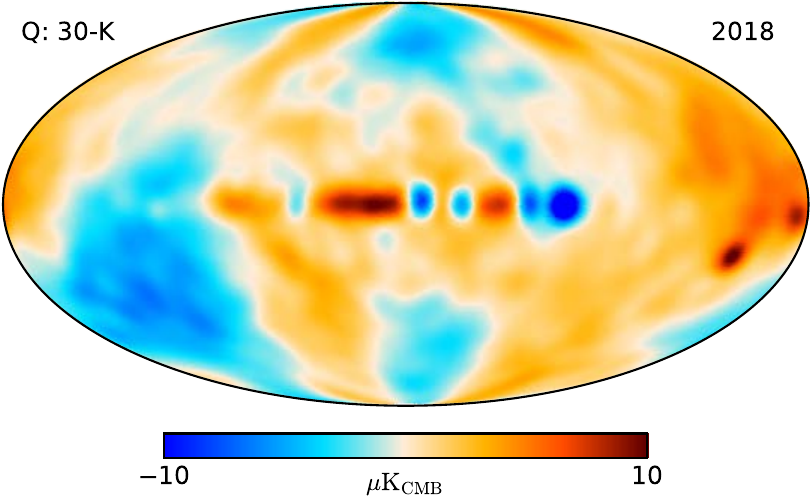}
    \includegraphics[width=8.8cm]{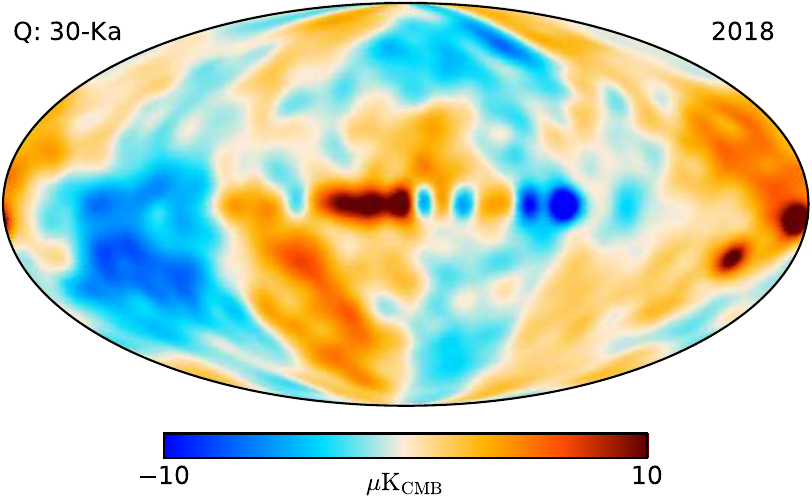}}
 \caption{Stokes $Q$ difference maps between \Planck\ 30\GHz\ 2015 (top) and 2018 (bottom) and \WMAP\ K-band (left)
 and Ka-band (right).  The \WMAP\ K and Ka band maps are rescaled to match the \Planck\ 30-GHz effective frequency assuming synchrotron emission with a spectral index of $-3$.  All maps used in this comparison have been smoothed with a 10\deg\ Gaussian beam.}
  \label{30Kwmap}
\end{figure*}

\begin{figure*}[!ht]
  \centerline{
    \includegraphics[width=8.8cm]{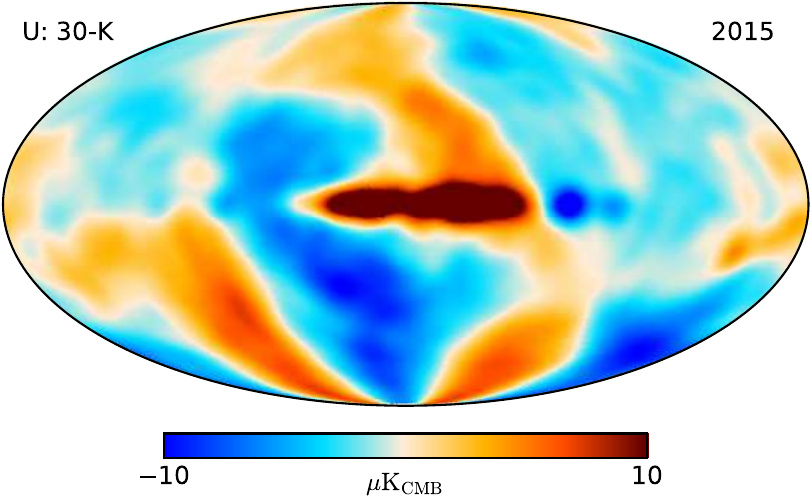}
    \includegraphics[width=8.8cm]{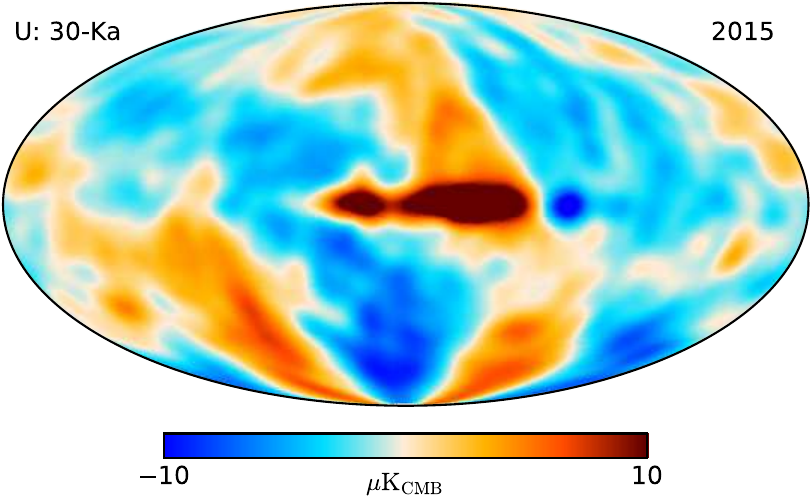}}
  \vspace{-1.cm}
  \centerline{
    \includegraphics[width=8.8cm]{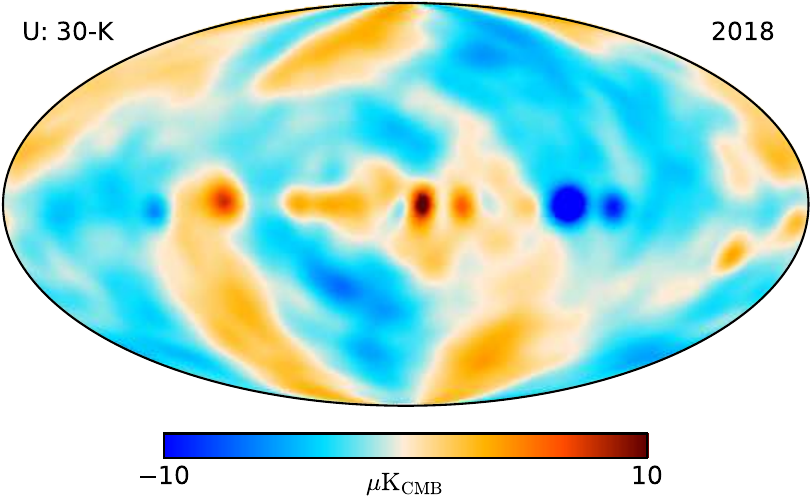}
    \includegraphics[width=8.8cm]{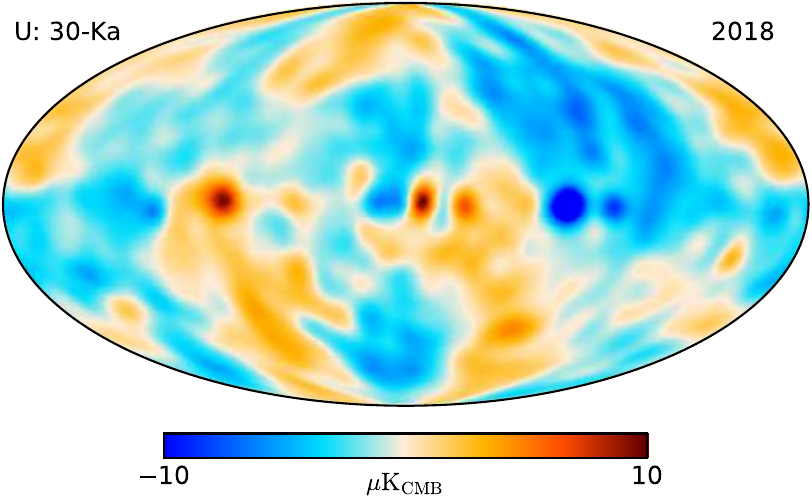}}
  \caption{Same as in Fig.~\ref{30Kwmap}, for Stokes $U$.}
  \label{30Kawmap}
\end{figure*}

In order to provide a further check on systematic effects,
we present here a comparison in polarization at large scales
between the LFI 30-GHz channel and the K and Ka bands from
\WMAP\ \citep{bennett2012}.  As a first step we subtract from the \Planck\ 2015
and 2018 maps the corresponding bandpass corrections. We then filter \Planck\
and \WMAP\ maps with a 10\deg-FWHM Gaussian beam in order to suppress
high-frequency noise and highlight the large-scale structures.
We rescale the K and Ka bands to the LFI 30-GHz effective frequency, assuming
synchrotron emission with a spectral index of $-3$.  Finally, we calculate $Q$
and $U$ differences between the \Planck\ 2015 and 2018 maps and the \WMAP\
maps, as shown in Figs.~\ref{30Kwmap} and \ref{30Kawmap}.

The amplitude of the large-scale structure in the difference maps decreases by roughly a factor of two between the 2015 and the 2018 maps, reflecting the improvements in the new calibration scheme.  This improved calibration has reduced or eliminated some of the features noted by \citet{weiland2018}  in their comparison of \WMAP\ and \Planck\ 2015 $Q$ and $U$ maps.  However, as pointed out in Sect.~\ref{sect:calresults}, the calibration could not be run to convergence for practical reasons, and we expect residuals in the 2018 maps with a pattern similar to that of the 2015 maps.   The fact that the difference maps shown in Figs.~\ref{30Kwmap} and \ref{30Kawmap} have some common features with the two \Planck\ releases is therefore not a surprise.  There is also some agreement between features in the difference maps and the 70-GHz gain correction template (Fig.~\ref{fig:gaintemp_70GHz}), which is based in part on the 30-GHz maps.  Despite the substantial improvement in the 2018 30-GHz polarization maps, residual systematics are still present, and included in our estimated error budget.   Additional improvements are planned in a future paper.

\section{Simulations of systematic effects}
\label{app_psm}

\subsection{Input Sky Model}

The input FFP10 sky includes the following components:
\begin{itemize}
\item Galactic thermal dust, spinning dust, synchrotron, free-free,
  and CO line emission;
\item the cosmic infrared background;
\item Galactic and extragalactic point sources (radio and infrared);
\item thermal and kinematic Sunyaev Zeldovich effects from Galaxy clusters; and
\item the CMB.
\end{itemize}

\subsubsection{Thermal dust}

Galactic thermal dust emission is modelled by scaling across frequencies a
polarized template of emission at 353\,GHz. The
intensity map has been obtained using the Generalized Needlet ILC
({\tt GNILC}) method of \citet{remazeilles2011}, applied to the 2015 (PR2)
release \Planck\ HFI maps,
as described in \citet{planck2016-XLVIII}.  The 353-GHz \Planck\ dust map
obtained in this way is colour-corrected to obtain a template at 353\,GHz.
%using as an emission law in each pixel the modified blackbody parameters
%obtained by \citet{planck2016-XLVIII} from frequencies higher than 353\,GHz,
%and using the same temperature for smaller frequencies.
An overall offset
of 0.13 \MJysr\ (corresponding to the residual CIB monopole) is subtracted
from the {\tt GNILC} dust map.

Polarization templates are generated using the prescription
of \citet{vansyngel2017} to generate random dust polarization maps.
The method relates the dust
polarization sky to the structure of the Galactic magnetic field that
is responsible for aligning
elongated emitting dust grains, building on the analysis of dust polarization
properties described in \citet{planck2016-XLIV}. The Galactic magnetic field is
modelled as a three-dimensional superposition of a mean uniform field and a
Gaussian random (turbulent) component with a power-law power spectrum of
exponent $-2.5$. Polarization maps are obtained from the superposition of
emission from four emitting layers that all share the same
intensity template.

In the vicinity of the Galactic plane,
the simulated polarization data are replaced by
real \Planck\ 353-GHz data. The transition between real data and simulations
is made using a Galactic mask with a 5\deg\ apodization (which leaves 68\,\% of the sky unmasked), taken from the set of \Planck\, common Galactic masks available in the
\Planck\, explanatory supplement.
\footnote{\url{http://wiki.cosmos.esa.int/planckpla2015/index.php/Frequency_Maps\#Masks}}
The first harmonic multipoles, corresponding to
$\ell <10$, come from the 353-GHz polarized sky map.

The scaling in frequency of the dust templates uses the prescription used
for FFP8. A different modified black-body emission law is used in each of
the $N_{\mathrm{side}}=2048$ pixels. In addition, the dust spectral
index used for scaling in frequency is different at frequencies
above and below 353\GHz: for frequency above 353\GHz\, the modified blackbody
parameters obtained in \citet{planck2016-XLVIII} are used; below 353\GHz\, the temperature
map is unchanged, but a map of emissivities computed as described in
\cite{planck2013-XVII} is used instead.

\subsubsection{Other Galactic emission}

Synchrotron intensity is modelled by scaling in frequency the 408-MHz template
map from \citet{haslam1982}, as reprocessed by
\citet{remazeilles2015}, using one single power law per pixel. The pixel-dependent
spectral index is derived in the analysis of the \WMAP\
data by \citet{miville2008}. The generation of synchrotron polarization
follows the prescription of the original PSM paper \citep{delabrouille2012}. 

Free-free, spinning dust models, and Galactic CO emission are essentially the same
as used for FFP8 simulations \citep{planck2014-a14}, but the actual synchrotron
and free-free maps used for FFP10 are obtained with a different realization of
small-scale fluctuations of the intensity. CO maps do not include small-scale
fluctuations, and are generated from the spectroscopic survey of \citet{dame2001}.
None of these three components is polarized in the FFP10 simulations.

\subsubsection{Cosmological parameters}

The generation of all extragalactic components in the FFP10 sky depends on the
assumed cosmological scenario, and in particular values for all cosmological
parameters that impact the CMB power spectra \citep[see][]{planck2014-a15}
and/or the statistical distribution
of extragalactic objects. We use the parameter values listed in Table~\ref{tab_PSMcosmo},
which are used as inputs to the Cosmic Linear Anisotropy Solving
System ({\tt CLASS}) code \citep{blas2011,didio2013}
for generating CMB $TT$, $EE$, $BB$ and lensing $\Phi$ power spectra,
as well as density-contrast shells in auto- and cross-spectra at a set of
redshifts (used to model CIB emission).
We assume 2.0328 massless neutrinos, and one massive neutrino with a
mass of 60\,meV, compatible with a standard neutrino hierarchy. The pivot scale
for the scalar perturbations is 0.002, and the tensor spectral index and the
running of the scalar spectral index are zero.

%TABLE %%%%%%%%%
\begin{table}[htpb]
  \begingroup
  \newdimen\tblskip \tblskip=5pt
  \caption{Cosmological parameters adopted in the PSM.}
  \label{tab_PSMcosmo}
  \nointerlineskip
  \vskip -3mm
  \footnotesize
  \setbox\tablebox=\vbox{
    \newdimen\digitwidth
    \setbox0=\hbox{\rm 0}
  \digitwidth=\wd0
  \catcode`*=\active
  \def*{\kern\digitwidth}
  \newdimen\signwidth
  \setbox0=\hbox{+}
  \signwidth=\wd0
  \catcode`!=\active
  \def!{\kern\signwidth}
  \halign{\hbox to 1.31in{#\leaderfil}\tabskip=1em&
    \hfil#\hfil\tabskip=0pt\cr
    \noalign{\doubleline}
%    \noalign{\vskip 3pt\hrule\vskip 5pt}
    $T_\mathrm{CMB} \mathrm{[K]}$& 2.7255\cr
    $h$& 0.6701904\cr
    $\Omega_{\rm m}$& 0.26782\cr
    $\Omega_{\rm b}$& 0.0493498\cr
    $\Omega_{\rm k}$& 0\cr
    $\sigma_8$& 0.8162\cr
    $n_{\rm s}$& 0.9636\cr
    $r$& 0.01\cr
    $\tau$& 0.060\cr
    $Y_{\rm P}$& 0.2453\cr
    $w$& $-1$\cr
    $10^9A_{\rm s}$& 2.119\cr
    \noalign{\vskip 5pt\hrule\vskip 3pt}
  }}
  \endPlancktable
  \endgroup
\end{table}

\subsubsection{CMB}

The CMB is a stationary Gaussian random field on the sphere, generated
from {\tt CLASS} output power spectra using the {\tt HEALPix}
package, and lensed using the {\tt Ilens} software
of \citet{basak2009}. The joint generation of density contrast shells
used for the CIB simulation and of the lensing potential map provides 
correlation between the CMB lensing and the CIB maps. This correlation
has been used to generate 25 independent CMB and CIB realizations.

\subsubsection{Unresolved sources and the cosmic infrared background}

Catalogues of individual radio and low-redshift infrared sources are
generated in the same way as for the FFP8 simulations
\citep{planck2014-a14}, but use a different seed for random number generation.
The generation of the cosmic infrared background (CIB), due to the integrated
emission of tens of billions of distant dusty galaxies, was substantially revised
to allow for the simulation of correlations between the lensing potential maps
and the CIB emission. Number counts for three types of galaxies --- early-type
proto-spheroids, along with spiral and starburst galaxies --- are based on
the model of \citet{cai2013}.  The entire Hubble volume up to $z=6$ is
cut into 64 spherical shells. For each shell, we generate a map of density
contrast integrated along the line of sight between $z_{\rm min}$ and $z_{\rm max}$,
such that the statistics of these density contrast maps (power spectrum of linear
density fluctuations, and cross-spectra between adjacent shells and with the CMB
lensing potential) agree with those computed by {\tt CLASS}. For each type of galaxy (spiral,
starburst, proto-spheroid), a catalogue of randomly-generated galaxies is generated
for each shell, following the appropriate number counts.  These galaxies are then
distributed in the shell to generate a single intensity map at a given reference
frequency, which is scaled across frequencies using the prototype galaxy spectral energy distribution at
the appropriate redshift. 

\subsubsection{Galaxy clusters}

A full-sky catalogue of galaxy clusters is generated based on number counts,
following the method of \citet{delabrouille2002}. The mass function of
\citet{tinker2008} is used to predict number counts. Clusters are distributed in
redshift shells proprotionally to the denstity contrast in each pixel with
a bias $b(z,M)$ in agreement with the linear bias model of \citet{MoWhite}.
For each cluster,
we assign a universal profile based on XMM observations, as described in
\citet{arnaud2010}. Relativistic corrections are included to first order,
following the expansion of \citet{wangl1998}. We use a mass
bias $M_{\mathrm{X-Ray}}/M_{\mathrm{mass-fn}}=0.63$ to match actual cluster
number counts observed by \Planck\ for the best-fit cosmological model based on CMB observations.

The kinematic SZ effect is computed assigning to each cluster a radial
velocity that is randomly drawn from a centred Gaussian distribution, with a
redshift dependent standard deviation computed from the power spectrum of
density fluctuations. This neglects correlations between cluster motions such
as bulk flows or pairwise velocities of nearby clusters.

\subsection{Bandpass integration}

The model of sky emission is finally integrated in frequency according to
\Planck\, bandpasses from the HFI and LFI Reduced Instrument Models (RIMO,
version R2.00 for HFI and R2.50 for LFI).

\subsection{From TOI to gains and maps}

\begin{figure}[htpb]
  \centerline{
    \includegraphics[width=9cm]{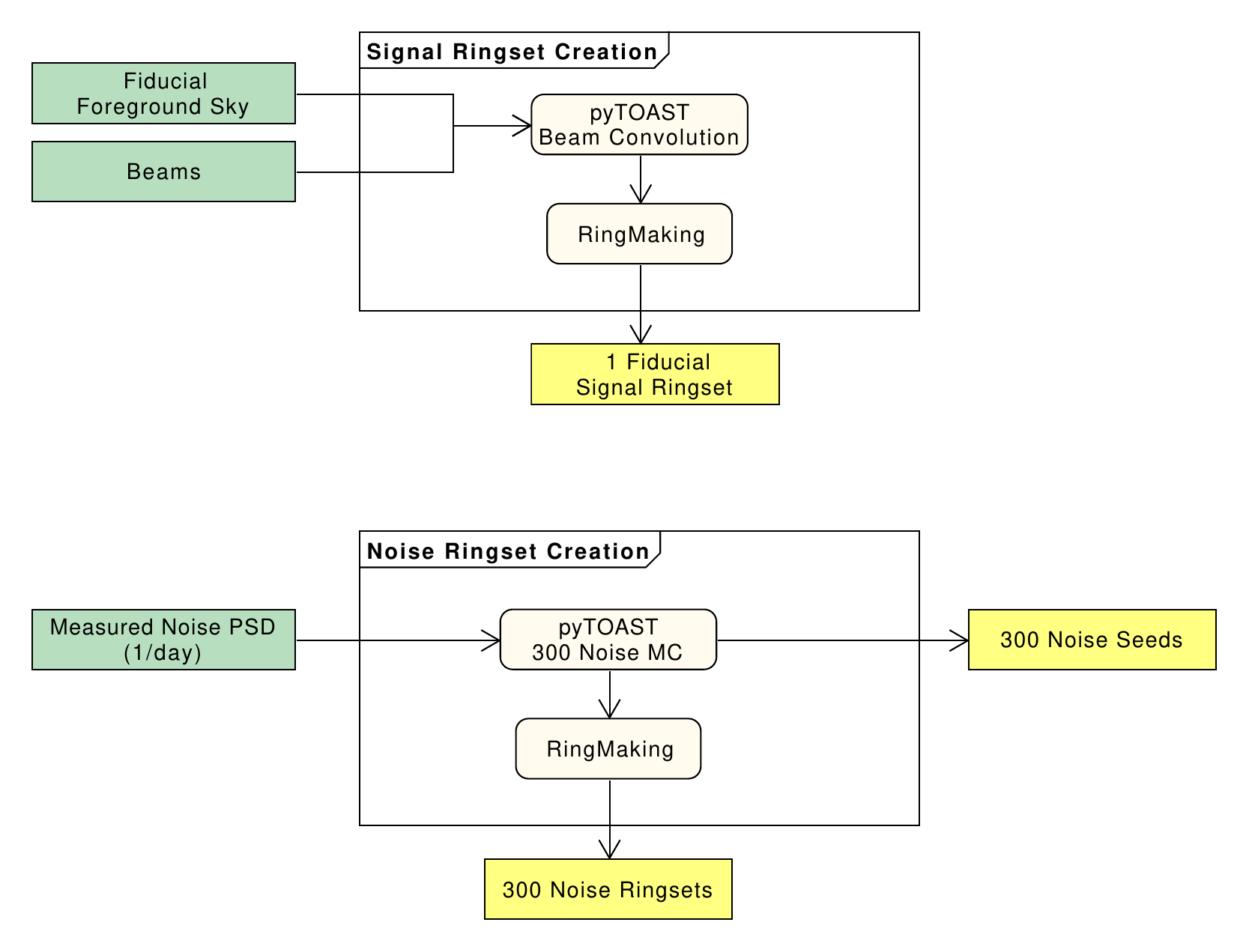}}
  \caption{Signal and noise ring-set creation pipelines.}
  \label{sim_step1}
\end{figure}

The LFI systematic effect simulations are done partially at time-line and
partially at ring-set level, with the goal of being as modular as possible,
in order to create a reusable set of simulations.  From the input sky model
and according to the pointing information, we create single-channel ring-sets
of the pure sky convolved with a suitable instrumental beam.  To these we
add pure noise (white and $1/f$) ring-sets generated from the noise power
spectrum distributions measured from real data one day at a time. The
overall scheme is given in Fig.~\ref{sim_step1}.

\begin{figure}[htpb]
  \centerline{
    \includegraphics[width=9cm]{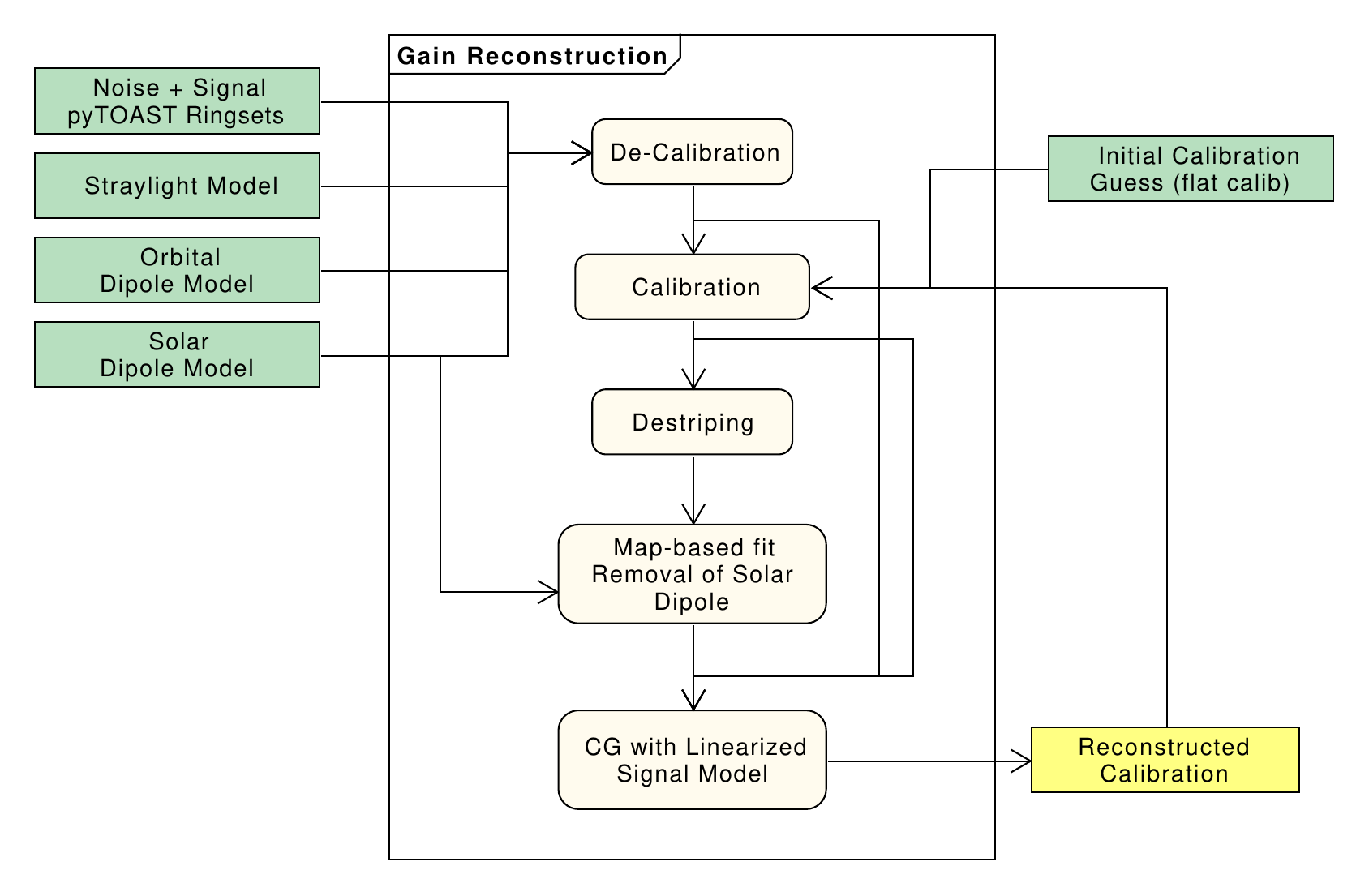}}
  \caption{Gain reconstruction pipeline.}
  \label{sim_step2}
\end{figure}

In the same manner, we create ring-sets for each of the specific systematic
effects we would like to measure.  We add together signal, noise, and systematic
ring-sets, and, given models for straylight (based on the {\tt{GRASP}} beams) and
the orbital dipole, we create ``perfectly-calibrated'' ring-sets (calibration constant = 1).
We use the gains estimate from the 2018 data release to ``de-calibrate''
these timelines, in other words, to convert them from kelvins to volts.  At this point
the calibration pipeline starts, and produces the reconstructed gains that
will be different from the ones used in the de-calibration process due to
the presence of simulated systematic effects.  The calibration pipeline is
algorithmically exactly the same as that used at the DPC for product creation,
but with a different implementation (based principally on {\tt{python}}).
The gain-smoothing algorithm is the same as used for the data, and has
been tuned to the actual data.  This means that there will be cases where
reconstructed gains from simulations differ significantly from the input
ones. We have verified that this indeed happens, but only for very few
pointing periods, and we therefore decided not to consider them in the
following analysis. The overall process for estimating gains is given in Fig.~\ref{sim_step2}.

\begin{figure}[htpb]
  \centerline{
    \includegraphics[width=9cm]{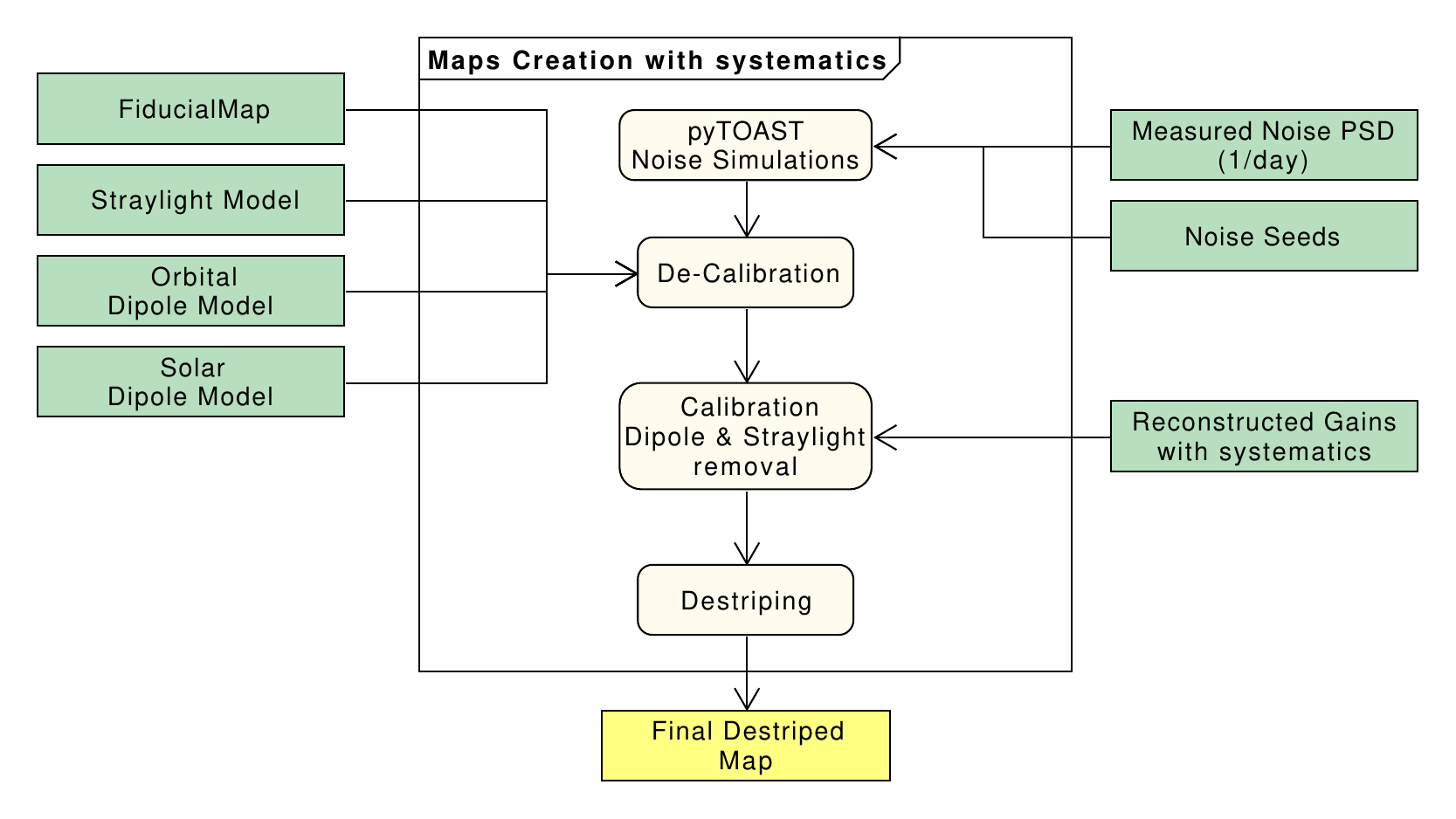}}
    \caption{Final simulation step, where calibrated maps are created from
      reconstructed gains, including the impact of systematic effects.}
    \label{sim_step3}
\end{figure}

At this point we are able to generate maps for full mission, half-ring,
and odd-even-year splits) that include the effects of systematic errors
on calibration.  In the final step, we produce timelines (which are never
stored) starting from the same fiducial sky map, using the same model for
straylight and the orbital dipole as in the previous steps, and from
generated noise-only timelines created with the same seeds and noise model
used before.  We then apply the official gains to ``de-calibrate'' the timelines,
which are immediately calibrated with the reconstructed gains in the previous step.
The nominal destriping mapmaking algorithm is then used to create final maps.
The complete data flow is given in Fig.~\ref{sim_step3}.

\end{appendix}

\end{document}